\documentclass[aps,prl,twocolumn,showpacs,superscriptaddress,letterpaper,showkeys,superscriptaddress,floatfix,nofootinbib]{revtex4-1}
\usepackage{graphicx}  
\usepackage{dcolumn}   
\usepackage{bm}        
\usepackage{adjustbox}
\usepackage{amssymb}   
\usepackage[caption=false]{subfig}
\usepackage{xcolor}
\usepackage{subfig}
\usepackage{amsmath}
\usepackage{subfiles}
\usepackage{placeins}
\usepackage{array}
\usepackage[normalem]{ulem}
\usepackage{xspace}
\usepackage{lipsum}
\usepackage{lineno}
\usepackage{float}
\usepackage{hyperref}

\makeatletter
\newcommand{\figwidth}{%
  \ifdim\columnwidth=\textwidth
    0.35\columnwidth
  \else
    0.49\columnwidth
  \fi
}
\makeatother

\newcommand{\showSections}[0]{0}

\newcommand{\supp}{the supplemental material}

\newif\ifPRLandSupp
\PRLandSupptrue

\newif\ifColorFigures
\ColorFigurestrue

\makeatletter
\renewcommand{\p@subsection}{}
\makeatother

\begin{document}
\newcommand{\minerva}{MINERvA\xspace}
\newcommand{\qelike}{quasielastic-like}
\newcommand{\tune}{\minerva~tune v1}
\newcommand{\ccnopi}{CC0$\pi$\xspace}
\newcommand{\enuqe}{$E_{\nu,QE}$\xspace}
\newcommand{\pt}{$p_{t}$\xspace}
\newcommand{\ptmu}{$p_{T \mu}$\xspace}
\newcommand{\pz}{$p_{||}$\xspace}
\newcommand{\pzmu}{$p_{|| \mu}$\xspace}
\newcommand{\eavail}{$E_{available}$\xspace}
\newcommand{\sumtp}{$\Sigma T_{p}$\xspace}
\newcommand{\qqqe}{$Q^{2}_{QE}$\xspace}
\newcommand{\addcitation}{[ADD CITATION]}
\newcommand{\Emu}{\ensuremath{E_{\mu}}\xspace}
\newcommand{\pmu}{\ensuremath{p_{\mu}}\xspace}
\newcommand{\thmu}{\ensuremath{\theta_{\mu}}\xspace}
\newcommand{\Ehad}{\ensuremath{E_{\mathrm{had}}}\xspace}
\newcommand{\Erec}{\ensuremath{E_{recoil}}\xspace}
\newcommand{\Tpi}{\ensuremath{T_{\pi}}\xspace}
\newcommand{\thpi}{\ensuremath{\theta_{\pi}}\xspace}
\newcommand{\W}{\ensuremath{W}\xspace}
\newcommand{\Wexp}{\ensuremath{W_{\mathrm{exp}}}\xspace}
\newcommand{\Enu}{\ensuremath{E_{\nu}}\xspace}
\newcommand{\Qsq}{\ensuremath{Q^{2}}\xspace}
\newcommand{\Qsqexp}{\ensuremath{Q^{2}_{exp}}\xspace}
\newcommand{\datapot}{\ensuremath{\mbox{9.99}\times\mbox{10}^{999}}\xspace}

\newcommand{\nue}{\ensuremath{\nu_{e}}\xspace}
\newcommand{\numu}{\ensuremath{\nu_{\mu}}\xspace}
\newcommand{\nutau}{\ensuremath{\nu_{\tau}}\xspace}
\newcommand{\nuebar}{\ensuremath{\overline{\nu}_{e}}\xspace}
\newcommand{\numubar}{\ensuremath{\overline{\nu}_{\mu}}\xspace}
\newcommand{\nutaubar}{\ensuremath{\overline{\nu}_{\tau}}\xspace}
\newcommand{\nuone}{\ensuremath{\nu_{1}}\xspace}
\newcommand{\nutwo}{\ensuremath{\nu_{2}}\xspace}
\newcommand{\nuthree}{\ensuremath{\nu_{3}}\xspace}
\newcommand{\pion}{\ensuremath{\pi}\xspace}
\newcommand{\piplus}{\ensuremath{\pi^{+}}\xspace}
\newcommand{\pip}{\ensuremath{\pi^{+}}\xspace}
\newcommand{\piminus}{\ensuremath{\pi^{-}}\xspace}
\newcommand{\pim}{\ensuremath{\pi^{-}}\xspace}
\newcommand{\pipm}{\ensuremath{\pi^{\pm}}\xspace}
\newcommand{\pizero}{\ensuremath{\pi^{0}}\xspace}
\newcommand{\kplus}{\ensuremath{\mbox{K}^{+}}\xspace}
\newcommand{\kap}{\ensuremath{\mbox{K}^{+}}\xspace}
\newcommand{\kminus}{\ensuremath{\mbox{K}^{-}}\xspace}
\newcommand{\kam}{\ensuremath{\mbox{K}^{-}}\xspace}
\newcommand{\kpm}{\ensuremath{\mbox{K}^{\pm}}\xspace}
\newcommand{\kzero}{\ensuremath{\mbox{K}^{0}}\xspace}
\newcommand{\kzeroL}{\ensuremath{\mbox{K}^{0}_{\mathrm{L}}}\xspace}
\newcommand{\prot}{\ensuremath{\mbox{p}}\xspace}
\newcommand{\neut}{\ensuremath{\mbox{n}}\xspace}
\newcommand{\muplus}{\ensuremath{\mu^{+}}\xspace}
\newcommand{\mup}{\ensuremath{\mu^{+}}\xspace}
\newcommand{\muminus}{\ensuremath{\mu^{-}}\xspace}
\newcommand{\mum}{\ensuremath{\mu^{-}}\xspace}
\newcommand{\mupm}{\ensuremath{\mu^{\pm}}\xspace}
\newcommand{\muon}{\ensuremath{\mu}\xspace}
\newcommand{\eplus}{\ensuremath{\mbox{e}^{+}}\xspace}
\newcommand{\eminus}{\ensuremath{\mbox{e}^{-}}\xspace}
\newcommand{\deltabaryonp}{\ensuremath{\Delta^{+}}\xspace}
\newcommand{\deltabaryonpp}{\ensuremath{\Delta^{++}}\xspace}

\newcommand{\kg}{\ensuremath{\mbox{kg}}\xspace}

\newcommand{\kA}{\ensuremath{\mbox{kA}}\xspace}

\newcommand{\rad}{\ensuremath{\mbox{rad}}\xspace}
\newcommand{\mrad}{\ensuremath{\mbox{mrad}}\xspace}
\newcommand{\degs}{\textdegree\xspace}

\newcommand{\eV}{\ensuremath{\mbox{eV}}\xspace}
\newcommand{\keV}{\ensuremath{\mbox{keV}}\xspace}
\newcommand{\MeV}{\ensuremath{\mathrm{MeV}}\xspace}
\newcommand{\GeV}{\ensuremath{\mathrm{GeV}}\xspace}
\newcommand{\GeVsq}{\ensuremath{\mathrm{GeV}^2}\xspace}
\newcommand{\eVc}{\ensuremath{\mathrm{eV/\textit{c}}}\xspace}
\newcommand{\MeVc}{\ensuremath{\mathrm{MeV/\textit{c}}}\xspace}
\newcommand{\GeVc}{\ensuremath{\mathrm{GeV/\textit{c}}}\xspace}
\newcommand{\GeVcsq}{\ensuremath{\mathrm{GeV/\textit{c}}^2}\xspace}
\newcommand{\GeVsqcsq}{\ensuremath{\mathrm{GeV}^2/\mathrm{\textit{c}}^2}\xspace}

\newcommand{\eVsq}{\ensuremath{\mbox{eV}^2}\xspace}
\newcommand{\eVsqbf}{\ensuremath{\mathbf{\mbox{eV}^2}}\xspace}

\newcommand{\T}{\ensuremath{\mbox{T}}\xspace}
\newcommand{\mmsq}{\ensuremath{\mbox{mm}^2}\xspace}
\newcommand{\cmsq}{\ensuremath{\mbox{cm}^2}\xspace}
\newcommand{\cmcube}{\ensuremath{\mbox{cm}^3}\xspace}
\newcommand{\gcmcube}{\ensuremath{\mbox{g/cm}^3}\xspace}
\newcommand{\msq}{\ensuremath{\mbox{m}^2}\xspace}
\newcommand{\mcube}{\ensuremath{\mbox{m}^3}\xspace}

\newcommand{\micron}{\ensuremath{\mu \mbox{m}}\xspace}
\newcommand{\nm}{\ensuremath{\mbox{nm}}\xspace}
\newcommand{\mm}{\ensuremath{\mbox{mm}}\xspace}
\newcommand{\cm}{\ensuremath{\mbox{cm}}\xspace}
\newcommand{\m}{\ensuremath{\mbox{m}}\xspace}
\newcommand{\km}{\ensuremath{\mbox{km}}\xspace}

\newcommand{\ps}{\ensuremath{\mbox{ps}}\xspace}
\newcommand{\ns}{\ensuremath{\mbox{ns}}\xspace}
\newcommand{\micros}{\ensuremath{\mu \mbox{s}}\xspace}
\newcommand{\ms}{\ensuremath{\mbox{ms}}\xspace}
\newcommand{\s}{\ensuremath{\mbox{s}}\xspace}

\newcommand{\bkgdtype}{_MENU1PI}
\newcommand{\Rutgers}{Rutgers, The State University of New Jersey, Piscataway, New Jersey 08854, USA}
\newcommand{\Hampton}{Hampton University, Dept. of Physics, Hampton, VA 23668, USA}
\newcommand{\Dortmund}{Institute of Physics, Dortmund University, 44221, Germany }
\newcommand{\Otterbein}{Department of Physics, Otterbein University, 1 South Grove Street, Westerville, OH, 43081 USA}
\newcommand{\JMU}{James Madison University, Harrisonburg, Virginia 22807, USA}
\newcommand{\Florida}{University of Florida, Department of Physics, Gainesville, FL 32611}
\newcommand{\UCIrvine}{Department of Physics and Astronomy, University of California, Irvine, Irvine, California 92697-4575, USA}
\newcommand{\CBPF}{Centro Brasileiro de Pesquisas F\'{i}sicas, Rua Dr. Xavier Sigaud 150, Urca, Rio de Janeiro, Rio de Janeiro, 22290-180, Brazil}
\newcommand{\PUCP}{Secci\'{o}n F\'{i}sica, Departamento de Ciencias, Pontificia Universidad Cat\'{o}lica del Per\'{u}, Apartado 1761, Lima, Per\'{u}}
\newcommand{\INRM}{Institute for Nuclear Research of the Russian Academy of Sciences, 117312 Moscow, Russia}
\newcommand{\Jlab}{Jefferson Lab, 12000 Jefferson Avenue, Newport News, VA 23606, USA}
\newcommand{\Pittsburgh}{Department of Physics and Astronomy, University of Pittsburgh, Pittsburgh, Pennsylvania 15260, USA}
\newcommand{\Guanajuato}{Campus Le\'{o}n y Campus Guanajuato, Universidad de Guanajuato, Lascurain de Retana No. 5, Colonia Centro, Guanajuato 36000, Guanajuato M\'{e}xico.}
\newcommand{\Athens}{Department of Physics, University of Athens, GR-15771 Athens, Greece}
\newcommand{\Tufts}{Physics Department, Tufts University, Medford, Massachusetts 02155, USA}
\newcommand{\WM}{Department of Physics, William \& Mary, Williamsburg, Virginia 23187, USA}
\newcommand{\FNAL}{Fermi National Accelerator Laboratory, Batavia, Illinois 60510, USA}
\newcommand{\Purdue}{Department of Chemistry and Physics, Purdue University Calumet, Hammond, Indiana 46323, USA}
\newcommand{\MCLA}{Massachusetts College of Liberal Arts, 375 Church Street, North Adams, MA 01247}
\newcommand{\UMD}{Department of Physics, University of Minnesota -- Duluth, Duluth, Minnesota 55812, USA}
\newcommand{\Northwestern}{Northwestern University, Evanston, Illinois 60208}
\newcommand{\UNI}{Facultad de Ciencias, Universidad Nacional de Ingenier\'{i}a, Apartado 31139, Lima, Per\'{u}}
\newcommand{\Rochester}{Department of Physics and Astronomy, University of Rochester, Rochester, New York 14627 USA}
\newcommand{\Austin}{Department of Physics, University of Texas, 1 University Station, Austin, Texas 78712, USA}
\newcommand{\USM}{Departamento de F\'{i}sica, Universidad T\'{e}cnica Federico Santa Mar\'{i}a, Avenida Espa\~{n}a 1680 Casilla 110-V, Valpara\'{i}so, Chile}
\newcommand{\Geneva}{University of Geneva, 1211 Geneva 4, Switzerland}
\newcommand{\Chicago}{Enrico Fermi Institute, University of Chicago, Chicago, IL 60637 USA}
\newcommand{\hired}{}
\newcommand{\OregonState}{Department of Physics, Oregon State University, Corvallis, Oregon 97331, USA}
\newcommand{\oxford}{Oxford University, Department of Physics, Oxford, OX1 3PJ United Kingdom}
\newcommand{\umiss}{University of Mississippi, Oxford, Mississippi 38677, USA}
\newcommand{\upenn}{Department of Physics and Astronomy, University of Pennsylvania, Philadelphia, PA 19104}
\newcommand{\AMU}{AMU Campus, Aligarh, Uttar Pradesh 202001, India}
\newcommand{\wroclaw}{University of Wroclaw, plac Uniwersytecki 1, 50-137 Wroa\l{}aw, Poland}
\newcommand{\Mohali}{Department of Physical Sciences, IISER Mohali, Knowledge City, SAS Nagar, Mohali - 140306, Punjab, India}
\newcommand{\CINVESTAV}{Departamento de Fisica Col. San Pedro Zacatenco, 07360 Mexico, DF, Av. Instituto PolitÃ©cnico Nacional, Mexico}
\newcommand{\york}{York University, Department of Physics and Astronomy, Toronto, Ontario, M3J 1P3 Canada}
\newcommand{\ND}{Department of Physics, University of Notre Dame, Notre Dame, Indiana 46556, USA}
\newcommand{\ICL}{The Blackett Laboratory,  Imperial College London,  London SW7 2BW, United Kingdom}
\newcommand{\warwick}{Department of Physics, University of Warwick, Coventry, CV4 7AL, UK}
\newcommand{\mascencioThanks}{Now at Iowa State University, Ames, IA 50011, USA}
\newcommand{\ricfregianThanks}{now at Department of Physics and Astronomy, University of California at Davis, Davis, CA 95616, USA}
\newcommand{\finerThanks}{Now at Los Alamos National Laboratory, Los Alamos, New Mexico 87545, USA}
\newcommand{\kleykampThanks}{now at Department of Physics and Astronomy, University of Mississippi, Oxford, MS 38677}
\newcommand{\bamThanks}{Now at University of Minnesota, Minneapolis, Minnesota 55455, USA}
\newcommand{\byaeggyThanks}{Now at Department of Physics, University of Cincinnati,  Cincinnati, Ohio 45221, USA}
\newcommand{\everThanks}{Now at Florida State University, Tallahassee, Florida 32306, USA}
\newcommand{\qmul}{G O Jones Building, Queen Mary University of London, 327 Mile End Road, London E1 4NS, UK}
\newcommand{\LLNL}{Nuclear and Chemical Sciences Division, Lawrence Livermore National Laboratory, Livermore, CA 94550, USA}

\newcommand{\aolivierThanks}{now at Argonne National Laboratory, Lemont, IL 60439 USA}
\newcommand{\lazazuetareyesThanks}{now at Syracuse University, Syracuse, NY 13244, USA}


\title{High Statistics Measurements of \numu Charged-Current Single \pip Production with Zero Pion Kinetic Energy Threshold in \minerva}

\author{E.~Granados} \thanks
{\everThanks} \affiliation{\Guanajuato}
\author{B.~Messerly} \thanks{\bamThanks} \affiliation{\Pittsburgh}
\author{S.~Akhter}                        \affiliation{\AMU}
\author{M.~Sajjad~Athar}                  \affiliation{\AMU}
\author{S.A.~Dytman}                      \affiliation{\Pittsburgh}
\author{J.~Felix}                         \affiliation{\Guanajuato}
\author{L.~Fields} \affiliation{\ND}
\author{P.K.Gaur}                         \affiliation{\AMU}
\author{S.M.~Gilligan}                    \affiliation{\OregonState}
\author{R.~Gran}                          \affiliation{\UMD}
\author{D.A.~Harris}                      \affiliation{\york}  \affiliation{\FNAL}
\author{A.L.~Hart}                        \affiliation{\qmul}
\author{J.~Kleykamp}\thanks{\kleykampThanks}  \affiliation{\Rochester}
\author{A.~Klustov\'{a}}                  \affiliation{\ICL}
\author{M.~Kordosky}                      \affiliation{\WM}
\author{D.~Last}                          \affiliation{\Rochester}  \affiliation{\upenn}
\author{S.~Manly}                         \affiliation{\Rochester}
\author{W.A.~Mann}                        \affiliation{\Tufts}
\author{K.S.~McFarland}                   \affiliation{\Rochester}
\author{O.~Moreno}                        \affiliation{\WM}  \affiliation{\Guanajuato}
\author{J.G.~Morf\'{i}n}                  \affiliation{\FNAL}
\author{A.~Olivier}\thanks{\aolivierThanks}  \affiliation{\ND}  \affiliation{\Rochester}
\author{V.~Paolone}                       \affiliation{\Pittsburgh}
\author{G.N.~Perdue}                      \affiliation{\FNAL}  \affiliation{\Rochester}
\author{C.~Pernas}                        \affiliation{\WM}
\author{M.A.~Ram\'{i}rez}                 \affiliation{\upenn}  \affiliation{\Guanajuato}
\author{N.~Roy}                           \affiliation{\york}
\author{D.~Ruterbories}                   \affiliation{\Rochester}
\author{C.J.~Solano~Salinas}              \affiliation{\UNI}
\author{M.~Sultana}                       \affiliation{\Rochester}
\author{N.H.~Vaughan}                     \affiliation{\OregonState}
\author{A.V.~Waldron}                     \affiliation{\qmul}  \affiliation{\ICL}
\author{M.O.~Wascko}                      \affiliation{\oxford}  \affiliation{\ICL}
\author{B.~Yaeggy}\thanks{\byaeggyThanks}  \affiliation{\USM}
\author{L.~Zazueta}\thanks{\lazazuetareyesThanks}  \affiliation{\WM}

\collaboration{The MINERvA Collaboration}\ \noaffiliation

\date{\today}
\begin{abstract}
This Letter presents measurements of single-differential cross sections of \numu-induced charged-current 1\pip production on scintillator using the MINERvA detector at Fermilab. These  measurements use traditional track-based pion reconstruction as well as pions identified solely via Michel electron decays, allowing measurement of kinetic energies from 0 to 350 \MeV. In total, 91,843 events were selected with \Wexp $<$ 1.4 GeV/c. Differential cross sections as a function of pion and muon kinematic variables are presented and compared with the predictions of several neutrino event generators. Overall, modern pion production models tend to agree with data at the ends of the kinematic regions probed, but are discrepant with the main regions of the phase space probed by up to 15\% in muon observables and up to 20\% in pion observables. No model describes any of the variables well, and this result highlights model areas that require improvement for the next generation of neutrino oscillation experiments. 

\end{abstract}

\maketitle

\if\showSections1 \section{Motivation} \fi

Pion production via resonances dominates the neutrino-induced charged-current (CC) interaction cross section between \Enu $\sim$ 1.5--4.5 \GeV~\cite{Formaggio:2013kya}.   Accurate models of these interactions are necessary in order for experiments such as NOvA \cite{PhysRevD.106.032004}, T2K \cite{Abe:2011ks}, HyperK \cite{Hyper-Kamiokande:2018ofw}, and DUNE \cite{DUNE:2015lol} to extract neutrino oscillation parameters from their datasets.  These models must account for both the primary neutrino-nucleon interactions as well as a variety of effects created by the dense nuclear environment. Their verity hinges on availability of precise measurements and their implementation requires validation and tuning. This Letter presents new measurements of \numu-induced CC 1\pip production on scintillator using the \minerva detector~\cite{Aliaga:2013uqz}. It follows previous measurements of this process by \minerva~\cite{Eberly:2014mra,McGivern:2016bwh,Bercellie.131.011801} but adds pions reconstructed solely via Michel electron decays, which enable a measurement without a pion energy threshold for the first time, and substantially reduces statistical uncertainties.

\if\showSections1 \section{\minerva Detector and Beam} \fi

This measurement uses $1.06\times 10^{21}$ protons on target (POT) collected in the NuMI neutrino beam~\cite{Anderson:1998zza,Adamson:2015dkw}.  The beam was in the ``Medium Energy" configuration, where 120 \GeVc protons strike a 1.2 \m long graphite target, producing a neutrino spectrum peaked at \Enu$\sim$ 6--7 \GeV at the on-axis \minerva detector located ~1 km downstream of the target.  

The \minerva detector~\cite{Aliaga:2013uqz} is composed of 208 hydrocarbon scintillator planes (16.6 mm thick) interspersed with various target materials. This analysis uses interactions produced in the central tracking region which contains only scintillator. Downstream of this region are an electromagnetic and hadronic calorimeter consisting of scintillator planes separated by thin layers of lead and iron, respectively. A target region, located upstream of the central tracking and containing passive helium, carbon, iron, lead, and water targets, is not used for this measurement, but pion production on these targets has been published in \cite{Bercellie.131.011801}. The MINOS near detector~\cite{MINOS:2008hdf} sits 2.1 meters downstream of \minerva. It serves as a spectrometer for muons exiting \minerva, and is used to determine whether the interaction originated from a neutrino or an antineutrino.
 
\if\showSections1 \section{Simulation} \fi

A detailed simulation of the \minerva experiment is used to estimate backgrounds and correct for detector efficiency, acceptance, and smearing. This simulation begins with a Geant4-based simulation~\cite{1610988,Agostinelli2003250} of the NuMI beamline and neutrino flux, which are tuned by external hadron-production data~\cite{Aliaga:2016oaz} and by \minerva measurements of neutrino-electron elastic scattering and inverse muon decay~\cite{MINERvA:2021dhf,PhysRevD.107.012001}.  The \minerva detector simulation is also based on Geant4, and the predicted detector response is constrained by data from a scaled-down version of \minerva operated in the Fermilab Test Beam Facility~\cite{Aliaga:2015aqe}.
\begin{figure*}[!bhtp]
    
    \centering
    \includegraphics[width=.90\linewidth]{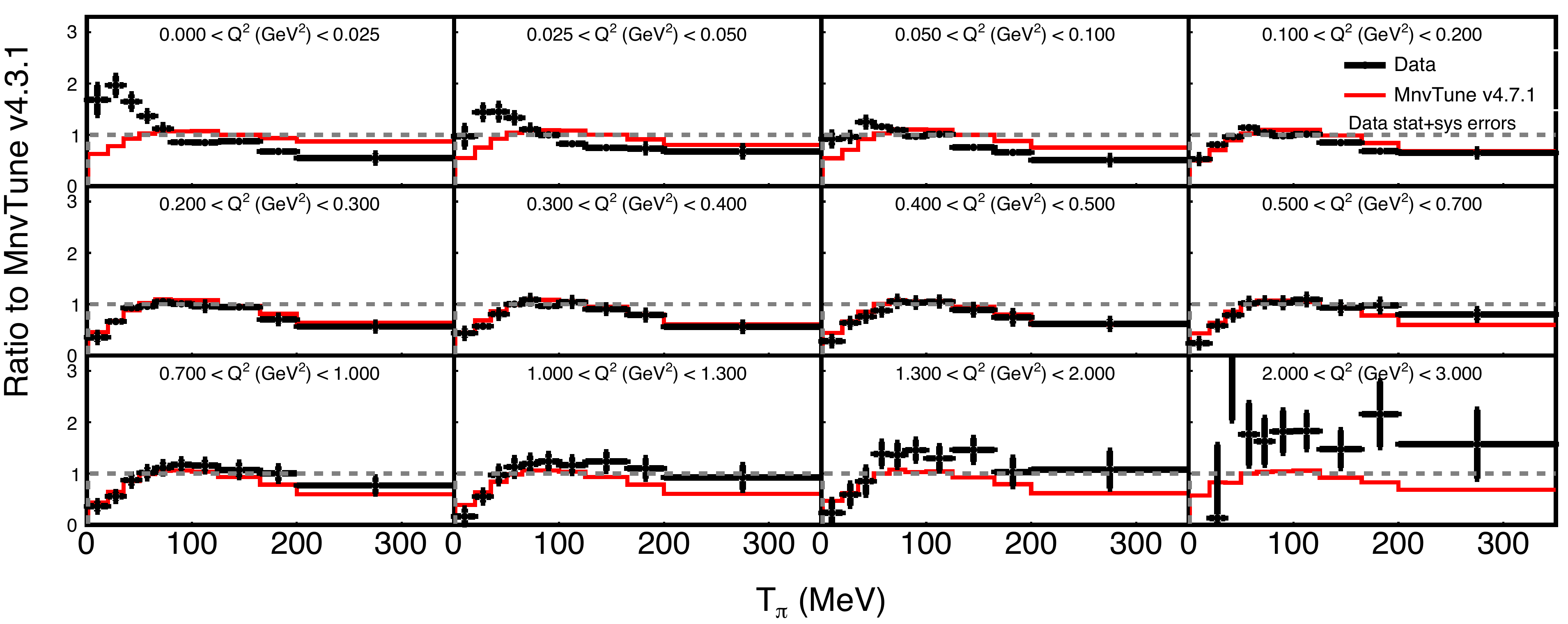}
    \caption{Ratios of predicted event distributions after background subtraction for data (black) and MnvTune v4.7.1 (red) to MnvTune v4.3.1.  } 
    \label{fig:tune_impact}
\end{figure*}

Neutrino interactions are simulated with a GENIE 2.12.6 base model~\cite{Andreopoulos201087,Andreopoulos:2009rq}, then modified with an empirical \minerva tune, MnvTune v4.7.1, which has been tuned to a variety of MINERvA data, including the data analyzed in this paper. The nuclear base model uses a Relativistic Fermi Gas (RFG)~\cite{Hassanabadi:2014kba} representation of the nuclear ground state with the Bodek-Ritchie high-momentum tail \cite{PhysRevD.23.1070,PhysRevD.24.1400}. Intranuclear rescattering is simulated by INTRANUKE-hA~\cite{dytman2008hadronicshowerenergyscale}. The quasielastic (QE) and 2-particle 2-hole (2p2h) scattering models employed are described in~\cite{MINERvA:2023kuz}; these processes produce pions only through Final State Interactions (FSI) and are predicted to constitute $<$ 1\% of the selected sample. The Bodek-Yang model~\cite{Bodek:2004pc} simulates Deep Inelastic Scattering (DIS) for events with invariant mass ($W$) beyond 1.4 GeV, including contributions to low square of the momentum transferred (\Qsq). Coherent pion production constitutes a small portion of this analysis's signal and is simulated using the Rein-Sehgal model~\cite{Rein:1982pf}.

Resonant pion production, the process to which this analysis is primarily sensitive, is simulated by using the Rein-Sehgal~\cite{Rein:1980wg} model with an axial mass of 0.94 \GeVcsq---based on re-analyzed deuterium bubble chamber data~\cite{Rodrigues:2016xjj,Mislivec:2017qfz}. From the same data, non-resonant pion production normalization is reduced to 43\% of the base model value~\cite{Rodrigues:2016xjj}, and resonant pion production normalization is increased by 15\%. A mistake present in GENIE's model of elastic FSI in pions and protons is also bypassed~\cite{Harewood:2019rzy} by removing elastic FSI from the model. The energy and angle distributions of coherent pions are adjusted to match \minerva data~\cite{PhysRevLett.131.051801}. The coherent events in the simulation are weighted up by 43.7\% to account for diffractive charged pion production on hydrogen, based on the Kopeliovich \cite{Kopeliovich:2012tu} model. A weight as a function of the four-momentum transfer squared (\Qsq) is applied to non-coherent pion production on non-hydrogen nuclei to suppress rates in the lowest bins of \Qsq based on the MINERvA data described in~\cite{Bercellie.131.011801}. This weight has the impact of aligning pion production simulation more closely with modern versions of generators such as GENIE 3~\cite{Andreopoulos201087} and NEUT~\cite{Hayato:2009zz}.

The simulation described up to this point specifies MnvTune v4.3.1 and was employed in \minerva's previous CC 1\pip result~\cite{Bercellie.131.011801}. This analysis differs from~\cite{Bercellie.131.011801} by extending the signal pion kinetic energy (\Tpi) region down to \Tpi $\sim$ 0. With the inclusion of these low-\Tpi events, substantial discrepancies between data and MnvTune v4.3.1 were observed, and thus an additional weight is applied that modifies 1 pion production as a function of true \Tpi and \Qsq. The result is MnvTune v4.7.1. Its impact, relative to v4.3.1, is shown in Fig.\ref{fig:tune_impact}.

\if\showSections1 \section{Signal Definition} \fi

The signal process of this analysis is defined as a \numu CC interaction inside of the inner tracking volume, producing a single \muminus, a single \pip, any number of baryons, and no other mesons in the final state. To isolate a primarily $\Delta(1232)$ resonance signal, the invariant mass \Wexp is required to be less than 1.4 \GeVcsq.  This is derived assuming the nucleon is at rest from experimentally observable quantities:
\begin{eqnarray}
E_{\nu} &=& \Ehad + E_{\mu}
    \label{eq:nu}\\
    Q^2 &=& 2E_\nu\left(E_\mu - \left|\vec{p}_\mu\right|\cos\theta_{\mu\nu}\right) - m_\mu^2 \label{eq:q2} \\
    \Wexp^2 &=& m_N^2 - \Qsq + 2m_N\Ehad \label{eq:wexp}.
\end{eqnarray}
Here, \Enu is the incoming neutrino energy, \Qsq is the square of the momentum transferred from the neutrino to the hadronic system, $m_N$ is the average of the proton and neutron masses, \Ehad is the total energy of final state hadrons, and $E_\mu$, $\vec{p}_\mu$, $\theta_{\mu\nu}$, and $m_\mu$ are the energy, momentum, angle with respect to the neutrino beam, and mass of the final state muon, respectively.

Kinematic constraints corresponding to detector acceptance limitations are further imposed on the signal process. Muon momentum is limited to 1.5 $<$ \pmu $<$ 20 \GeVc and muon angle to \thmu $<$ 20\degs. High energy pions tend to exit or interact in the detector, degrading their energy measurement, and thus \Tpi is required to be less than 350 \MeV.

For the \thpi differential cross section measurement only, an additional signal constraint requiring \Tpi $>$ 20 \MeV is imposed. Below this energy, tracks are too short to be reconstructed, pions are identified from their Michel electron candidate signatures alone, and \thpi cannot be resolved.

\if\showSections1 \section{Event Selection and Reconstruction} \fi

Within a 10 $\mu$s NuMI spill, multiple neutrino interactions may occur in the MINERvA detector, each producing energy deposition (``hits'') in MINERvA scintillator strips from charged particles. Hits within a single spill are grouped into time slices that range from 24 \ns to 100 \ns, corresponding to energy depositions from a single neutrino interaction. Within each time slice, hits on neighboring scintillator strips are grouped into clusters and tracking algorithms are used to reconstruct charged particle tracks. Tracks that are matched to corresponding tracks in the MINOS detector are assumed to be muons.  Muon tracks are required to have reconstructed momentum 1.5 $ < p <$ 20 \GeVc and angle $\theta <$ 20\degs, matching the geometric acceptance of the MINERvA and MINOS detectors. Each event candidate considered in this analysis corresponds to a single time slice containing one muon.

Pion identification and reconstruction begins with a decay Michel electron candidate from the process $\pi^+\rightarrow\mu^+\rightarrow e^+$.  The \minerva detector collects data for approximately 4 \micros after the end of a beam spill, corresponding to two to six muon lifetimes. Michel electron candidates are first identified as groups of hits with appropriate energies that occur in a time slice after the neutrino interaction. The number of planes traversed and the quality of a positional fit of the hits to a straight line are calculated as subsequent reconstruction quality metrics. For events with tracked pion candidates, each Michel electron candidate is paired with its closest track endpoint up to a maximum distance of between 7.5 \cm and 50 \cm, chosen based on the Michel electron candidate quality metrics, and optimized to maximize efficiency and purity and to minimize systematic uncertainties. After a successful track pairing, the track's energy deposition ($dE/dx$) is required to be consistent with that of a stopping pion. Pion energy is reconstructed using a fit to the track $dE/dx$ and the pion angle is reconstructed using a Kalman filter~\cite{Kalman1,Kalman2}.

\begin{figure}[bhtp]
    \centering
    \includegraphics[width=0.98\linewidth]{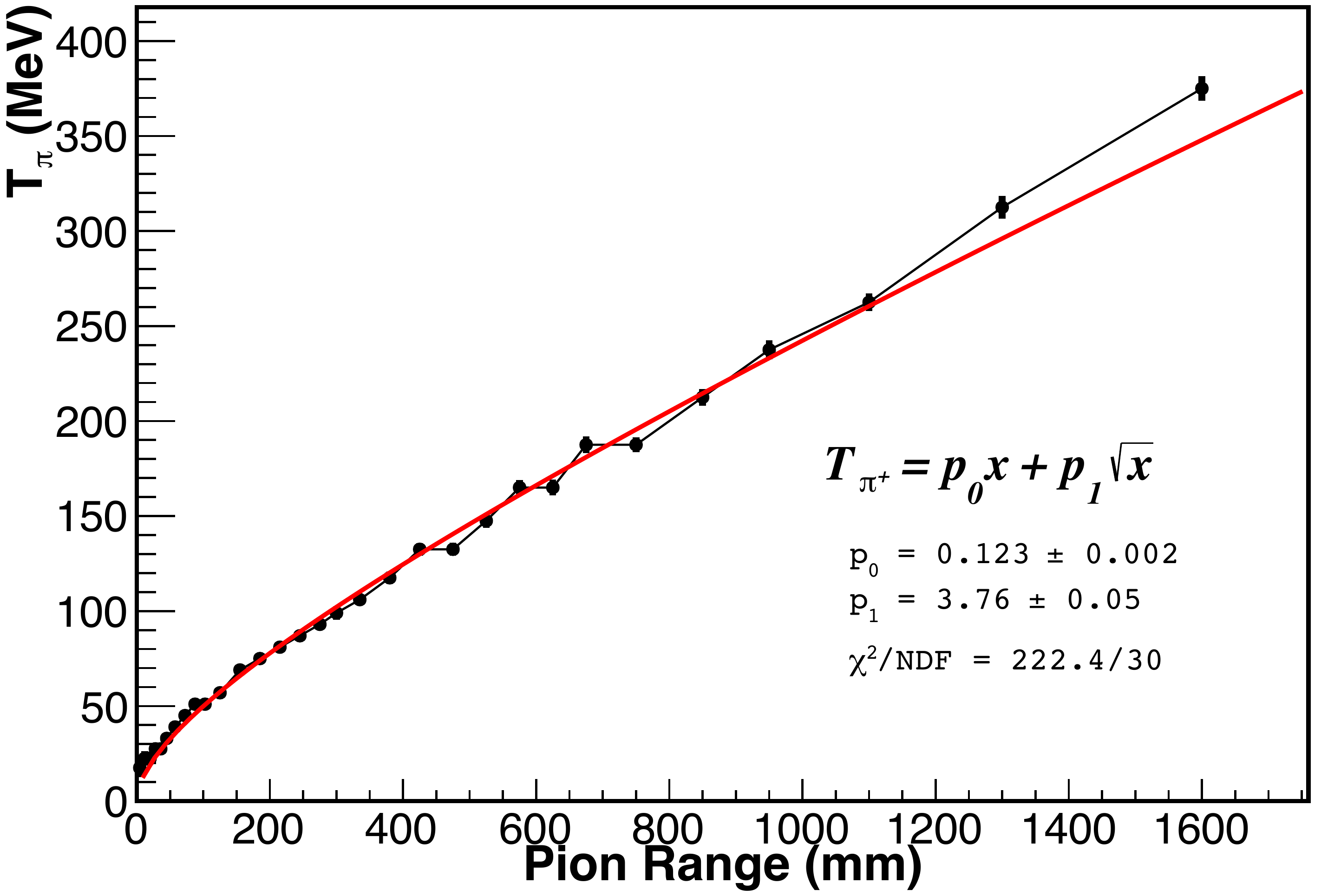}
    \caption{Correlation (black) between true \Tpi and measured Michel electron candidate distance from the interaction vertex or track endpoint for simulated pions and the fit (red) that is used to reconstruct \Tpi based on pion range in data. Error bars are statistics-only.}
    \label{fig:tpi_estimator}
\end{figure}

Michel electron candidates that fail any of the track pairing requirements (or candidates that occur after an event that lacks a reconstructed pion track altogether) are given another opportunity to be paired with any track endpoint with no $dE/dx$ constraints or to the interaction vertex, but with significantly more stringent distance and Michel electron reconstruction quality constraints. Events passing these cuts are considered to contain an untracked pion, and the analysis uses their reconstructed location relative to the paired track endpoint/interaction vertex for kinetic energy and angle determination.  As shown in Fig. \ref{fig:tpi_estimator}, the kinetic energy is extracted from the fitted correlation between Michel-to-endpoint (or Michel-to-vertex) distance (``Pion Range'') and pion kinetic energy \Tpi for simulated pion events. The pion angle \thpi is defined as the angle between the neutrino beam direction and a vector from the neutrino interaction vertex to the Michel electron candidate position. Further detail on this reconstruction method can be found in~\cite{SultanaThesis}, and we note a similar method was employed in recent \nue CC \pip results from T2K~\cite{klhv-7t6h}.


\begin{figure}
    \centering
    \includegraphics[width=0.99\linewidth]{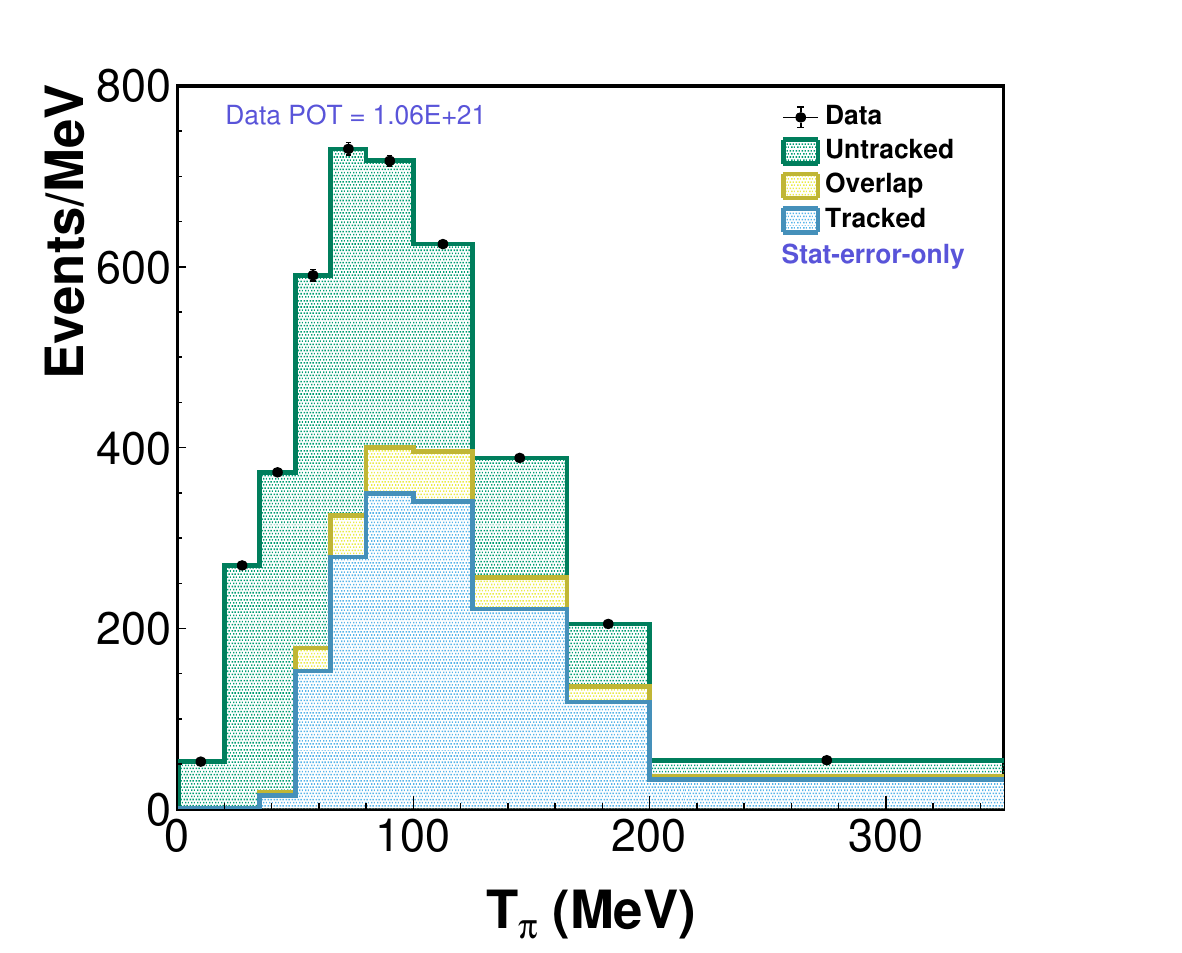}
    \caption{Selected data events in \Tpi, separated by pion reconstruction method. The new ``Untracked'' Michel-only reconstruction method significantly increases acceptance and efficiency on top of the ``Tracked'' method used in previous \minerva CC$\pipm$ analyses, especially at low \Tpi. In the ``Overlap'' category are events passing selection criteria for both reconstruction techniques, in which case the ``Tracked'' reconstruction is used for subsequent analysis. }
    \label{fig:pion_reco}
\end{figure}

This untracked pion selection method expands \minerva's acceptance of charged pions for the first time to include 0 $<$ \Tpi $<$ 35 \MeV, while also nearly doubling the efficiency (from 6\% to 11\%) across the \Tpi spectrum. Simulation studies indicate that the more stringent match distance and Michel electron candidate quality cuts identify pion endpoints with high reliability. Fig. \ref{fig:pion_reco} shows selected reconstructed pion kinetic energy before background subtraction in data, subdivided by reconstruction method. Relatively few events satisfy both reconstruction methods (``Overlap'' category), which only occurs when a high quality Michel electron candidate was matched to a pion-like track endpoint with the tighter match range requirements. For these events, kinematics from the ``Tracked'' method are used for subsequent analysis.  

After charged pion identification, a cut requiring at most one isolated energy deposition is used to remove events with $\pi^0$s. Simulation predicts that these events constitute $<$ 3\% of the selected sample after all cuts. Finally, an experimental invariant hadronic mass cut \Wexp $<$ 1.4 $\GeV/c^2$ isolates the $\Delta$ (1232) resonance and removes high-multiplicity events which are difficult to resolve. Hadronic energy (\Ehad) in Eq.~\eqref{eq:wexp} is measured as the sum of the reconstructed pion energy and the remaining detector energy, which is corrected by data-driven calorimetry~\cite{Aliaga:2015aqe}.

In all, 45,182 events were selected by the untracked technique, 46,661 events by the tracked technique, for a total of 91,843 total events accounting for overlap.

\if\showSections1 \section{Cross Section Measurement} \fi

A sideband composed of events with reconstructed \Wexp $>$ 1.5 $\GeV/c^2$ is employed to constrain the background model. This region is dominated by events with true \Wexp $>$ 1.4 $\GeV/c^2$,  high pion multiplicity, and unseen neutral pion energy, which constitute this analysis's dominant background. The sideband region is broken up into four samples: (1) signal events with reconstructed \Wexp $>$ 1.5 \GeVcsq, (2) background with true \Wexp $<$ 1.4 \GeVcsq (lower sideband background), (3) background with 1.4 $<$ true \Wexp $<$ 1.8  $\GeV/c^2$ and (4) background with 1.8 $\GeV/c^2$ $<$ true \Wexp. The normalizations for samples (3) and (4) are simultaneously fit to match the data. The fitted weights are 0.99 and 1.03 in the lower and upper sideband backgrounds, respectively. These weights are close to unity, indicating that the model background prediction closely matches the data. This good agreement is in part due to the \Qsq and \Tpi tunes of the signal model, without which the weights are $\sim$10--15\%.

After background subtraction, iterative D'Agostini unfolding~\cite{D'Agostini:1994zf,DAgostini:2010xxxxx,Adye:2011gm} is used to correct for the detector resolution. The number of iterations used for unfolding was determined by the stability of unfolding with fake data from alternate models. Four iterations were used for \pmu, \pzmu, and \thmu, and ten iterations for \Tpi, \thpi, \ptmu, and \Qsq. Uncertainty on the unfolding matrix itself is considered by manually inflating the statistical uncertainty until nominal fake data is unfolded to simulated unsmeared values with $\chi^2$/ndf $\sim$ 1. 

\begin{figure}
    \centering
    \includegraphics[width=\figwidth]{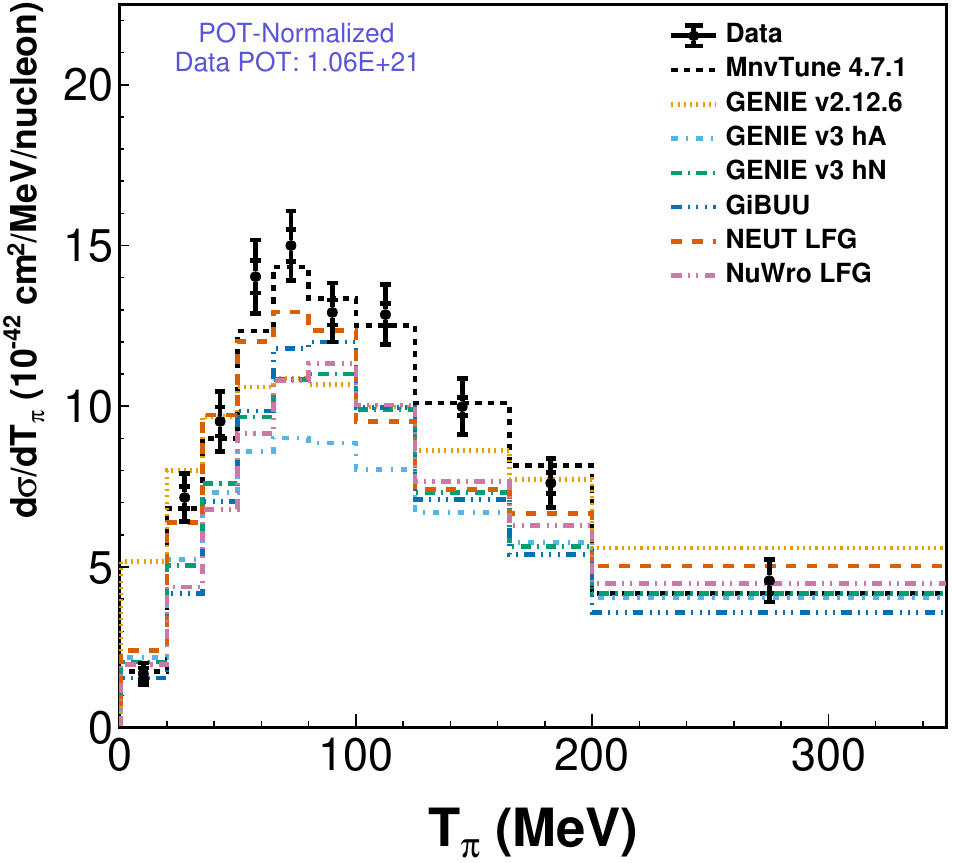}
    \includegraphics[width=\figwidth]{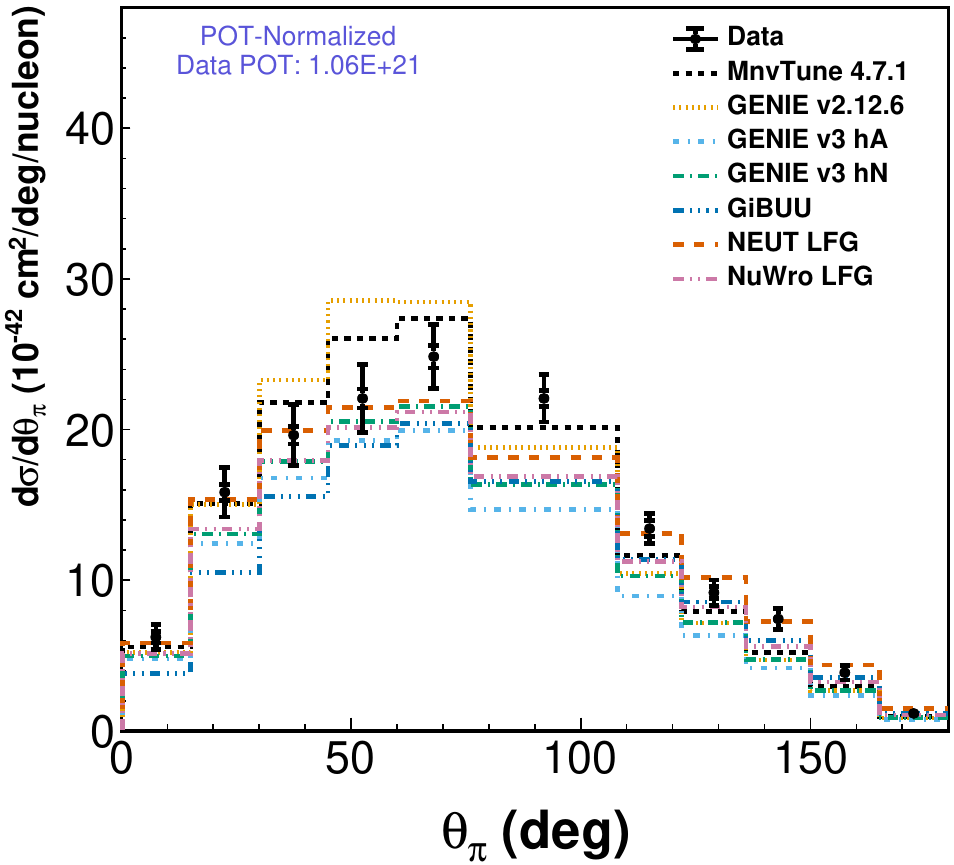}
    \includegraphics[width=\figwidth]{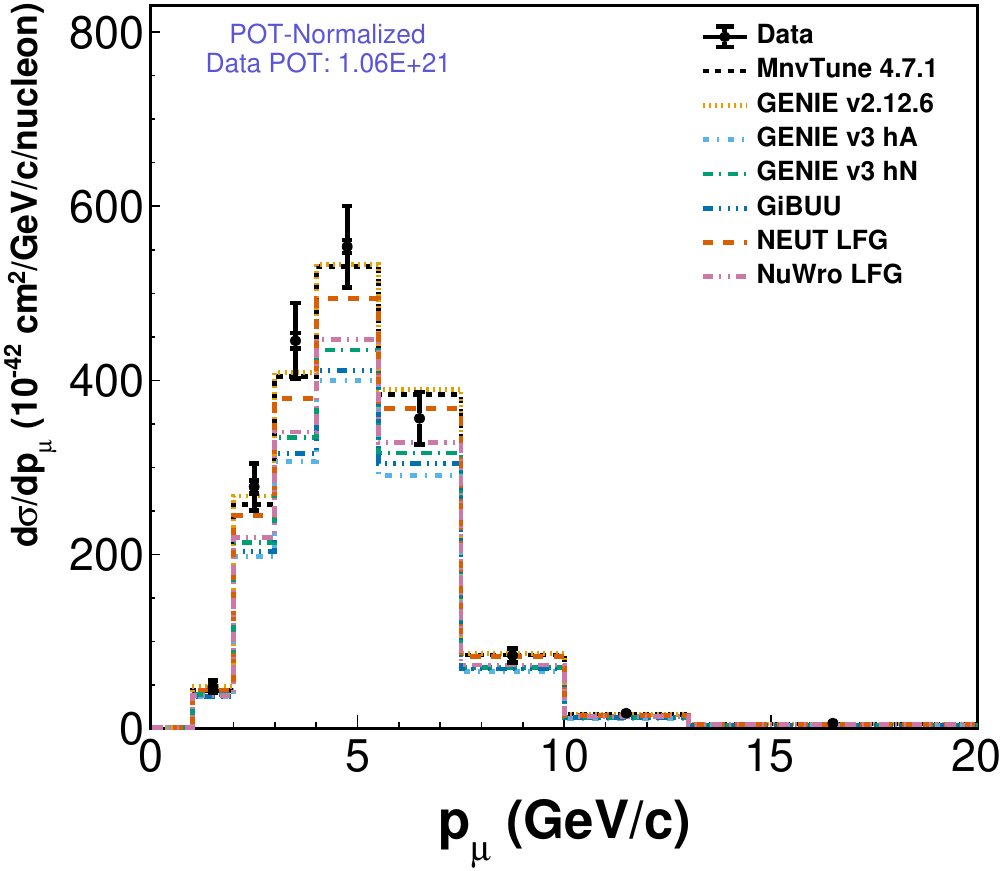}
    \includegraphics[width=\figwidth]{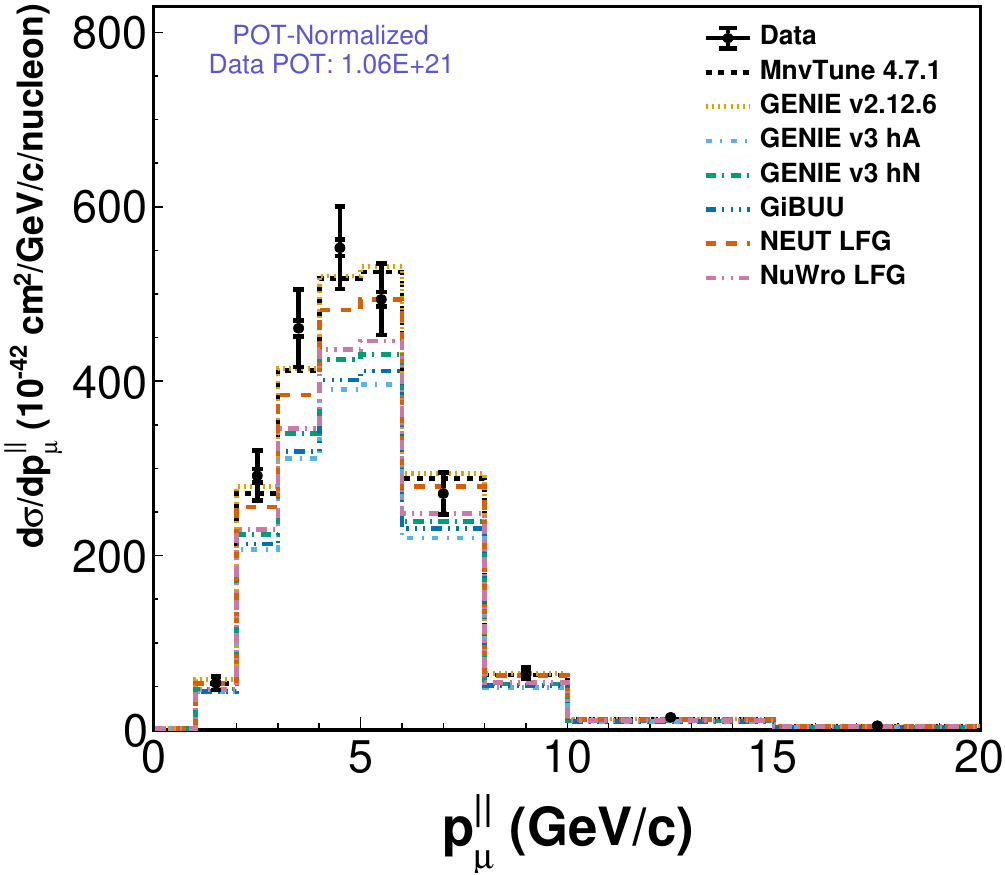}
    \includegraphics[width=.47\linewidth]{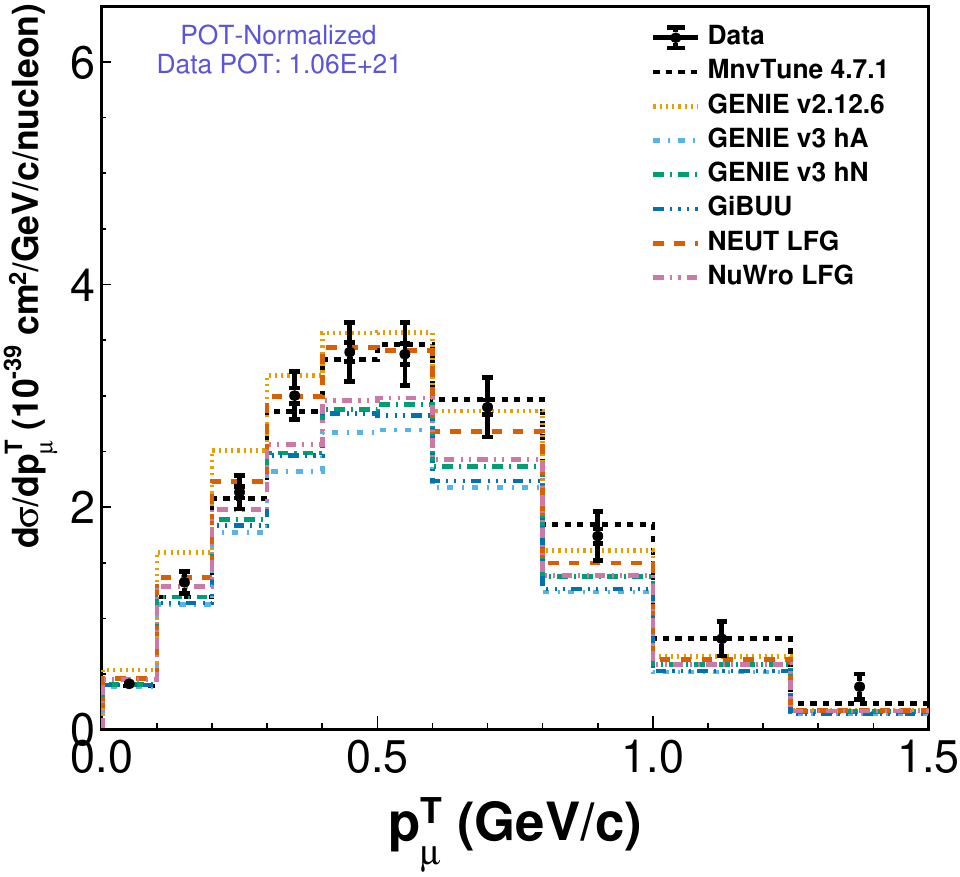}
    \includegraphics[width=\figwidth]{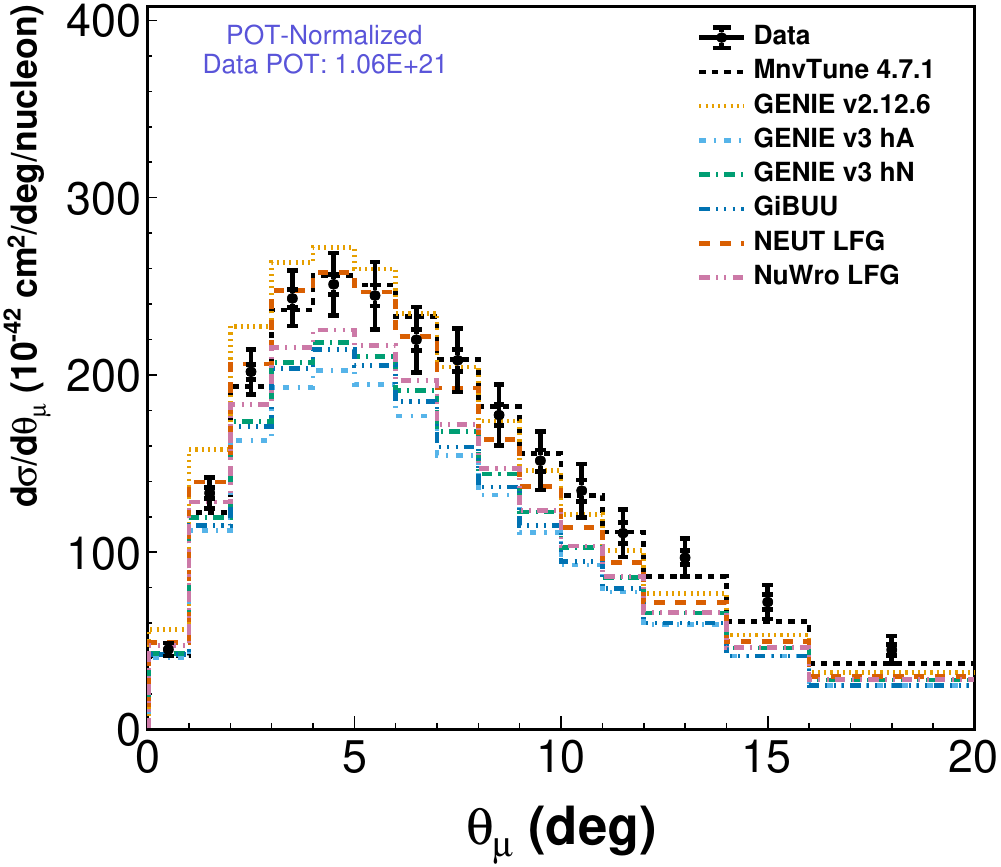}
    \includegraphics[width=\figwidth]{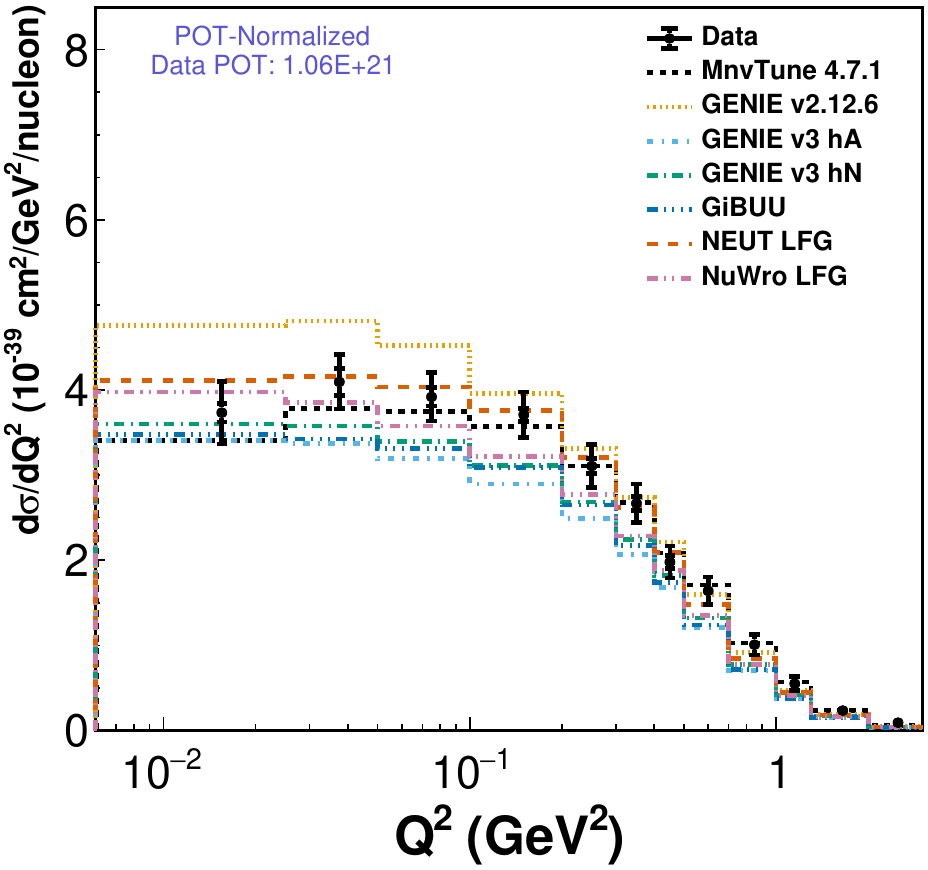}
    \caption{Differential cross sections compared with several generators for \Tpi, \thpi, \pmu, \pzmu, \ptmu, \thmu, and \Qsq with statistical (inner) and total (outer) error bars. Overall, GENIE v3 $hN$ has the best agreement with data, but no model captures the full range of variables and kinematics.}
    \label{fig:body_xsecs}
\end{figure}

Differential cross sections are obtained by correcting for efficiency and acceptance and normalizing by the integrated neutrino flux (0-100 \GeV) and the number of target nucleons ($\sim 3.47\times10^{30}$). Fig. \ref{fig:body_xsecs} shows differential cross sections as a function of \Tpi, \thpi, \pmu, \pzmu, \ptmu, \thmu, and \Qsq with total (statistical + systematic) uncertainties and various generator predictions.

Systematic uncertainties are estimated by repeating the cross section measurement in many different universes wherein a parameter or group of parameters associated with modeling of the neutrino interaction, beam flux, detector response, or reconstruction method are varied within their tolerances. From these measurements a covariance matrix is constructed for each variable, the diagonal of which contains the standard deviations of universe measurements, which are taken to be the systematic uncertainties. The procedure is further detailed in~\cite{Messerly_2021}. \minerva's procedure for flux uncertainty estimation is described in~\cite{Aliaga:2016oaz}. Uncertainties on the signal model are estimated using GENIE event reweighting functionality as well as additional uncertainties assessed on each of the modifications MINERvA makes to the GENIE model.  To address outstanding disagreement between the data and MnvTune v4.7.1 simulation that is not covered by these signal model uncertainties, a single systematic universe is implemented in which the model is further weighted in true \Tpi-\Qsq space to match the background-subtracted data.   

Total uncertainties are close to 6\% in most regions of phase space. The dominant source of uncertainty varies across the kinematic regions reported. Statistical uncertainties are smaller than systematic uncertainties in most bins. Individual systematic contributions vary around 3\% with largest contributions from neutrino flux and muon energy reconstruction.

MnvTune v4.7.1 weights, sideband distributions, efficiencies, error distributions, cross section tables, data-MC ratios, covariance matrices, and $\chi^2$ scores describing model agreement with data for all variables can be found in \supp.

\if\showSections1 \section{Results and Conclusion} \fi

GENIE v3.0.6 $hN$~\cite{andreopoulos2015genieneutrinomontecarlo} performs best overall, but no model captures the full range of variables and kinematics. In \Tpi, models and data agree in the first bin (0 -- 20 \MeV) and last bin (200 -- 350 \MeV), but agreement is poor elsewhere, with GENIE v3 $hN$ best across the entire spectrum. NEUT 5.4.1~\cite{Hayato_2021} and NuWro  19.02~\cite{Golan:2012wx} do significantly better in \thpi than GENIE and GiBUU 2021~\cite{BUSS20121}. The variable best-described by all models is \Qsq, though all models diverge from data at high \Qsq, strongly underpredicting. All models agree better in muon variables compared to the pion variables, perhaps due to the greater availability of prior data in muon variables, which have informed models.  

In summary, this analysis improves upon previous \minerva CC 1\pip measurements by significantly increasing statistics with untracked pions reconstructed from their Michel electron signature alone. For model comparisons to CH, the new results and signal definition supersede the one in \cite{Bercellie.131.011801}. The previous results remain the best for the A-dependent ratios. This method also extends the measured phase space to \Tpi $\sim$ 0, a region where there is a dearth of published results. The differential cross sections are also compared with several generator predictions.  While some generators perform better than others, none accurately predicts the data across all of phase space, indicating a need for further model development informed by these data.

\begin{acknowledgments}

This document was prepared by members of the MINERvA Collaboration using the resources of the Fermi National Accelerator Laboratory (Fermilab), a U.S. Department of Energy, Office of Science, Office of High Energy Physics HEP User Facility. Fermilab is managed by Fermi Forward Discovery Group, LLC, acting under Contract No. 89243024CSC000002.
These resources included support for the MINERvA construction project, and support
for construction also
was granted by the United States National Science Foundation under
Award No. PHY-0619727 and by the University of Rochester. Support for
participating scientists was provided by NSF and DOE (USA); by CAPES
and CNPq (Brazil); by CoNaCyT (Mexico); by ANID PIA / APOYO AFB180002, CONICYT PIA ACT1413, and Fondecyt 3170845 and 11130133 (Chile); 
by CONCYTEC (Consejo Nacional de Ciencia, Tecnolog\'ia e Innovaci\'on Tecnol\'ogica), DGI-PUCP (Direcci\'on de Gesti\'on de la Investigaci\'on  - Pontificia Universidad Cat\'olica del Peru), and VRI-UNI (Vice-Rectorate for Research of National University of Engineering) (Peru); NCN Opus Grant No. 2016/21/B/ST2/01092 (Poland); by Science and Technology Facilities Council (UK); by EU Horizon 2020 Marie Skłodowska-Curie Action; by a Cottrell Postdoctoral Fellowship from the Research Corporation for Scientific Advancement; by an Imperial College London President's PhD Scholarship.  We thank the MINOS Collaboration for use of its near detector data. Finally, we thank the staff of
Fermilab for support of the beam line, the detector, and computing infrastructure.

%
%
%
%
%
%
%
%
%
%
%
%
\end{acknowledgments}

\bibliography{MINERvAPion}

\ifPRLandSupp

\clearpage

\onecolumngrid
\pagebreak
\section{Supplemental Material}
\renewcommand\thefigure{Supp.\arabic{figure}}
\setcounter{figure}{0}


\renewcommand\thetable{Supp.\Roman{table}}
\setcounter{table}{0}

\FloatBarrier

\subsection{Error Summary Plots}
\label{sec:error_plots}

\begin{figure}[!h]
    \includegraphics[width=0.32\linewidth]{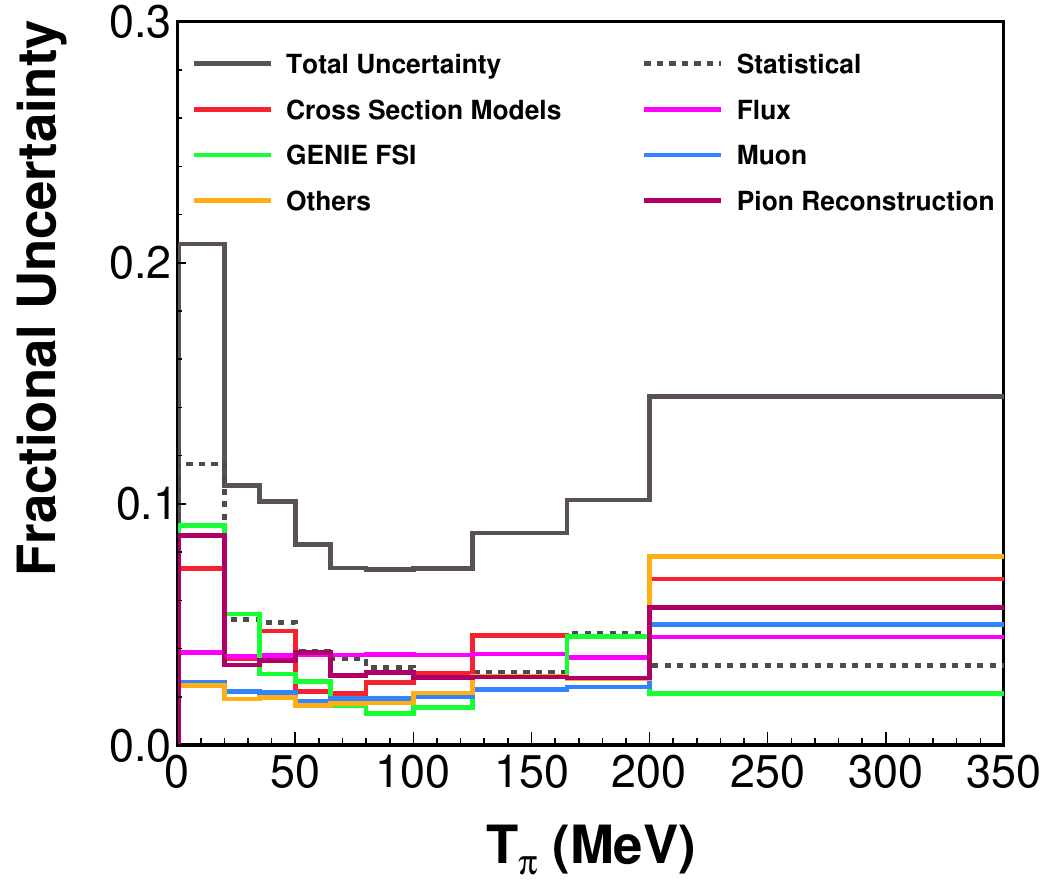}
    \includegraphics[width=0.32\linewidth]{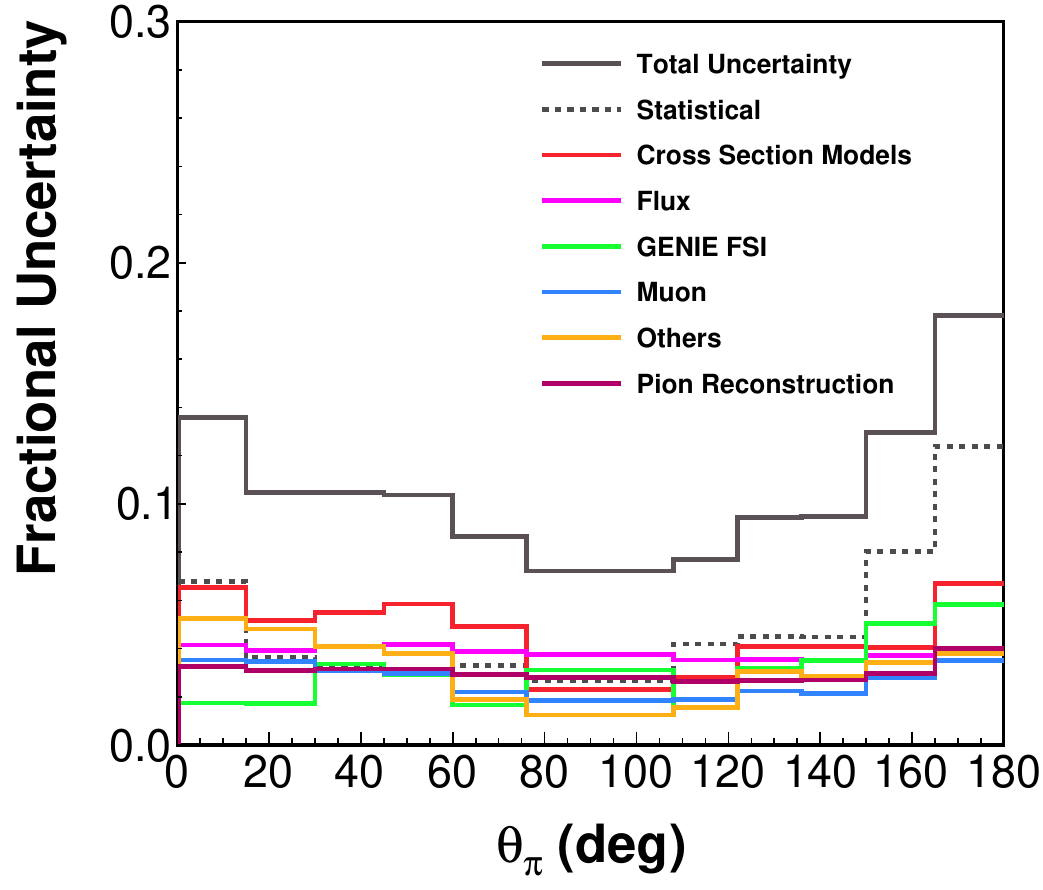}
    \includegraphics[width=0.32\linewidth]{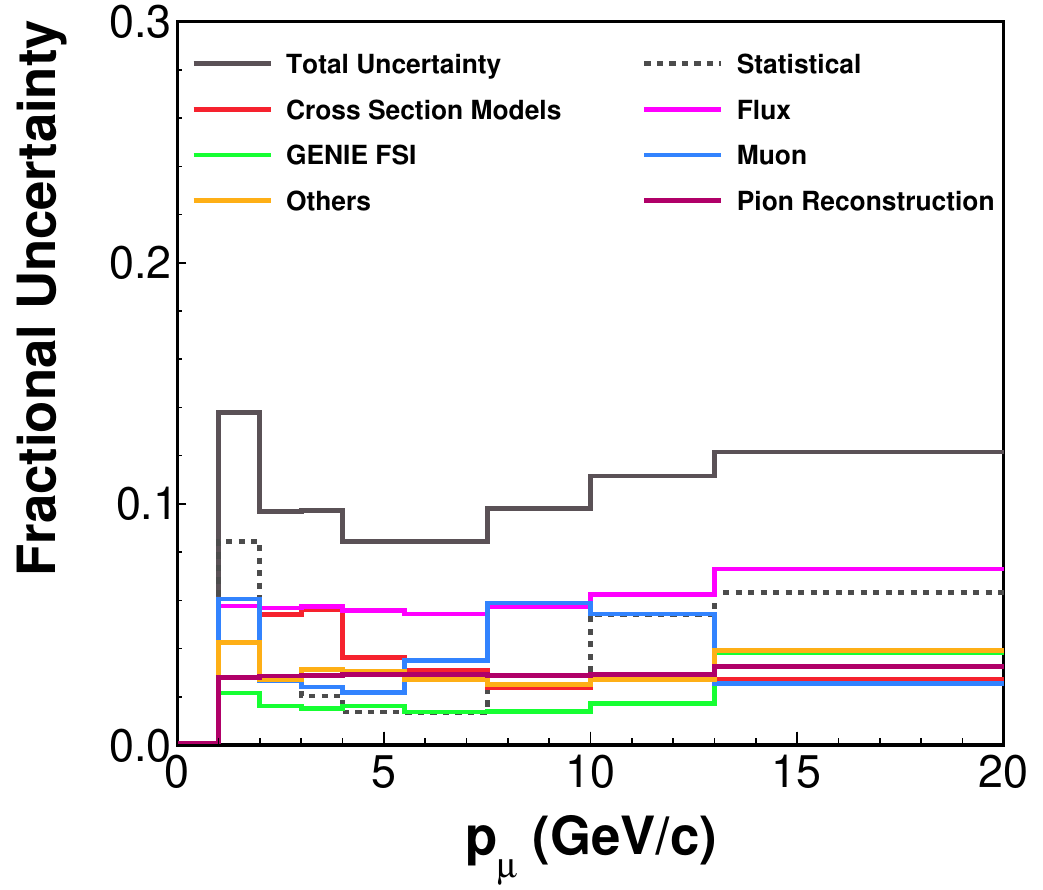}
    \includegraphics[width=0.32\linewidth]{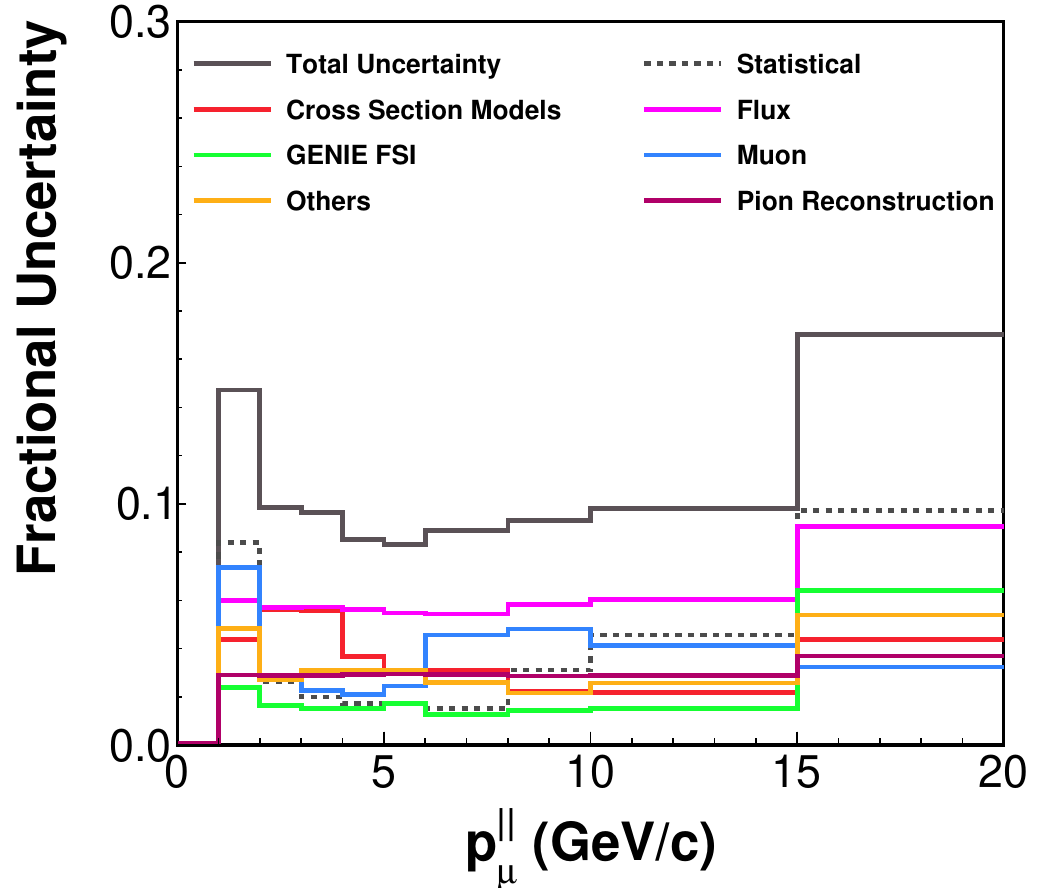}
    \includegraphics[width=0.32\linewidth]{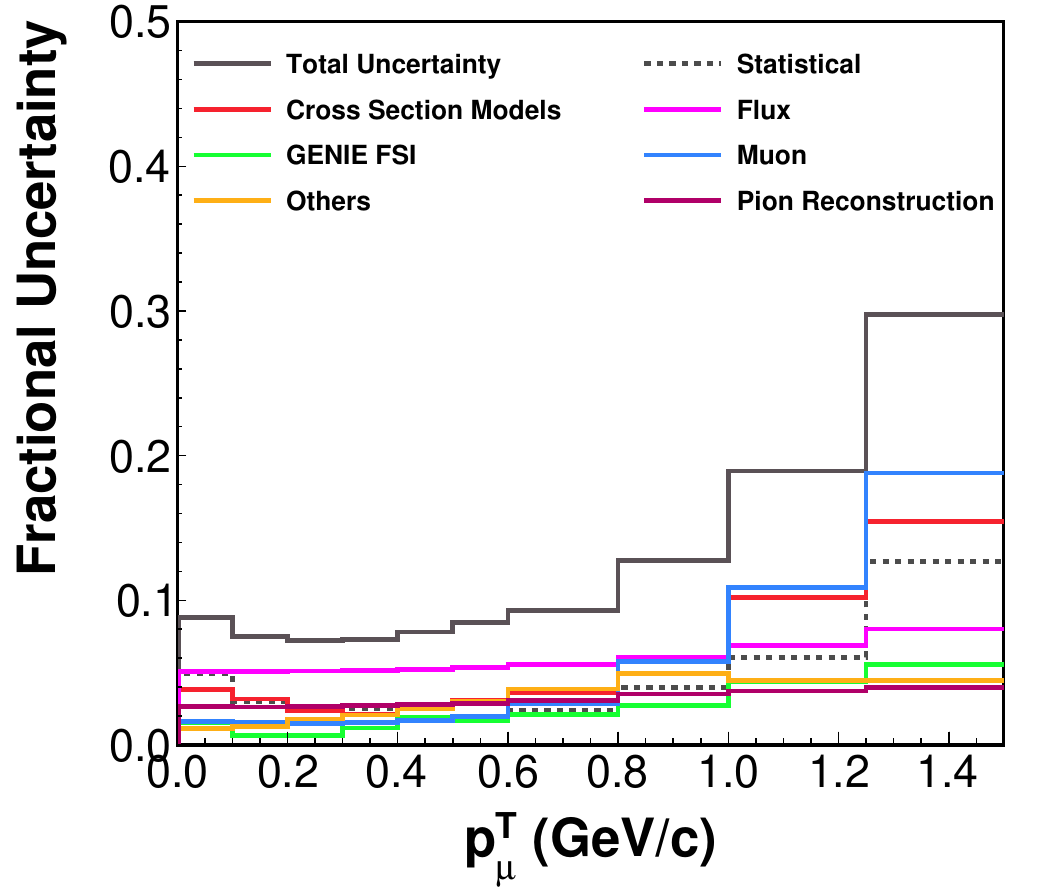}
    \includegraphics[width=0.32\linewidth]{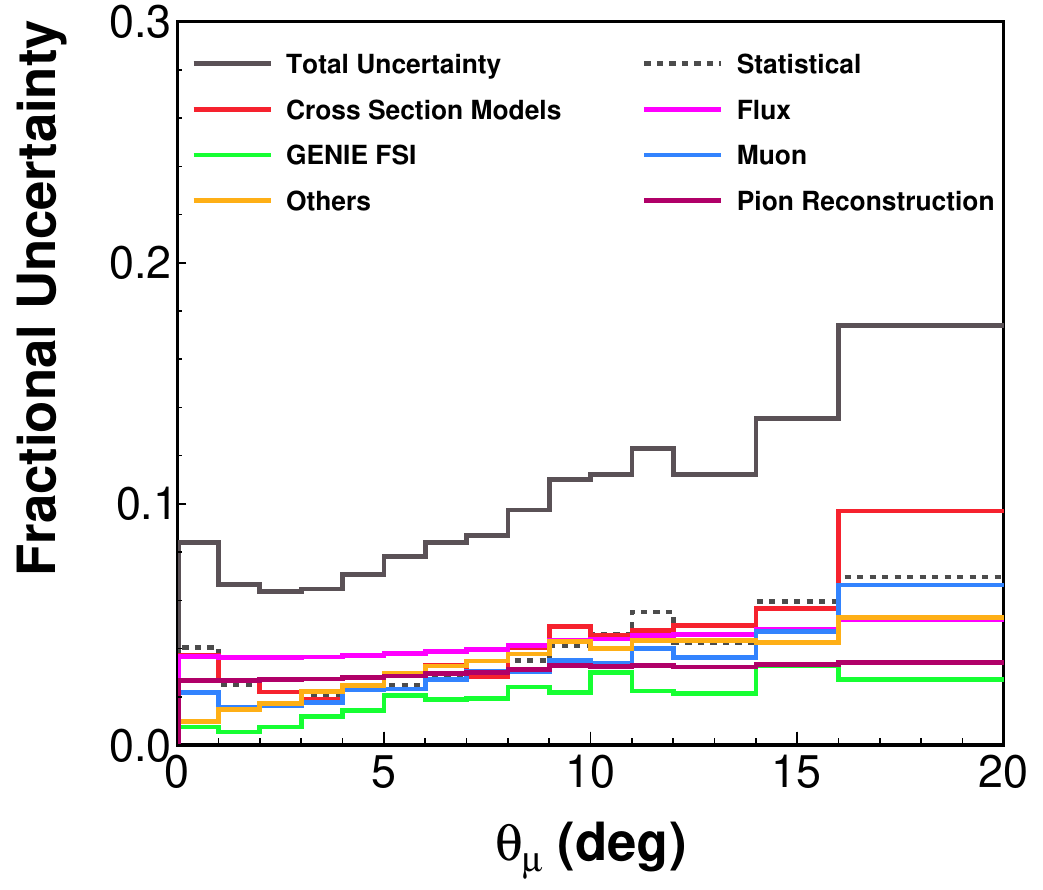}
    \includegraphics[width=0.32\linewidth]{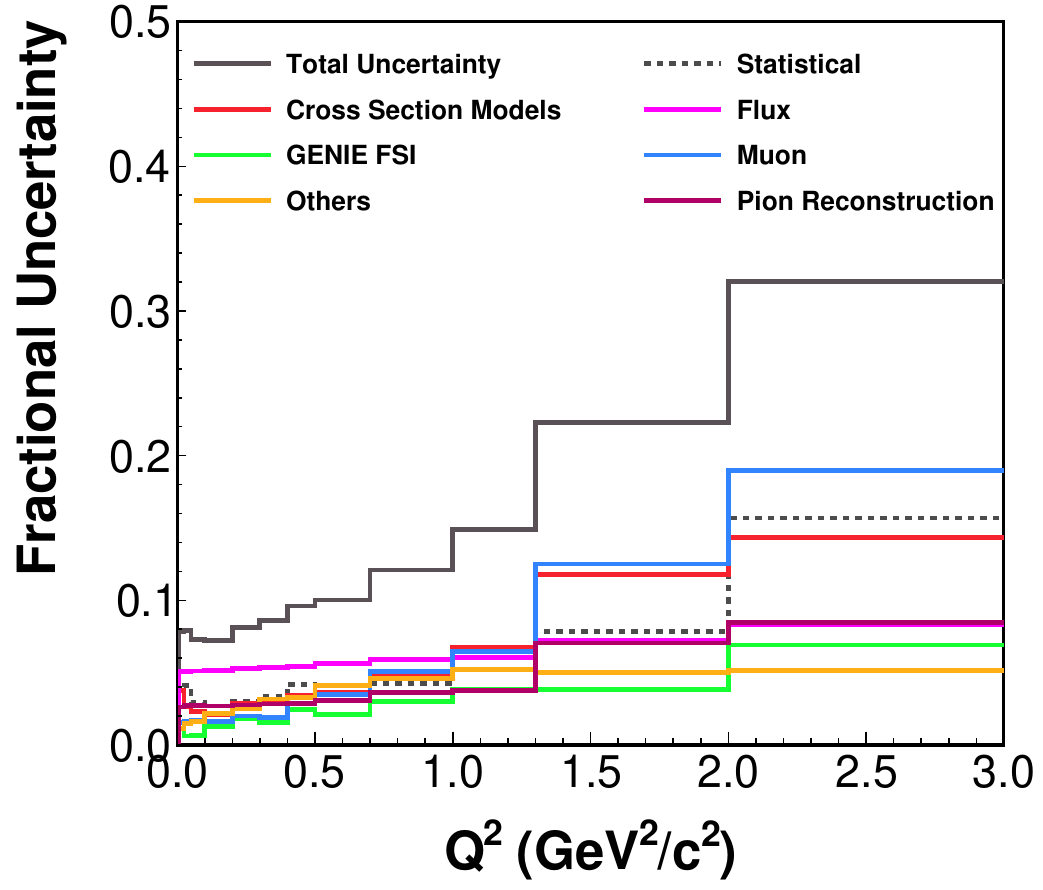}
    \caption{Differential cross section error summaries. The signal model uncertainty (within ``Cross Section Model'') is $\sim$2-4\% and enters via the efficiency correction stage. The ``Others'' category includes primarily hadron response and Geant effects, and the ``Muon'' category includes muon reconstruction effects and is dominated by reconstruction efficiency. }
    \label{fig:supp:cross_section_errors}
\end{figure}

\newpage
\FloatBarrier

\subsection{Cross Section Data-MC Ratio Plots} \label{sec:cross_section_ratio_plots}

\begin{figure}[!h]
    \includegraphics[width=0.32\linewidth]{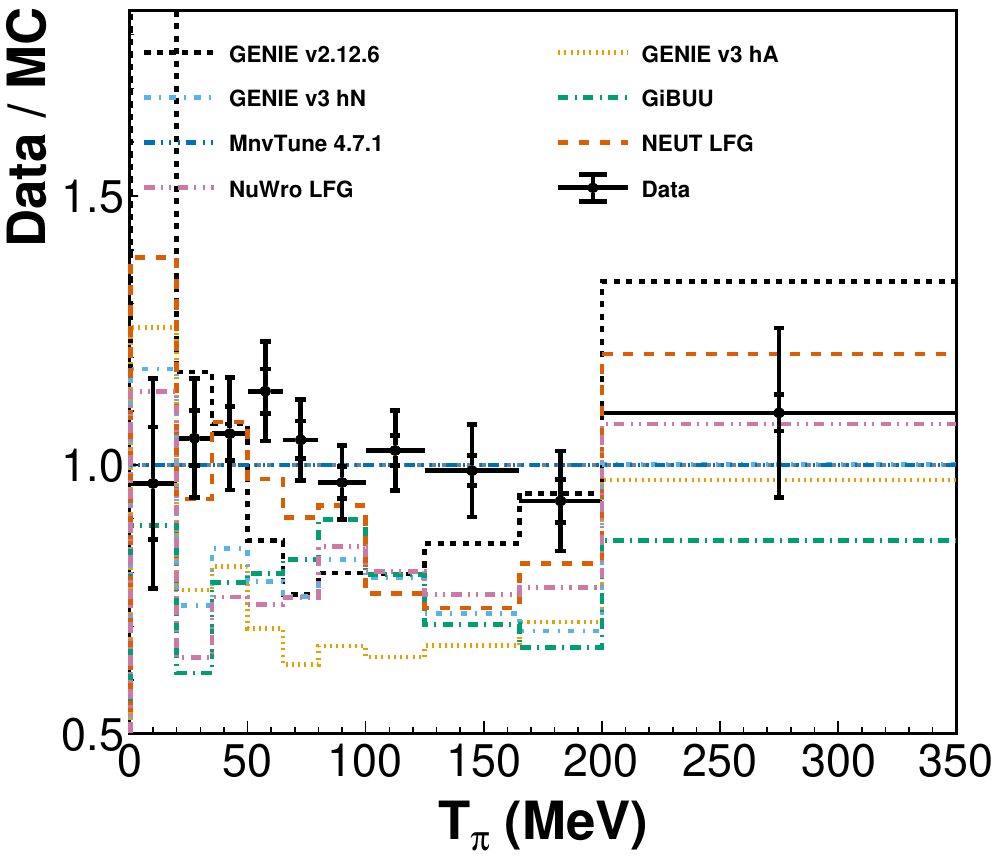}
    \includegraphics[width=0.32\linewidth]{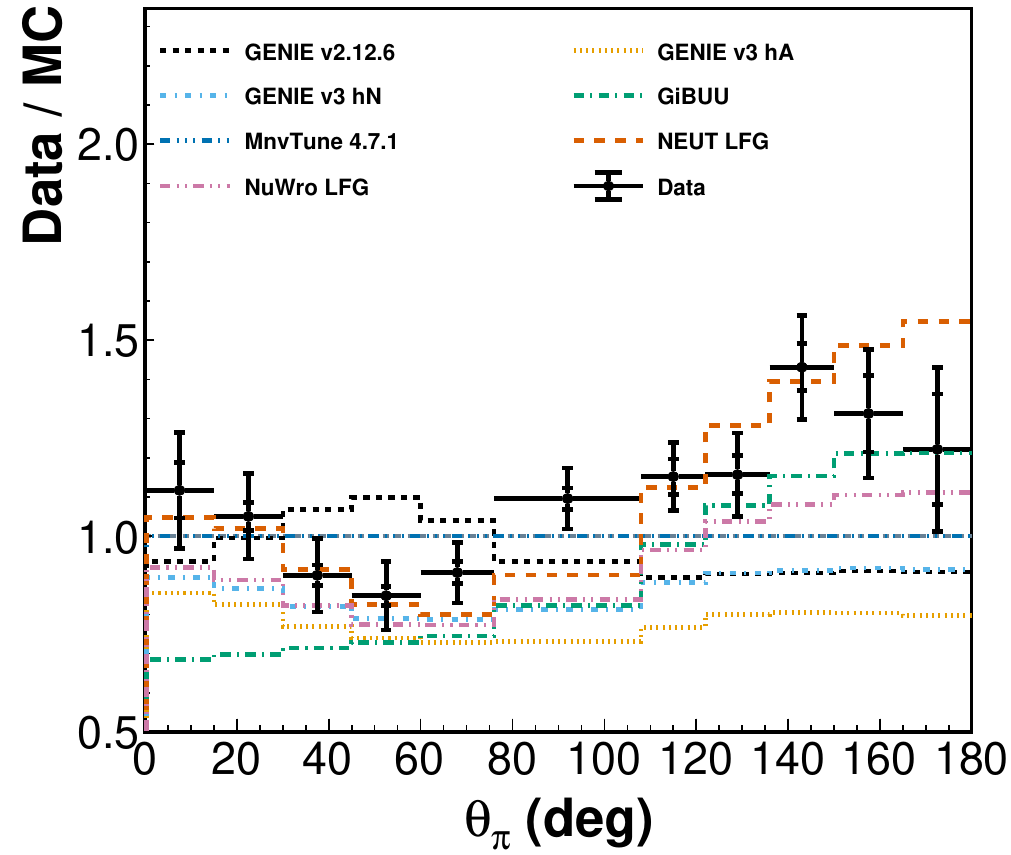}
    \includegraphics[width=0.32\linewidth]{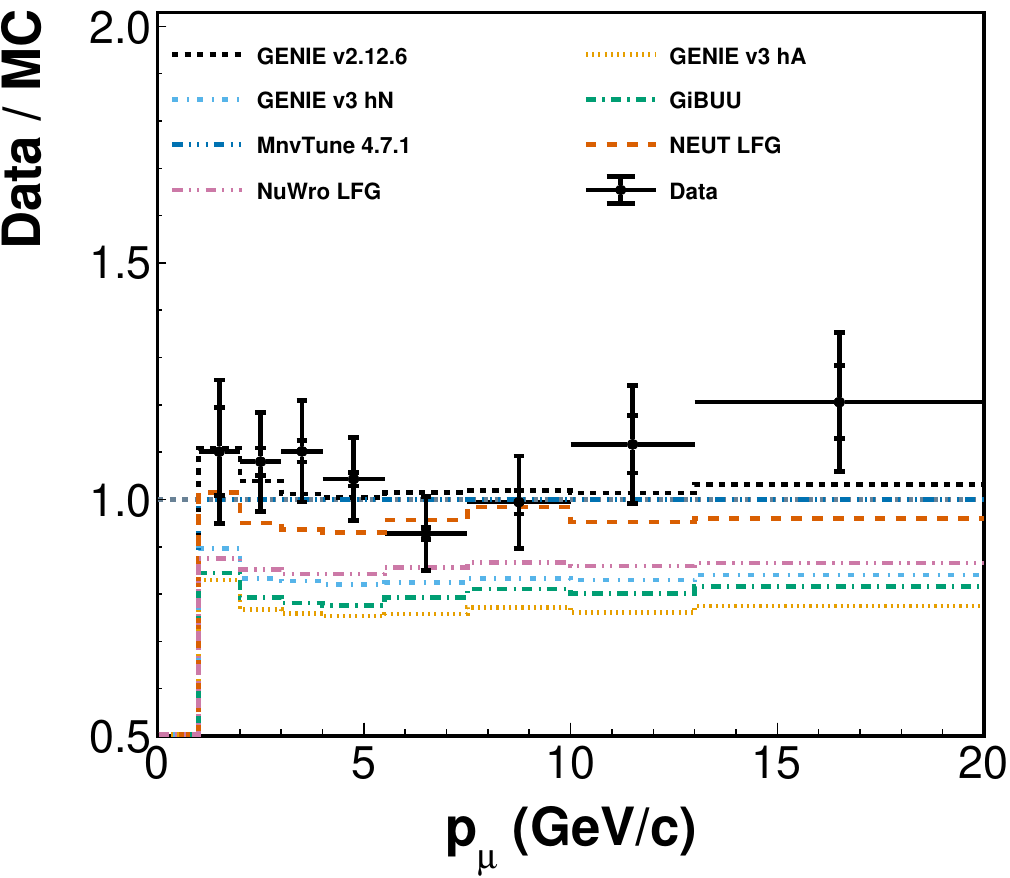}
    \includegraphics[width=0.32\linewidth]{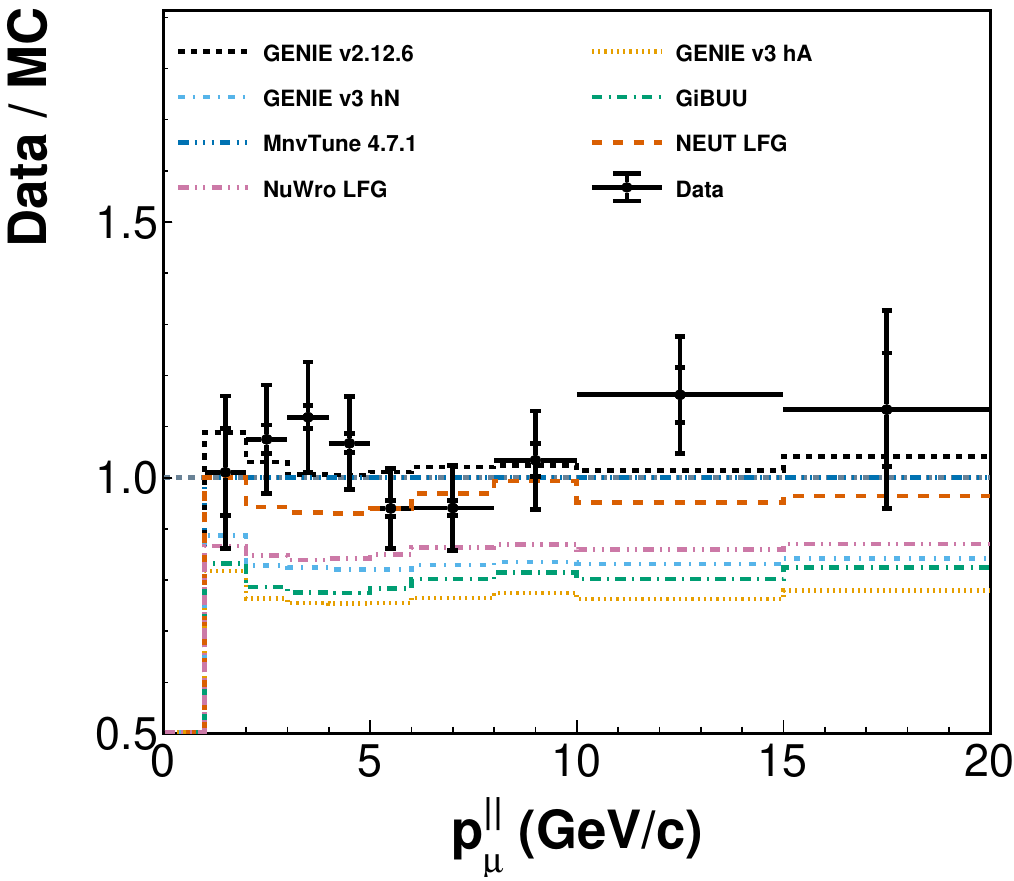}
    \includegraphics[width=0.32\linewidth]{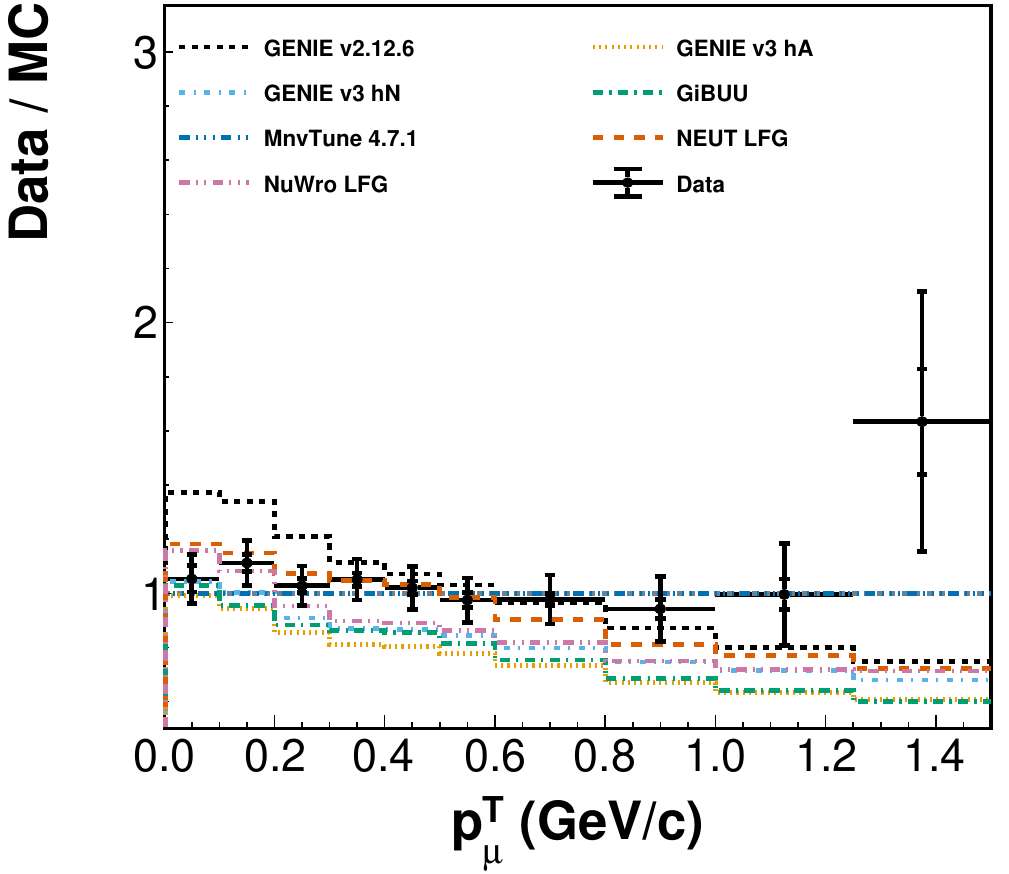}
    \includegraphics[width=0.32\linewidth]{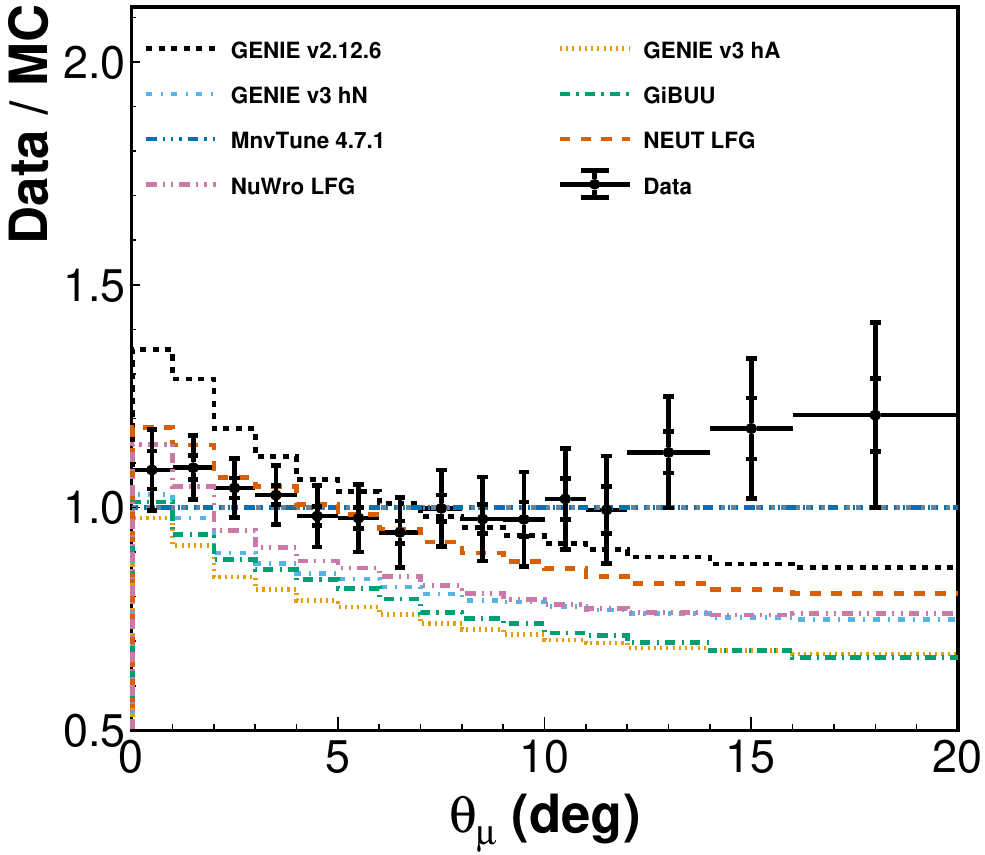}
    \includegraphics[width=0.32\linewidth]{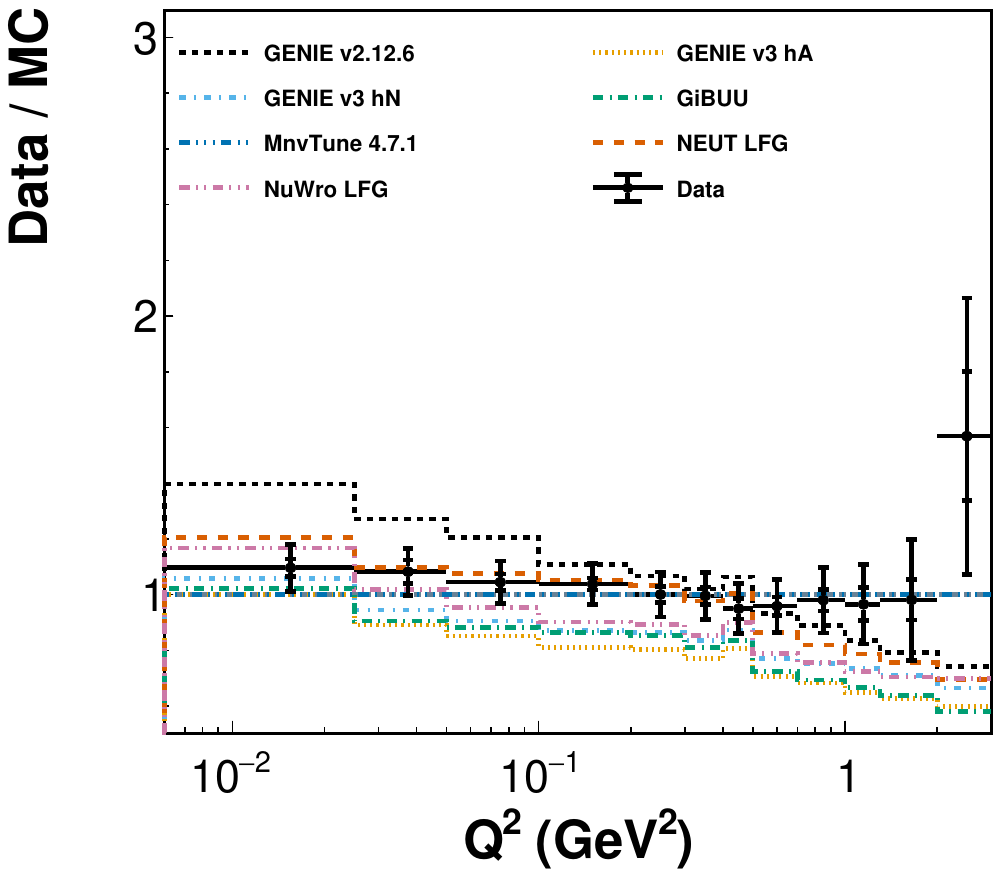}
    \caption{Ratios of differential cross section measurements to MnvTune v4.7.1 MC. The ratios for the pion variables show a nonuniform  discrepancy between data and the predictions. For muon variables, the ratios show statistical agreement between data and MnvTune v4.7.1; for other models, shapes are consistent with data. In \Qsq, a large disagreement is observed between data and all the models, particularly in the last bin.}
    \label{fig:supp:cross_section_ratios}
\end{figure}

\newpage
\FloatBarrier

\subsection{Cross Section Tables \label{sec:cross_section_tables}}

\begin{table}[H]
  \centering
  \renewcommand{\arraystretch}{1.15}
  \caption{Measured cross section as a function of \Tpi on scintillator, in units of $10^{-42}$ \cmsq/\MeV/nucleon with absolute and fractional cross section uncertainties.}
    \begin{tabular}{c||c|c||c|c|c}
      Bin edges (\MeV) &  $\sigma$  &  Abs. Unc.  &  Frac. Stat. Unc.  &  Frac. Sys. Unc.  &  Frac. Flux Unc. \\
      \hline
      \hline
        0 -- 20 & 1.67915 & 0.33884 & 0.11 & 0.17 & 0.038\\
        20 -- 35 & 7.16007 & 0.75347 & 0.048 & 0.093 & 0.037\\
        35 -- 50 & 9.52944 & 0.93958 & 0.047 & 0.087 & 0.037\\
        50 -- 65 & 14.0306 & 1.1426 & 0.036 & 0.073 & 0.037\\
        65 -- 80 & 15.0006 & 1.0772 & 0.033 & 0.064 & 0.037\\
        80 -- 100 & 12.918 & 0.92088 & 0.03 & 0.065 & 0.037\\
        100 -- 125 & 12.8516 & 0.92806 & 0.027 & 0.067 & 0.037\\
        125 -- 165 & 9.98917 & 0.86816 & 0.028 & 0.082 & 0.038\\
        165 -- 200 & 7.60749 & 0.75796 & 0.043 & 0.09 & 0.036\\
        200 -- 350 & 4.57352 & 0.65765 & 0.031 & 0.14 & 0.045\\
    \end{tabular}
  \label{tbl:CrossSectiontpi}
\end{table}

\begin{table}[H]
  \centering
  \renewcommand{\arraystretch}{1.15}
  \caption{Measured cross section as a function of \thpi on scintillator, in units of $10^{-42}$ \cmsq/degree/nucleon with absolute and fractional cross section uncertainties.}
    \begin{tabular}{c||c|c||c|c|c}
      Bin edges (degree) &  $\sigma$  &  Abs. Unc.  &  Frac. Stat. Unc.  &  Frac. Sys. Unc.  &  Frac. Flux Unc. \\
      \hline
      \hline
        0 -- 15 & 6.22647 & 0.82283 & 0.063 & 0.12 & 0.041\\
        15 -- 30 & 15.8456 & 1.6399 & 0.034 & 0.098 & 0.039\\
        30 -- 45 & 19.6293 & 2.0361 & 0.029 & 0.099 & 0.041\\
        45 -- 60 & 22.0659 & 2.2665 & 0.029 & 0.099 & 0.042\\
        60 -- 76 & 24.8393 & 2.1157 & 0.03 & 0.08 & 0.039\\
        76 -- 108 & 22.066 & 1.5699 & 0.025 & 0.067 & 0.038\\
        108 -- 122 & 13.43 & 1.0057 & 0.039 & 0.064 & 0.035\\
        122 -- 136 & 9.18089 & 0.84642 & 0.042 & 0.082 & 0.035\\
        136 -- 150 & 7.43138 & 0.6882 & 0.042 & 0.083 & 0.035\\
        150 -- 165 & 3.86764 & 0.48325 & 0.074 & 0.1 & 0.037\\
        165 -- 180 & 1.16235 & 0.19873 & 0.11 & 0.13 & 0.04\\
    \end{tabular}
  \label{tbl:CrossSectionthetapi}
\end{table}

\begin{table}[H]
  \centering
  \renewcommand{\arraystretch}{1.15}
  \caption{Measured cross section as a function of \pmu on scintillator, in units of $10^{-42}$ \cmsq/\GeVc/nucleon with absolute and fractional cross section uncertainties.}
    \begin{tabular}{c||c|c||c|c|c}
      Bin edges (\GeVc) &  $\sigma$  &  Abs. Unc.  &  Frac. Stat. Unc.  &  Frac. Sys. Unc.  &  Frac. Flux Unc. \\
      \hline
      \hline
        1.0 -- 2.0 & 48.4454 & 7.47 & 0.084 & 0.13 & 0.042\\
        2.0 -- 3.0 & 277.849 & 25.611 & 0.027 & 0.088 & 0.042\\
        3.0 -- 4.0 & 445.459 & 40.699 & 0.02 & 0.089 & 0.043\\
        4.0 -- 5.5 & 553.657 & 42.644 & 0.014 & 0.076 & 0.042\\
        5.5 -- 7.5 & 356.419 & 27.245 & 0.014 & 0.075 & 0.039\\
        7.5 -- 10.0 & 84.199 & 7.5974 & 0.025 & 0.087 & 0.036\\
        10.0 -- 13.0 & 17.8685 & 2.0062 & 0.054 & 0.098 & 0.041\\
        13.0 -- 20.0 & 6.4554 & 0.82748 & 0.063 & 0.11 & 0.061\\
    \end{tabular}
  \label{tbl:CrossSectionpmu}
\end{table}

\begin{table}[H]
  \centering
  \renewcommand{\arraystretch}{1.15}
  \caption{Measured cross section as a function of \pzmu on scintillator, in units of $10^{-42}$ \cmsq/\GeVc/nucleon with absolute and fractional cross section uncertainties.}
    \begin{tabular}{c||c|c||c|c|c}
      Bin edges (\GeVc) &  $\sigma$  &  Abs. Unc.  &  Frac. Stat. Unc.  &  Frac. Sys. Unc.  &  Frac. Flux Unc. \\
      \hline
      \hline
        0 -- 1 & 0.0 & 0.0 & 0.0 & 0.0 & 0.0\\
        1 -- 2 & 53.8178 & 8.7435 & 0.084 & 0.14 & 0.045\\
        2 -- 3 & 291.572 & 27.331 & 0.026 & 0.09 & 0.042\\
        3 -- 4 & 460.723 & 41.614 & 0.02 & 0.088 & 0.043\\
        4 -- 5 & 552.913 & 43.328 & 0.017 & 0.076 & 0.042\\
        5 -- 6 & 493.812 & 37.705 & 0.017 & 0.074 & 0.041\\
        6 -- 8 & 271.061 & 21.936 & 0.015 & 0.08 & 0.038\\
        8 -- 10 & 65.2803 & 5.6005 & 0.031 & 0.08 & 0.036\\
        10 -- 15 & 14.6478 & 1.4167 & 0.046 & 0.085 & 0.041\\
        15 -- 20 & 4.94936 & 0.92611 & 0.097 & 0.16 & 0.078\\
    \end{tabular}
  \label{tbl:CrossSectionpzmu}
\end{table}

\begin{table}[H]
  \centering
  \renewcommand{\arraystretch}{1.15}
  \caption{Measured cross section as a function of \ptmu on scintillator, in units of $10^{-42}$ \cmsq/\GeVc/nucleon with absolute and fractional cross section uncertainties.}
    \begin{tabular}{c||c|c||c|c|c}
      Bin edges (\GeVc) &  $\sigma$  &  Abs. Unc.  &  Frac. Stat. Unc.  &  Frac. Sys. Unc.  &  Frac. Flux Unc. \\
      \hline
      \hline
        0.0 -- 0.1 & 411.435 & 36.426 & 0.046 & 0.075 & 0.036\\
        0.1 -- 0.2 & 1326.28 & 91.951 & 0.028 & 0.063 & 0.037\\
        0.2 -- 0.3 & 2131.18 & 139.3 & 0.026 & 0.06 & 0.037\\
        0.3 -- 0.4 & 3004.04 & 196.69 & 0.024 & 0.061 & 0.037\\
        0.4 -- 0.5 & 3394.69 & 243.34 & 0.026 & 0.067 & 0.037\\
        0.5 -- 0.6 & 3367.63 & 264.77 & 0.03 & 0.072 & 0.038\\
        0.6 -- 0.8 & 2899.68 & 249.06 & 0.023 & 0.083 & 0.039\\
        0.8 -- 1.0 & 1745.75 & 214.3 & 0.039 & 0.12 & 0.043\\
        1.0 -- 1.25 & 811.522 & 152.72 & 0.061 & 0.18 & 0.047\\
        1.25 -- 1.5 & 341.796 & 116.8 & 0.14 & 0.31 & 0.058\\
    \end{tabular}
  \label{tbl:CrossSectionptmu}
\end{table}

\begin{table}[H]
  \centering
  \renewcommand{\arraystretch}{1.15}
  \caption{Measured cross section as a function of \thmu on scintillator, in units of $10^{-42}$ \cmsq/degree/nucleon with absolute and fractional cross section uncertainties.}
    \begin{tabular}{c||c|c||c|c|c}
      Bin edges (degree) &  $\sigma$  &  Abs. Unc.  &  Frac. Stat. Unc.  &  Frac. Sys. Unc.  &  Frac. Flux Unc. \\
      \hline
      \hline
        0 -- 1 & 45.1395 & 3.7456 & 0.039 & 0.073 & 0.037\\
        1 -- 2 & 133.403 & 8.8004 & 0.024 & 0.061 & 0.036\\
        2 -- 3 & 201.782 & 12.739 & 0.021 & 0.059 & 0.036\\
        3 -- 4 & 243.169 & 15.621 & 0.02 & 0.061 & 0.036\\
        4 -- 5 & 251.124 & 17.674 & 0.022 & 0.067 & 0.037\\
        5 -- 6 & 244.781 & 18.993 & 0.024 & 0.074 & 0.038\\
        6 -- 7 & 219.879 & 18.367 & 0.027 & 0.079 & 0.039\\
        7 -- 8 & 208.269 & 17.948 & 0.03 & 0.081 & 0.04\\
        8 -- 9 & 177.382 & 17.169 & 0.034 & 0.091 & 0.041\\
        9 -- 10 & 151.637 & 16.557 & 0.04 & 0.1 & 0.043\\
        10 -- 11 & 134.714 & 14.977 & 0.044 & 0.1 & 0.044\\
        11 -- 12 & 110.783 & 13.459 & 0.053 & 0.11 & 0.045\\
        12 -- 14 & 96.9158 & 10.776 & 0.041 & 0.1 & 0.046\\
        14 -- 16 & 71.9116 & 9.6244 & 0.058 & 0.12 & 0.048\\
        16 -- 20 & 44.9274 & 7.7396 & 0.068 & 0.16 & 0.052\\
    \end{tabular}
  \label{tbl:CrossSectionthetamu}
\end{table}

\begin{table}[H]
  \centering
  \renewcommand{\arraystretch}{1.15}
  \caption{Measured cross section as a function of \Qsq on scintillator, in units of $10^{-42}$ \cmsq/\GeVsq/nucleon with absolute and fractional cross section uncertainties.}
    \begin{tabular}{c||c|c||c|c|c}
      Bin edges (\GeVsq) &  $\sigma$  &  Abs. Unc.  &  Frac. Stat. Unc.  &  Frac. Sys. Unc.  &  Frac. Flux Unc. \\
      \hline
      \hline
        0.0 -- 0.025 & 3733.72 & 273.26 & 0.029 & 0.067 & 0.036\\
        0.025 -- 0.05 & 4095.02 & 304.4 & 0.039 & 0.063 & 0.037\\
        0.05 -- 0.1 & 3919.25 & 259.75 & 0.028 & 0.06 & 0.037\\
        0.1 -- 0.2 & 3709.28 & 237.78 & 0.021 & 0.061 & 0.037\\
        0.2 -- 0.3 & 3106.79 & 230.52 & 0.028 & 0.069 & 0.038\\
        0.3 -- 0.4 & 2668.38 & 210.66 & 0.032 & 0.072 & 0.038\\
        0.4 -- 0.5 & 1978.01 & 178.15 & 0.039 & 0.081 & 0.039\\
        0.5 -- 0.7 & 1639.55 & 154.3 & 0.034 & 0.088 & 0.04\\
        0.7 -- 1.0 & 1006.45 & 116.56 & 0.04 & 0.11 & 0.042\\
        1.0 -- 1.3 & 548.673 & 79.944 & 0.061 & 0.13 & 0.043\\
        1.3 -- 2.0 & 232.922 & 51.598 & 0.074 & 0.21 & 0.05\\
        2.0 -- 3.0 & 90.7911 & 29.736 & 0.15 & 0.29 & 0.057\\
    \end{tabular}
  \label{tbl:CrossSectionq2}
\end{table}

\FloatBarrier


\subsection{Selection tables}
\label{sec:selection_tables}

\begin{table}[H]
  \centering
  \renewcommand{\arraystretch}{1.15}
  \caption{Number of selected events in data before background subtraction versus \Tpi.}
    \begin{tabular}{c||c|c|c|c||c}
      Bin edges (\MeV) &  Total Sel. & Sel. Tracked & Sel. Untracked & Sel. Overlap & Frac. Stat. Unc. \\
      \hline
      \hline
        0 -- 20 & 1055 & 0 & 1055 & 0 & 0.031 \\
        20 -- 35 & 4046 & 6 & 4040 & 0 & 0.016 \\
        35 -- 50 & 5591 & 230 & 5312 & 49 & 0.013 \\
        50 -- 65 & 8860 & 2296 & 6183 & 381 & 0.011 \\
        65 -- 80 & 10957 & 4185 & 6078 & 694 & 0.0096  \\
        80 -- 100 & 14345 & 6980 & 6348 & 1017 & 0.0083 \\
        100 -- 125 & 15629 & 8519 & 5734 & 1376 & 0.008  \\
        125 -- 165 & 15546 & 8866 & 5288 & 1392 & 0.008 \\
        165 -- 200 & 7176 & 4151 & 2433 & 592 & 0.012 \\
        200 -- 350 & 8144 & 4899 & 2711 & 534 & 0.011 \\
    \end{tabular}
  \label{tbl:DataSelectiontpi}
\end{table}

\begin{table}[H]
  \centering
  \renewcommand{\arraystretch}{1.15}
  \caption{Number of selected events in data before background subtraction versus \thpi. }
    \begin{tabular}{c||c|c|c|c||c}
      Bin edges (degree) &  Total Sel. & Sel. Tracked & Sel. Untracked & Sel. Overlap & Frac. Stat. Unc. \\
      \hline
      \hline
        0 -- 15 & 4563 & 2611 & 1636 & 316 & 0.015  \\
        15 -- 30 & 12125 & 7222 & 3861 & 1042 & 0.0091  \\
        30 -- 45 & 16408 & 10011 & 4883 & 1514 & 0.0078  \\
        45 -- 60 & 16135 & 9206 & 5597 & 1332 & 0.0079 \\
        60 -- 76 & 10426 & 3576 & 6386 & 464 & 0.0098 \\
        76 -- 108 & 12772 & 326 & 12421 & 25 & 0.0088  \\
        108 -- 122 & 5569 & 1306 & 4061 & 202 & 0.013  \\
        122 -- 136 & 5718 & 2608 & 2635 & 475 & 0.013  \\
        136 -- 150 & 4569 & 2559 & 1551 & 459 & 0.015 \\
        150 -- 165 & 2082 & 1060 & 834 & 188 & 0.022 \\
        165 -- 180 & 421 & 140 & 262 & 19 & 0.049 \\
    \end{tabular}
  \label{tbl:DataSelectionthetapi}
\end{table}

\begin{table}[H]
  \centering
  \renewcommand{\arraystretch}{1.15}
  \caption{Number of selected events in data before background subtraction versus \pmu.}
    \begin{tabular}{c||c|c|c|c||c}
      Bin edges (\MeVc) &  Total Sel. & Sel. Tracked & Sel. Untracked & Sel. Overlap & Frac. Stat. Unc. \\
      \hline
      \hline
        1.0 -- 2.0 & 807 & 350 & 398 & 59 & 0.035 \\
        2.0 -- 3.0 & 7088 & 3096 & 3516 & 476 & 0.012 \\
        3.0 -- 4.0 & 14360 & 6088 & 7350 & 922 & 0.0083  \\
        4.0 -- 5.5 & 29562 & 12814 & 14793 & 1955 & 0.0058 \\
        5.5 -- 7.5 & 26635 & 11987 & 12834 & 1814 & 0.0061  \\
        7.5 -- 10 & 8643 & 4060 & 4024 & 559 & 0.011 \\
        10.0 -- 13.0 & 2665 & 1247 & 1265 & 153 & 0.019 \\
        13.0 -- 20.0 & 2083 & 981 & 1002 & 100 & 0.022 \\
    \end{tabular}
  \label{tbl:DataSelectionpmu}
\end{table}

\begin{table}[H]
  \centering
  \renewcommand{\arraystretch}{1.15}
  \caption{Number of selected events in data before background subtraction versus \pzmu.}
    \begin{tabular}{c||c|c|c|c||c}
      Bin edges (\MeVc) &  Total Sel. & Sel. Tracked & Sel. Untracked & Sel. Overlap & Frac. Stat. Unc. \\
      \hline
      \hline
        0 -- 1 & 0 & 0 & 0 & 0 & 0 \\
        1 -- 2 & 947 & 422 & 463 & 62 & 0.032  \\
        2 -- 3 & 7532 & 3258 & 3764 & 510 & 0.012 \\
        3 -- 4 & 14742 & 6260 & 7549 & 933 & 0.0082  \\
        4 -- 5 & 19562 & 8529 & 9715 & 1318 & 0.0071 \\
        5 -- 6 & 18762 & 8183 & 9333 & 1246 & 0.0073 \\
        6 -- 8 & 20037 & 9168 & 9511 & 1358 & 0.0071 \\
        8 -- 10 & 5570 & 2609 & 2600 & 361 & 0.013 \\
        10 -- 15 & 3524 & 1648 & 1674 & 202 & 0.017 \\
        15 -- 20 & 1167 & 546 & 573 & 48 & 0.029 \\
    \end{tabular}
  \label{tbl:DataSelectionpzmu}
\end{table}

\begin{table}[H]
  \centering
  \renewcommand{\arraystretch}{1.15}
  \caption{Number of selected events in data before background subtraction versus \ptmu.}
    \begin{tabular}{c||c|c|c|c||c}
      Bin edges (\MeVc) &  Total Sel. & Sel. Tracked & Sel. Untracked & Sel. Overlap & Frac. Stat. Unc. \\
      \hline
      \hline
        0.0 -- 0.1 & 1711 & 888 & 662 & 161 & 0.024 \\
        0.1 -- 0.2 & 5245 & 2522 & 2293 & 430 & 0.014 \\
        0.2 -- 0.3 & 8652 & 4132 & 3789 & 731 & 0.011 \\
        0.3 -- 0.4 & 11071 & 4963 & 5276 & 832 & 0.0095 \\
        0.4 -- 0.5 & 12002 & 5321 & 5838 & 843 & 0.0091 \\
        0.5 -- 0.6 & 11486 & 4852 & 5877 & 757 & 0.0093 \\
        0.6 -- 0.8 & 19370 & 8261 & 9916 & 1193 & 0.0072 \\
        0.8 -- 1.0 & 12163 & 4944 & 6528 & 691 & 0.0091 \\
        1.0 -- 1.25 & 7121 & 3004 & 3781 & 336 & 0.012 \\
        1.25 -- 1.5 & 2060 & 954 & 1046 & 60 & 0.022 \\
    \end{tabular}
  \label{tbl:DataSelectionptmu}
\end{table}

\begin{table}[H]
  \centering
  \renewcommand{\arraystretch}{1.15}
  \caption{Number of selected events in data before background subtraction versus \thmu.}
    \begin{tabular}{c||c|c|c|c||c}
      Bin edges (degree) &  Total Sel. & Sel. Tracked & Sel. Untracked & Sel. Overlap & Frac. Stat. Unc. \\
      \hline
      \hline
        0 -- 1 & 1778 & 923 & 707 & 148 & 0.024 \\
        1 -- 2 & 5218 & 2479 & 2293 & 446 & 0.014 \\
        2 -- 3 & 7813 & 3596 & 3605 & 612 & 0.011 \\
        3 -- 4 & 9432 & 4286 & 4450 & 696 & 0.01 \\
        4 -- 5 & 9825 & 4424 & 4709 & 692 & 0.01 \\
        5 -- 6 & 9496 & 4205 & 4611 & 680 & 0.01 \\
        6 -- 7 & 8664 & 3775 & 4287 & 602 & 0.011 \\
        7 -- 8 & 7927 & 3468 & 3960 & 499 & 0.011 \\
        8 -- 9 & 6682 & 2841 & 3431 & 410 & 0.012 \\
        9 -- 10 & 5647 & 2418 & 2924 & 305 & 0.013 \\
        10 -- 11 & 4634 & 2047 & 2342 & 245 & 0.015 \\
        11 -- 12 & 3707 & 1601 & 1914 & 192 & 0.016 \\
        12 -- 14 & 5245 & 2154 & 2831 & 260 & 0.014 \\
        14 -- 16 & 3160 & 1343 & 1678 & 139 & 0.018 \\
        16 -- 20 & 2615 & 1063 & 1440 & 112 & 0.02 \\
    \end{tabular}
  \label{tbl:DataSelectionthetamu}
\end{table}

\begin{table}[H]
  \centering
  \renewcommand{\arraystretch}{1.15}
  \caption{Number of selected events in data before background subtraction versus \Qsq.}
    \begin{tabular}{c||c|c|c|c||c
}
      Bin edges (\GeVsqcsq) &  Total Sel. & Sel. Tracked & Sel. Untracked & Sel. Overlap & Frac. Stat. Unc. \\
      \hline
      \hline
        0.0 -- 0.025 & 3887 & 1905 & 1640 & 342 & 0.016 \\
        0.025 -- 0.05 & 4026 & 1961 & 1729 & 336 & 0.016 \\
        0.05 -- 0.1 & 7925 & 3762 & 3501 & 662 & 0.011  \\
        0.1 -- 0.2 & 13816 & 6163 & 6595 & 1058 & 0.0085 \\
        0.2 -- 0.3 & 11243 & 4925 & 5571 & 747 & 0.0094 \\
        0.3 -- 0.4 & 9002 & 3818 & 4586 & 598 & 0.011 \\
        0.4 -- 0.5 & 7400 & 3181 & 3741 & 478 & 0.012 \\
        0.5 -- 0.7 & 10956 & 4672 & 5618 & 666 & 0.0096 \\
        0.7 -- 1.0 & 10157 & 4150 & 5433 & 574 & 0.0099 \\
        1.0 -- 1.3 & 5715 & 2322 & 3084 & 309 & 0.013 \\
        1.3 -- 2.0 & 5445 & 2381 & 2829 & 235 & 0.014 \\
        2.0 -- 3.0 & 1597 & 804 & 760 & 33 & 0.025 \\
    \end{tabular}
  \label{tbl:DataSelectionq2}
\end{table}

\FloatBarrier


\subsection{Model $\chi^2$ Values}
\label{sec:chi2_tables}

\begin{table}[H]
  \centering
  \renewcommand{\arraystretch}{1.15}
  \caption{$\chi^2$ for each model for \Tpi (ndf = 10)}
    \begin{tabular}{c||c|c}
        Model & Conventional $\chi^2$ & Log-normal $\chi^2$ \\
        \hline
        \hline
        MnvTune v4.7.1 & 9.56547 & 9.95949\\
        GENIE v2.12.6 & 254.277 & 127.67\\
        GENIE v3 $hA$  & 72.5972 & 92.5025\\
        GENIE v3 $hN$  & 45.4694 & 54.4004\\
        GiBUU & 45.2752 & 65.9332\\
        NEUT LFG & 56.0228 & 63.851\\
        NuWro LFG & 57.6983 & 75.803\\

    \end{tabular}
  \label{tbl:chi2_tpi}
\end{table}

\begin{table}[H]
  \centering
  \renewcommand{\arraystretch}{1.15}
  \caption{ $\chi^2$ for each model for \thpi (ndf = 11)}
    \begin{tabular}{c||c|c|c}

      Model & Conventional $\chi^2$ & Log-normal $\chi^2$ \\
      \hline
      \hline
        MnvTune v4.7.1 & 28.9918 & 33.8245\\
        GENIE v2.12.6 & 59.3826 & 65.6561\\
        GENIE v3 $hA$  & 38.0883 & 54.3804\\
        GENIE v3 $hN$  & 26.2851 & 34.8697\\
        GiBUU & 31.8699 & 43.0243\\
        NEUT LFG & 24.6153 & 25.144\\
        NuWro LFG & 19.0918 & 22.852\\
        
    \end{tabular}
  \label{tbl:chi2_thpi}
\end{table}

\begin{table}[H]
  \centering
  \renewcommand{\arraystretch}{1.15}
  \caption{$\chi^2$ for each model for \pmu (ndf = 8)}
    \begin{tabular}{c||c|c}
        Model & Conventional $\chi^2$ & Log-normal $\chi^2$ \\
        \hline
        \hline
        MnvTune v4.7.1 & 29.712 & 29.1455\\
        GENIE v2.12.6 & 33.9112 & 32.4504\\
        GENIE v3 $hA$  & 24.597 & 38.2269\\
        GENIE v3 $hN$  & 23.9233 & 33.188\\
        GiBUU & 25.9 & 38.3365\\
        NEUT LFG & 31.8742 & 34.7631\\
        NuWro LFG & 26.223 & 34.5181\\
    \end{tabular}
  \label{tbl:chi2_pmu}
\end{table}

\begin{table}[H]
  \centering
  \renewcommand{\arraystretch}{1.15}
  \caption{$\chi^2$ for each model for \pzmu (ndf = 10)}
    \begin{tabular}{c||c|c}
        Model & Conventional $\chi^2$ & Log-normal $\chi^2$ \\
  	    \hline
  	    \hline
        MnvTune v4.7.1 & 38.8537 & 39.3392\\
        GENIE v2.12.6 & 45.2432 & 44.8812\\
        GENIE v3 $hA$  & 32.8062 & 53.1369\\
        GENIE v3 $hN$  & 31.8845 & 45.7788\\
        GiBUU & 34 & 52.5511\\
        NEUT LFG & 39.7324 & 45.1796\\
        NuWro LFG & 34.2089 & 46.9136\\
    \end{tabular}
  \label{tbl:chi2_pzmu}
\end{table}

\begin{table}[H]
  \centering
  \renewcommand{\arraystretch}{1.15}
  \caption{$\chi^2$ for each model for \ptmu (ndf = 10)}
    \begin{tabular}{c||c|c}
        Model & Conventional $\chi^2$ & Log-normal $\chi^2$ \\
        \hline
        \hline
            MnvTune v4.7.1 & 13.5832 & 17.6924\\
            GENIE v2.12.6 & 28.3886 & 29.4045\\
            GENIE v3 $hA$  & 17.3331 & 29.3266\\
            GENIE v3 $hN$  & 14.6532 & 24.3992\\ 
            GiBUU & 14.4482 & 24.5983\\ 
            NEUT LFG & 10.6076 & 15.8653\\
            NuWro LFG & 20.6059 & 28.4988\\ 
    \end{tabular}
  \label{tbl:chi2_ptmu}
\end{table}

\begin{table}[H]
  \centering
  \renewcommand{\arraystretch}{1.15}
  \caption{$\chi^2$ for each model for \thmu (ndf = 15)}
    \begin{tabular}{c||c|c}
        Model & Conventional $\chi^2$ & Log-normal $\chi^2$ \\
        \hline
        \hline
        MnvTune v4.7.1 & 10.9635 & 11.6781\\ 
        GENIE v2.12.6 & 32.9809 & 32.2083\\
        GENIE v3 $hA$  & 18.9206 & 26.7965\\
        GENIE v3 $$hN$$  & 14.3069 & 19.8863\\ 
        GiBUU & 17.9112 & 27.4868\\ 
        NEUT LFG & 18.7421 & 24.1093\\ 
        NuWro LFG & 17.5757 & 24.3697\\
    \end{tabular}
  \label{tbl:chi2_thmu}
\end{table}

\begin{table}[H]
  \centering
  \renewcommand{\arraystretch}{1.15}
  \caption{$\chi^2$ for each model for \Qsq (ndf = 12)}
    \begin{tabular}{c||c|c}
        Model & Conventional $\chi^2$ & Log-normal $\chi^2$ \\
        \hline
        \hline
        MnvTune v4.7.1 & 9.06689 & 11.2707\\
        GENIE v2.12.6 & 22.9818 & 21.028\\
        GENIE v3 $hA$  & 23.1635 & 24.2076\\
        GENIE v3 $hN$  & 18.8989 & 18.6167\\
        GiBUU & 18.6934 & 21.509\\
        NEUT LFG & 13.9915 & 17.2119\\
        NuWro LFG & 23.4356 & 17.8761\\
    \end{tabular}
  \label{tbl:chi2_q2}
\end{table}

\subsection{Sideband Background Tune}
\label{sec:sidebands}

\begin{figure}[!h]
    \includegraphics[width=0.45\linewidth]{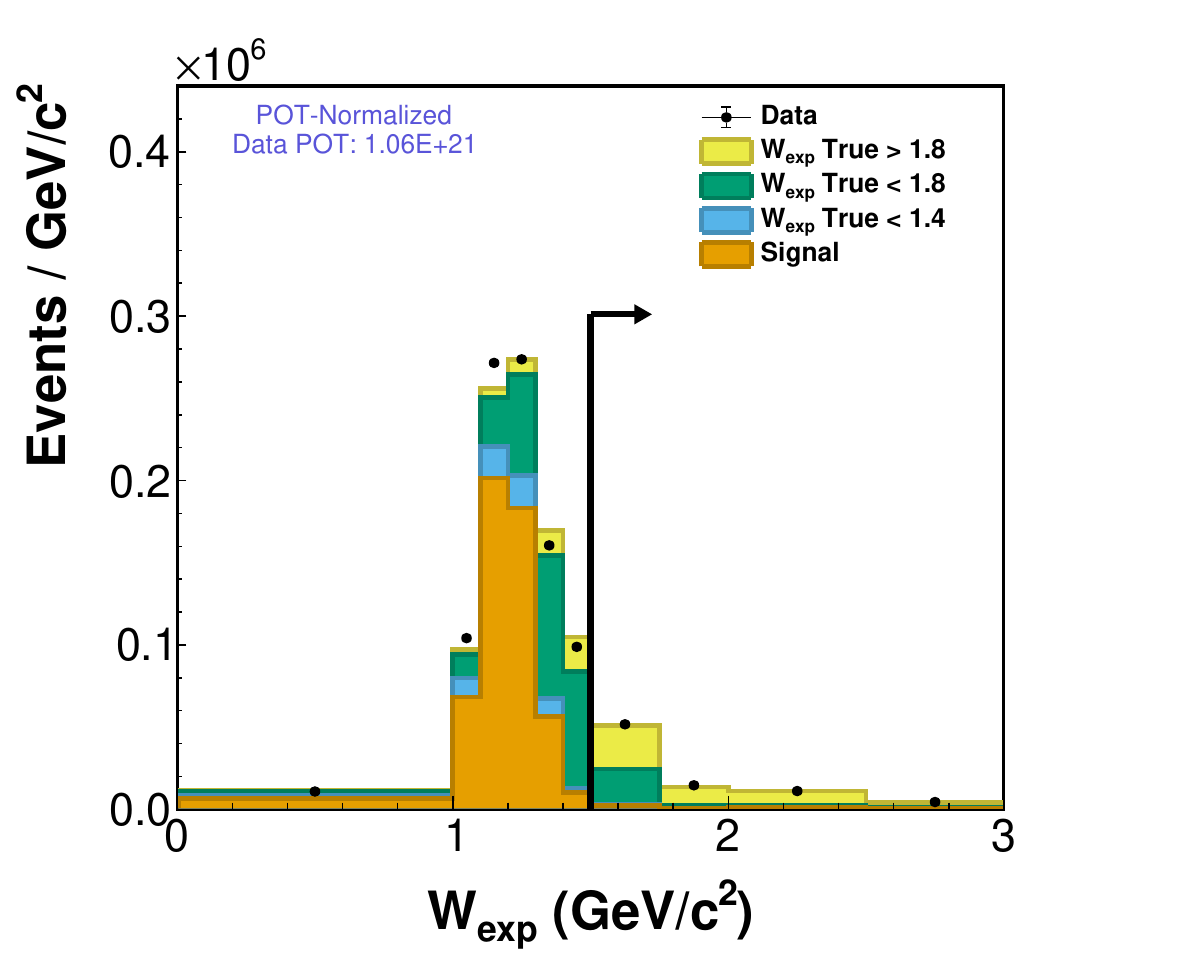}
    \includegraphics[width=0.45\linewidth]{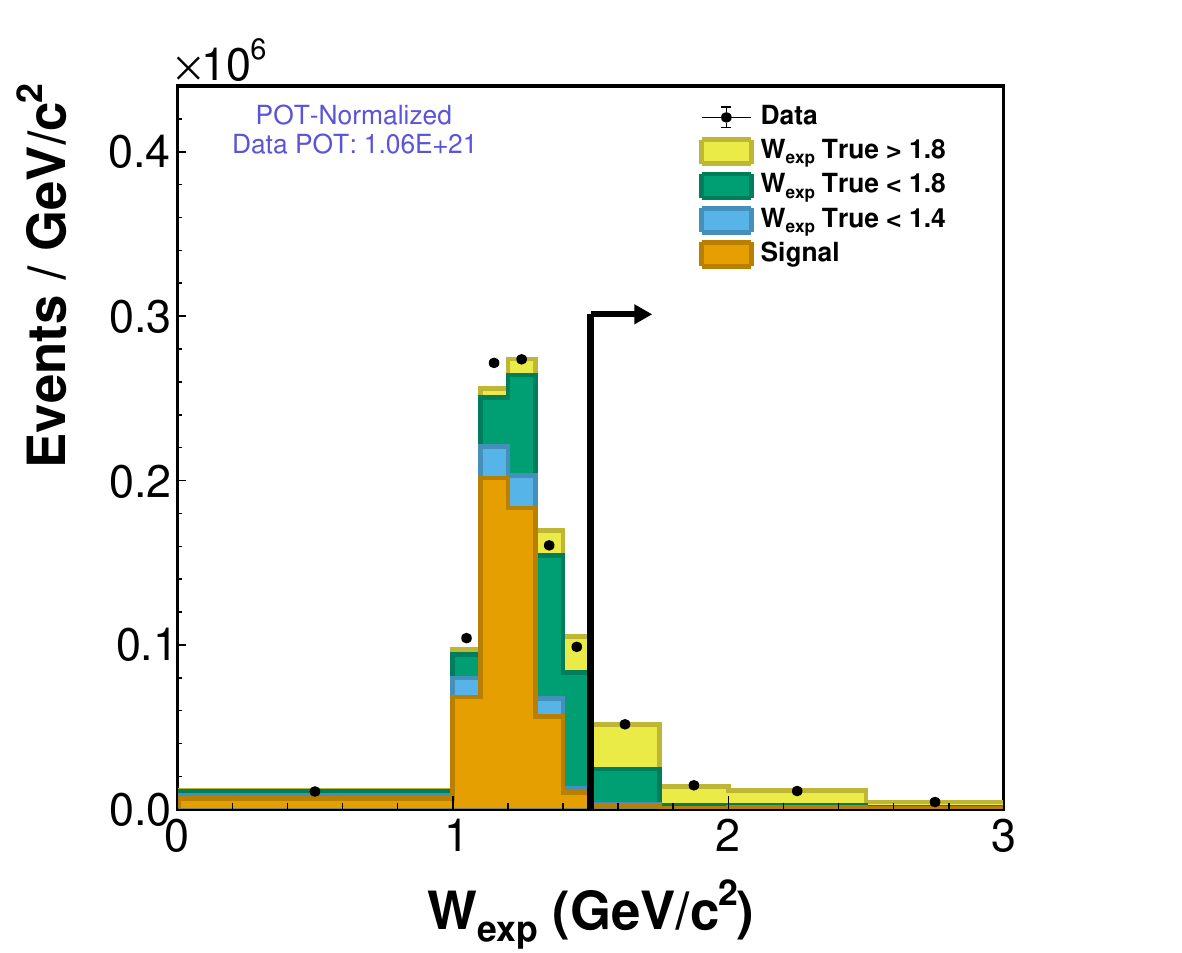}
    \caption{\Wexp distribution of selected events for the \Tpi, \pmu, \pzmu, \ptmu, \thmu, and \Qsq cross section analysis before (left) and after (right) constraining the simulation using background estimates from data. The arrow delineates the high \Wexp sideband region.  Since the fitted background weights are close to unity, the differences between the two distributions are small.  }
    \label{fig:supp:sidebands}
\end{figure}

\begin{figure}[H]
    \centering
    \includegraphics[width=0.45\linewidth]{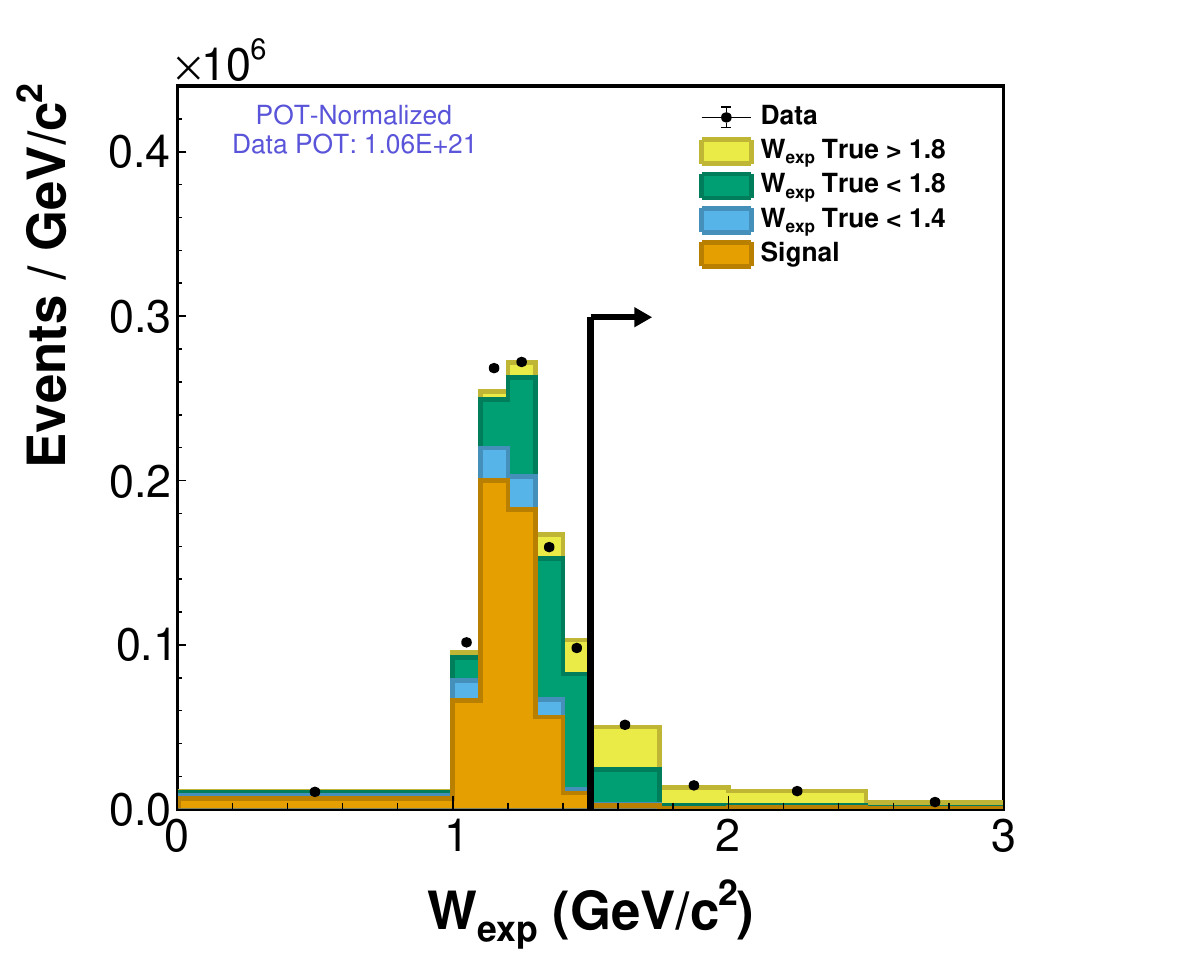}
    \includegraphics[width=0.45\linewidth]{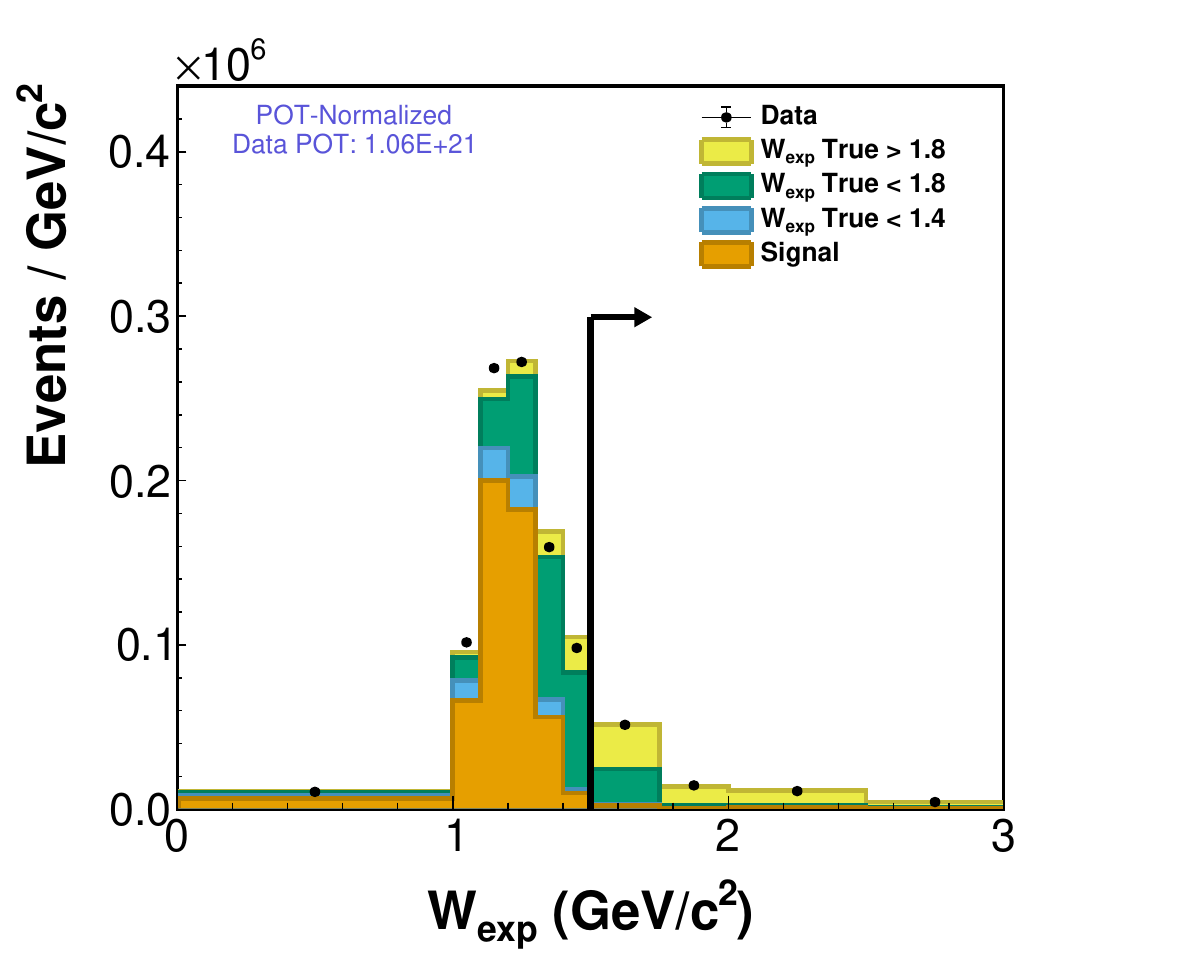}
    \caption{\Wexp distribution of selected events for the \thpi cross section analysis before (left) and after (right) constraining the simulation using background estimates from data. The arrow delineates the high \Wexp sideband region.  Since the fitted background weights are close to unity, the differences between the two distributions are small. }
    \label{fig:supp:sidebands_thpi}
\end{figure}

\subsection{Efficiencies}
\label{sec:efficiencies}

\begin{figure}[!h]
    \includegraphics[width=0.32\linewidth]{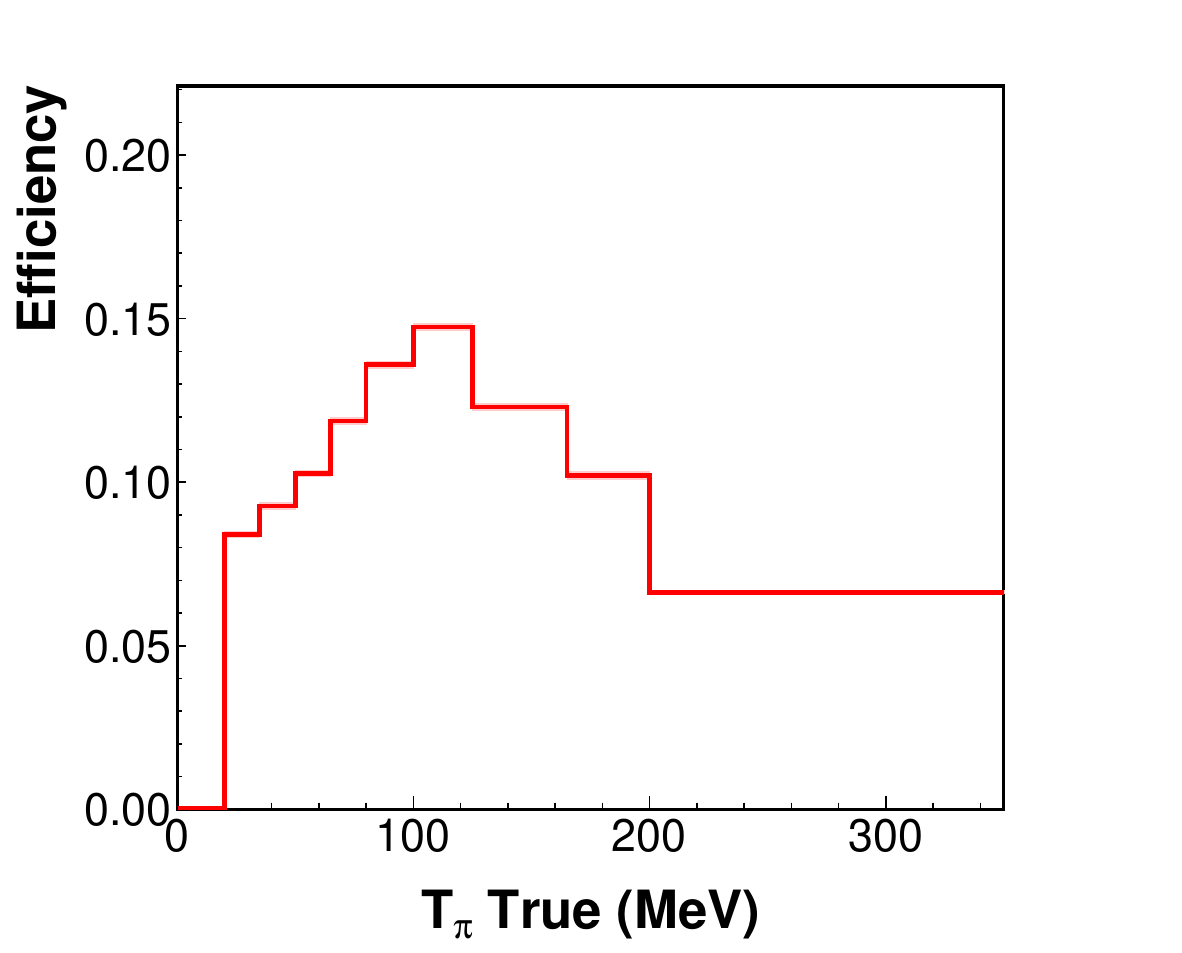}
    \includegraphics[width=0.32\linewidth]{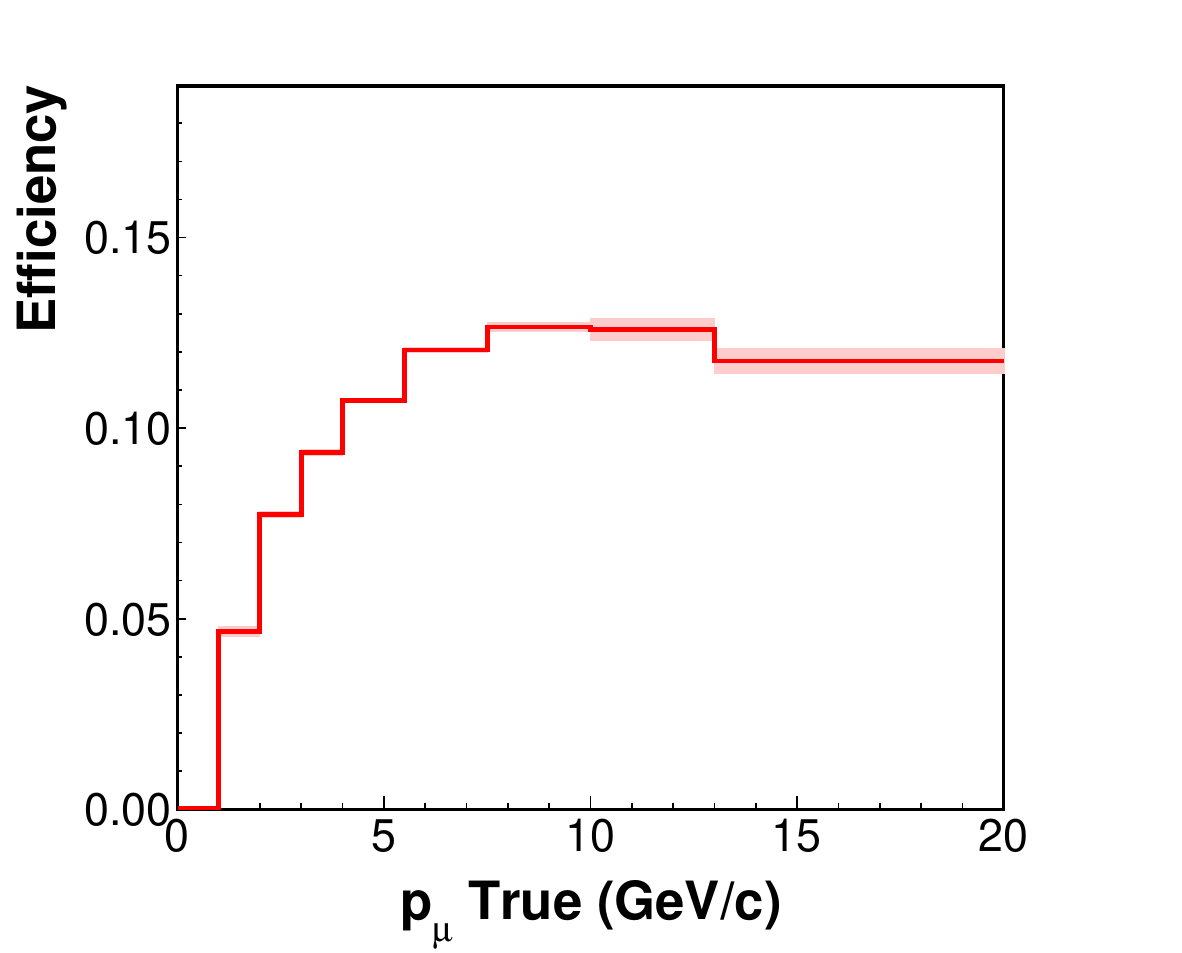}
    \includegraphics[width=0.32\linewidth]{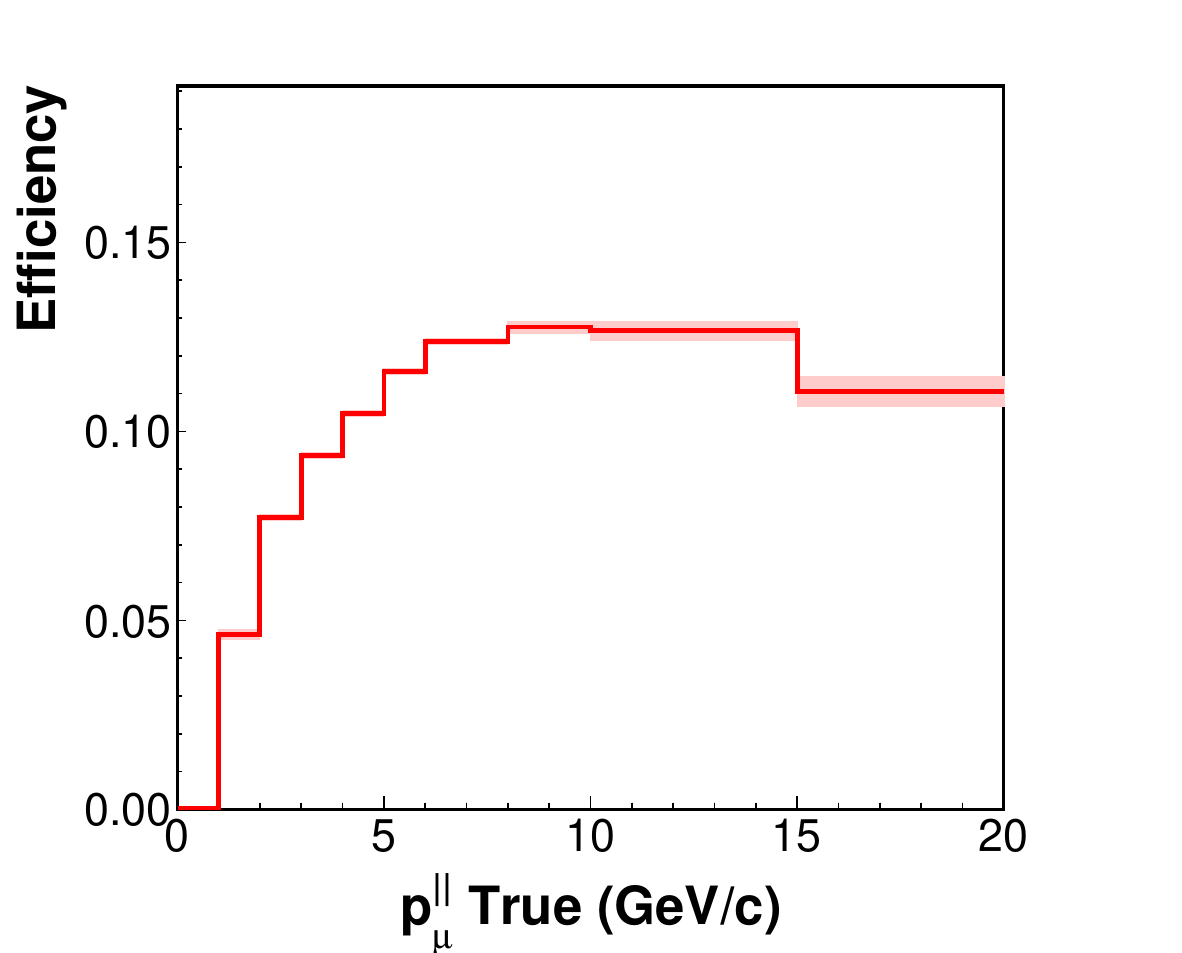}
    \includegraphics[width=0.32\linewidth]{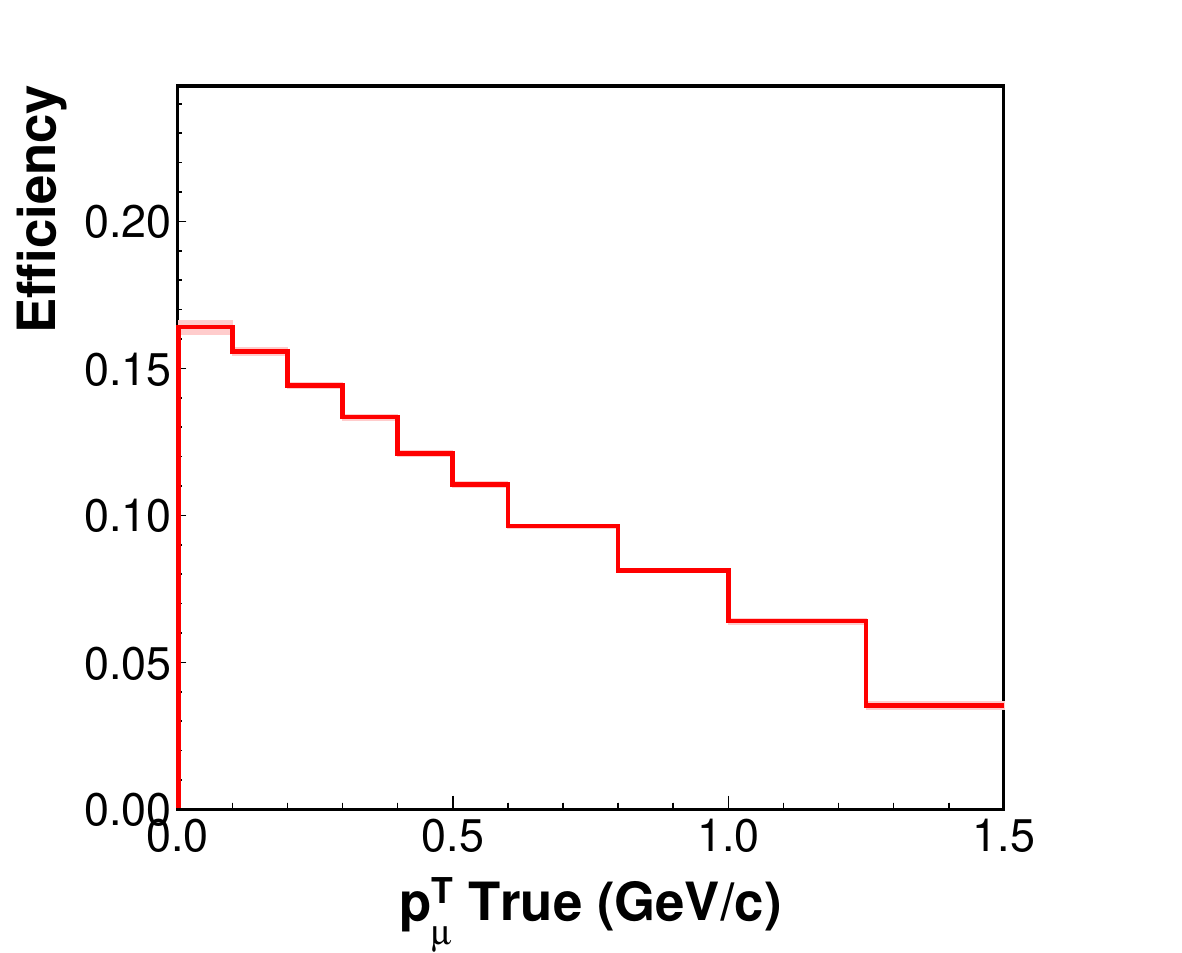}
    \includegraphics[width=0.32\linewidth]{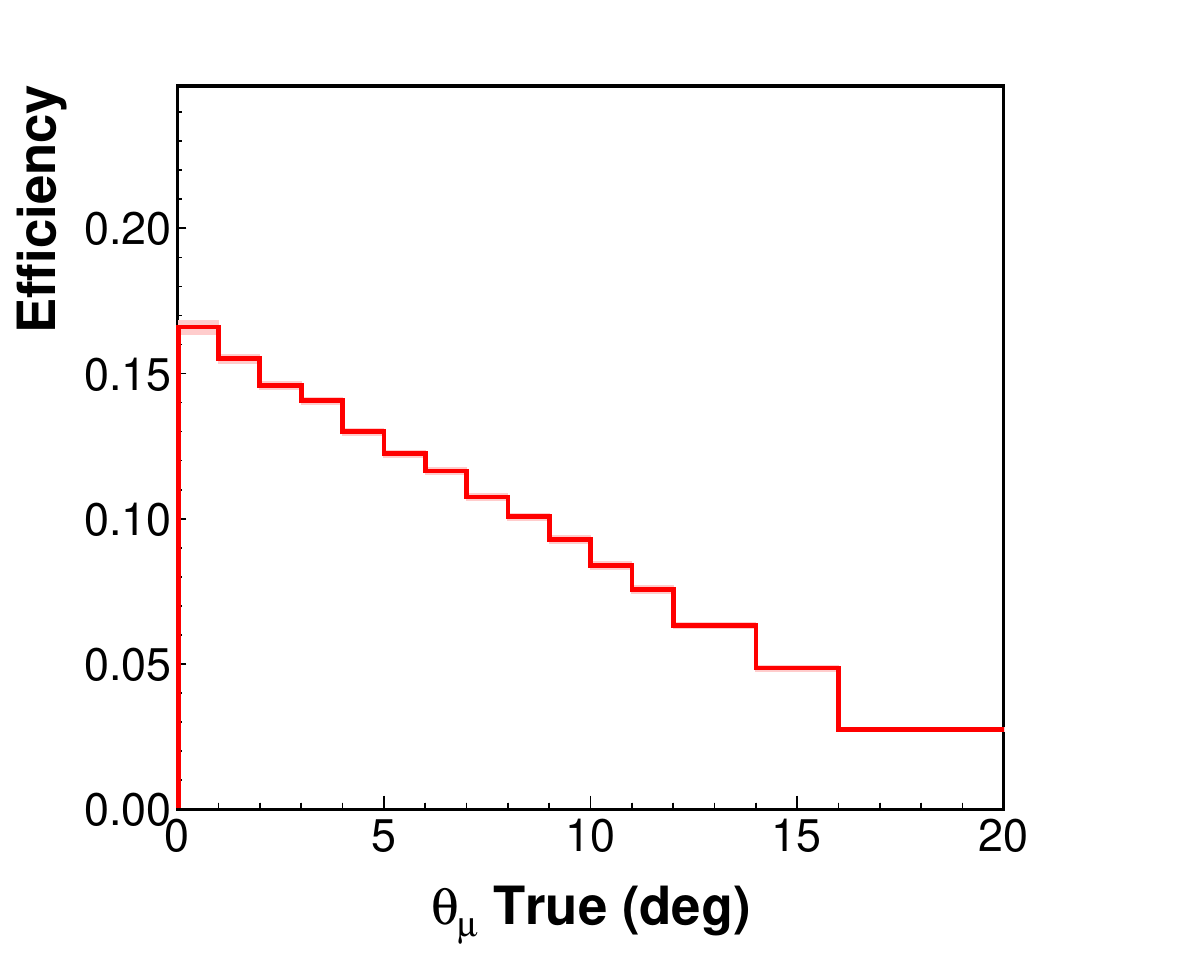}
    \includegraphics[width=0.32\linewidth]{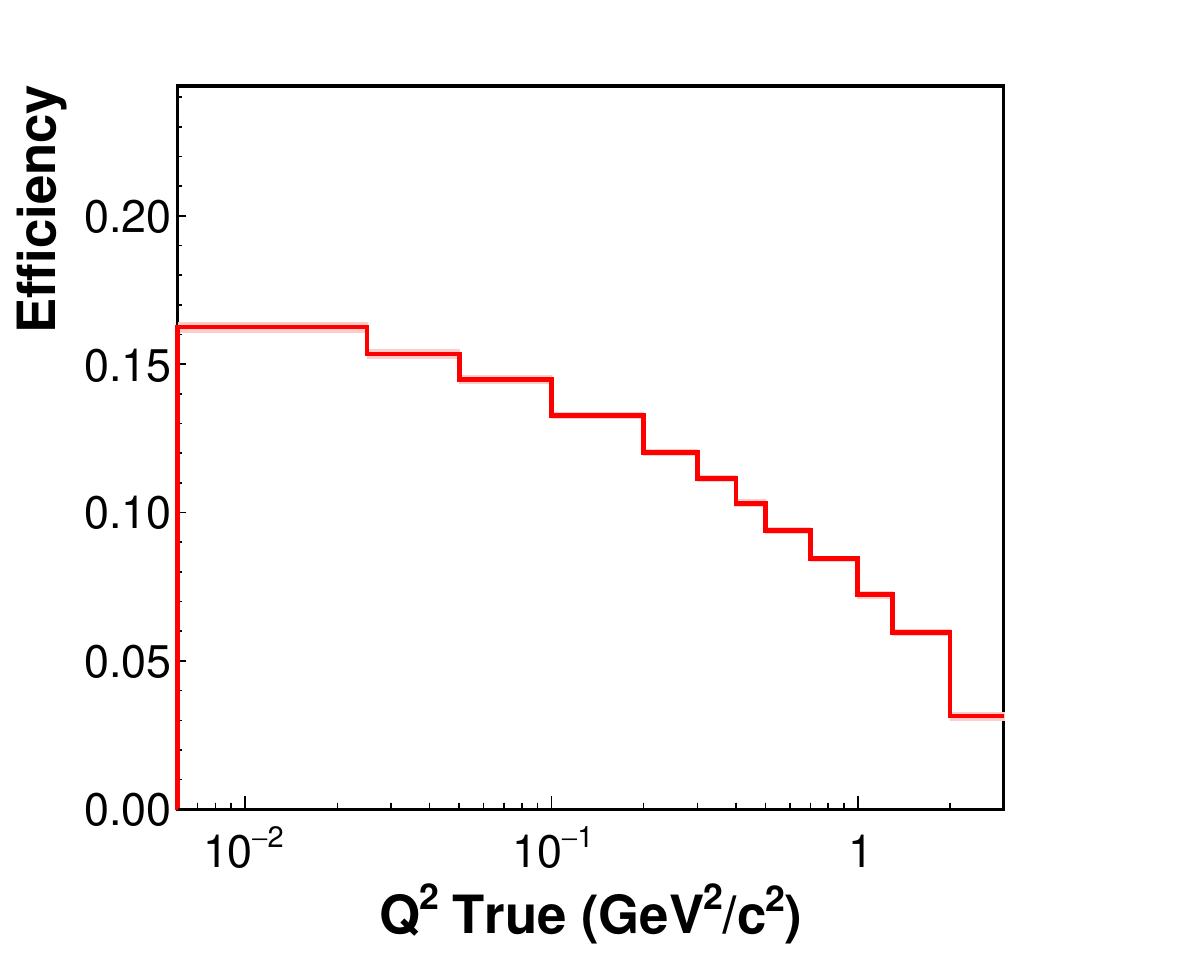}
    \caption{Efficiencies. The shaded error bands show total uncertainties, including systematics and simulated statistics. These plots show the efficiency of many regions above 10\%, compared with the results from \cite{Bercellie.131.011801} where the efficiency for most part of the bins is bellow 8\%. The combination of the two data selection techniques leads to the substantial efficiency improvement.}
    \label{fig:supp:efficiencies}
\end{figure}



\end{document}


\setlength\tabcolsep{0.5em}
\newcommand{\minerva}{MINERvA\xspace}
\newcommand{\qelike}{quasielastic-like}
\newcommand{\tune}{\minerva~tune v1}
\newcommand{\ccnopi}{CC0$\pi$\xspace}
\newcommand{\enuqe}{$E_{\nu,QE}$\xspace}
\newcommand{\pt}{$p_{t}$\xspace}
\newcommand{\ptmu}{$p_{T \mu}$\xspace}
\newcommand{\pz}{$p_{||}$\xspace}
\newcommand{\pzmu}{$p_{|| \mu}$\xspace}
\newcommand{\eavail}{$E_{available}$\xspace}
\newcommand{\sumtp}{$\Sigma T_{p}$\xspace}
\newcommand{\qqqe}{$Q^{2}_{QE}$\xspace}
\newcommand{\addcitation}{[ADD CITATION]}
\newcommand{\Emu}{\ensuremath{E_{\mu}}\xspace}
\newcommand{\pmu}{\ensuremath{p_{\mu}}\xspace}
\newcommand{\thmu}{\ensuremath{\theta_{\mu}}\xspace}
\newcommand{\Ehad}{\ensuremath{E_{\mathrm{had}}}\xspace}
\newcommand{\Erec}{\ensuremath{E_{recoil}}\xspace}
\newcommand{\Tpi}{\ensuremath{T_{\pi}}\xspace}
\newcommand{\thpi}{\ensuremath{\theta_{\pi}}\xspace}
\newcommand{\W}{\ensuremath{W}\xspace}
\newcommand{\Wexp}{\ensuremath{W_{\mathrm{exp}}}\xspace}
\newcommand{\Enu}{\ensuremath{E_{\nu}}\xspace}
\newcommand{\Qsq}{\ensuremath{Q^{2}}\xspace}
\newcommand{\Qsqexp}{\ensuremath{Q^{2}_{exp}}\xspace}
\newcommand{\datapot}{\ensuremath{\mbox{9.99}\times\mbox{10}^{999}}\xspace}

\newcommand{\nue}{\ensuremath{\nu_{e}}\xspace}
\newcommand{\numu}{\ensuremath{\nu_{\mu}}\xspace}
\newcommand{\nutau}{\ensuremath{\nu_{\tau}}\xspace}
\newcommand{\nuebar}{\ensuremath{\overline{\nu}_{e}}\xspace}
\newcommand{\numubar}{\ensuremath{\overline{\nu}_{\mu}}\xspace}
\newcommand{\nutaubar}{\ensuremath{\overline{\nu}_{\tau}}\xspace}
\newcommand{\nuone}{\ensuremath{\nu_{1}}\xspace}
\newcommand{\nutwo}{\ensuremath{\nu_{2}}\xspace}
\newcommand{\nuthree}{\ensuremath{\nu_{3}}\xspace}
\newcommand{\pion}{\ensuremath{\pi}\xspace}
\newcommand{\piplus}{\ensuremath{\pi^{+}}\xspace}
\newcommand{\pip}{\ensuremath{\pi^{+}}\xspace}
\newcommand{\piminus}{\ensuremath{\pi^{-}}\xspace}
\newcommand{\pim}{\ensuremath{\pi^{-}}\xspace}
\newcommand{\pipm}{\ensuremath{\pi^{\pm}}\xspace}
\newcommand{\pizero}{\ensuremath{\pi^{0}}\xspace}
\newcommand{\kplus}{\ensuremath{\mbox{K}^{+}}\xspace}
\newcommand{\kap}{\ensuremath{\mbox{K}^{+}}\xspace}
\newcommand{\kminus}{\ensuremath{\mbox{K}^{-}}\xspace}
\newcommand{\kam}{\ensuremath{\mbox{K}^{-}}\xspace}
\newcommand{\kpm}{\ensuremath{\mbox{K}^{\pm}}\xspace}
\newcommand{\kzero}{\ensuremath{\mbox{K}^{0}}\xspace}
\newcommand{\kzeroL}{\ensuremath{\mbox{K}^{0}_{\mathrm{L}}}\xspace}
\newcommand{\prot}{\ensuremath{\mbox{p}}\xspace}
\newcommand{\neut}{\ensuremath{\mbox{n}}\xspace}
\newcommand{\muplus}{\ensuremath{\mu^{+}}\xspace}
\newcommand{\mup}{\ensuremath{\mu^{+}}\xspace}
\newcommand{\muminus}{\ensuremath{\mu^{-}}\xspace}
\newcommand{\mum}{\ensuremath{\mu^{-}}\xspace}
\newcommand{\mupm}{\ensuremath{\mu^{\pm}}\xspace}
\newcommand{\muon}{\ensuremath{\mu}\xspace}
\newcommand{\eplus}{\ensuremath{\mbox{e}^{+}}\xspace}
\newcommand{\eminus}{\ensuremath{\mbox{e}^{-}}\xspace}
\newcommand{\deltabaryonp}{\ensuremath{\Delta^{+}}\xspace}
\newcommand{\deltabaryonpp}{\ensuremath{\Delta^{++}}\xspace}

\newcommand{\kg}{\ensuremath{\mbox{kg}}\xspace}

\newcommand{\kA}{\ensuremath{\mbox{kA}}\xspace}

\newcommand{\rad}{\ensuremath{\mbox{rad}}\xspace}
\newcommand{\mrad}{\ensuremath{\mbox{mrad}}\xspace}
\newcommand{\degs}{\textdegree\xspace}

\newcommand{\eV}{\ensuremath{\mbox{eV}}\xspace}
\newcommand{\keV}{\ensuremath{\mbox{keV}}\xspace}
\newcommand{\MeV}{\ensuremath{\mathrm{MeV}}\xspace}
\newcommand{\GeV}{\ensuremath{\mathrm{GeV}}\xspace}
\newcommand{\GeVsq}{\ensuremath{\mathrm{GeV}^2}\xspace}
\newcommand{\eVc}{\ensuremath{\mathrm{eV/\textit{c}}}\xspace}
\newcommand{\MeVc}{\ensuremath{\mathrm{MeV/\textit{c}}}\xspace}
\newcommand{\GeVc}{\ensuremath{\mathrm{GeV/\textit{c}}}\xspace}
\newcommand{\GeVcsq}{\ensuremath{\mathrm{GeV/\textit{c}}^2}\xspace}
\newcommand{\GeVsqcsq}{\ensuremath{\mathrm{GeV}^2/\mathrm{\textit{c}}^2}\xspace}

\newcommand{\eVsq}{\ensuremath{\mbox{eV}^2}\xspace}
\newcommand{\eVsqbf}{\ensuremath{\mathbf{\mbox{eV}^2}}\xspace}

\newcommand{\T}{\ensuremath{\mbox{T}}\xspace}
\newcommand{\mmsq}{\ensuremath{\mbox{mm}^2}\xspace}
\newcommand{\cmsq}{\ensuremath{\mbox{cm}^2}\xspace}
\newcommand{\cmcube}{\ensuremath{\mbox{cm}^3}\xspace}
\newcommand{\gcmcube}{\ensuremath{\mbox{g/cm}^3}\xspace}
\newcommand{\msq}{\ensuremath{\mbox{m}^2}\xspace}
\newcommand{\mcube}{\ensuremath{\mbox{m}^3}\xspace}

\newcommand{\micron}{\ensuremath{\mu \mbox{m}}\xspace}
\newcommand{\nm}{\ensuremath{\mbox{nm}}\xspace}
\newcommand{\mm}{\ensuremath{\mbox{mm}}\xspace}
\newcommand{\cm}{\ensuremath{\mbox{cm}}\xspace}
\newcommand{\m}{\ensuremath{\mbox{m}}\xspace}
\newcommand{\km}{\ensuremath{\mbox{km}}\xspace}

\newcommand{\ps}{\ensuremath{\mbox{ps}}\xspace}
\newcommand{\ns}{\ensuremath{\mbox{ns}}\xspace}
\newcommand{\micros}{\ensuremath{\mu \mbox{s}}\xspace}
\newcommand{\ms}{\ensuremath{\mbox{ms}}\xspace}
\newcommand{\s}{\ensuremath{\mbox{s}}\xspace}

\newcommand{\bkgdtype}{_MENU1PI}
\begin{table}
  \centering
  \renewcommand{\arraystretch}{1.15}
  \caption{Statistical covariance matrix of measured cross section as a function of $T_{\pi}$, in units of $10^{-84}$ $(\cmsq/\MeV/\mathrm{nucleon})^2$}
  \begin{adjustbox}{max width=\textwidth}
    \begin{tabular}{c||cccccccccc}
      Bin edges (\MeV) & 0-20 & 20-35 & 35-50 & 50-65 & 65-80 & 80-100 & 100-125 & 125-165 & 165-200 & 200-350\\
      \hline
      \hline
      0-20 & 0.044 & 0.011 & -0.005 & -0.003 & -0.001 & 0.000 & 0.000 & 0.000 & 0.000 & 0.000\\
      20-35 & 0.011 & 0.174 & 0.009 & -0.013 & -0.005 & -0.002 & -0.001 & -0.001 & -0.001 & 0.000\\
      35-50 & -0.005 & 0.009 & 0.278 & 0.019 & -0.015 & -0.005 & -0.002 & -0.001 & -0.001 & -0.001\\
      50-65 & -0.003 & -0.013 & 0.019 & 0.353 & 0.022 & -0.012 & -0.006 & -0.002 & -0.002 & -0.001\\
      65-80 & -0.001 & -0.005 & -0.015 & 0.022 & 0.339 & 0.015 & -0.011 & -0.004 & -0.002 & -0.002\\
      80-100 & 0.000 & -0.002 & -0.005 & -0.012 & 0.015 & 0.206 & 0.006 & -0.007 & -0.004 & -0.002\\
      100-125 & 0.000 & -0.001 & -0.002 & -0.006 & -0.011 & 0.006 & 0.169 & 0.001 & -0.007 & -0.003\\
      125-165 & 0.000 & -0.001 & -0.001 & -0.002 & -0.004 & -0.007 & 0.001 & 0.110 & 0.003 & -0.004\\
      165-200 & 0.000 & -0.001 & -0.001 & -0.002 & -0.002 & -0.004 & -0.007 & 0.003 & 0.140 & 0.000\\
      200-350 & 0.000 & 0.000 & -0.001 & -0.001 & -0.002 & -0.002 & -0.003 & -0.004 & 0.000 & 0.030
    \end{tabular}
  \end{adjustbox}
  \label{tbl:statcov_mixtpi_scintillator}
\end{table}

\begin{table}
  \centering
  \renewcommand{\arraystretch}{1.15}
  \caption{Systematic covariance matrix of measured cross section as a function of $T_{\pi}$, in units of $10^{-84}$ $(\cmsq/\MeV/\mathrm{nucleon})^2$}
  \begin{adjustbox}{max width=\textwidth}
    \begin{tabular}{c||cccccccccc}
      Bin edges (\MeV) & 0-20 & 20-35 & 35-50 & 50-65 & 65-80 & 80-100 & 100-125 & 125-165 & 165-200 & 200-350\\
      \hline
      \hline
      0-20 & 0.069 & 0.138 & 0.176 & 0.211 & 0.136 & 0.122 & 0.088 & 0.085 & 0.038 & 0.035\\
      20-35 & 0.138 & 0.390 & 0.400 & 0.534 & 0.435 & 0.349 & 0.258 & 0.222 & 0.107 & 0.141\\
      35-50 & 0.176 & 0.400 & 0.598 & 0.659 & 0.581 & 0.517 & 0.467 & 0.435 & 0.277 & 0.241\\
      50-65 & 0.211 & 0.534 & 0.659 & 0.945 & 0.750 & 0.658 & 0.554 & 0.475 & 0.293 & 0.214\\
      65-80 & 0.136 & 0.435 & 0.581 & 0.750 & 0.814 & 0.697 & 0.657 & 0.575 & 0.396 & 0.333\\
      80-100 & 0.122 & 0.349 & 0.517 & 0.658 & 0.697 & 0.637 & 0.615 & 0.538 & 0.369 & 0.281\\
      100-125 & 0.088 & 0.258 & 0.467 & 0.554 & 0.657 & 0.615 & 0.688 & 0.635 & 0.485 & 0.333\\
      125-165 & 0.085 & 0.222 & 0.435 & 0.475 & 0.575 & 0.538 & 0.635 & 0.641 & 0.497 & 0.338\\
      165-200 & 0.038 & 0.107 & 0.277 & 0.293 & 0.396 & 0.369 & 0.485 & 0.497 & 0.431 & 0.260\\
      200-350 & 0.035 & 0.141 & 0.241 & 0.214 & 0.333 & 0.281 & 0.333 & 0.338 & 0.260 & 0.401
    \end{tabular}
  \end{adjustbox}
  \label{tbl:syscov_mixtpi_scintillator}
\end{table}

\begin{table}
  \centering
  \renewcommand{\arraystretch}{1.15}
  \caption{Statistical covariance matrix of measured cross section as a function of $p_{\mu}$, in units of $10^{-84}$ $(\cmsq/\GeVc/\mathrm{nucleon})^2$}
  \begin{adjustbox}{max width=\textwidth}
    \begin{tabular}{c||ccccccccc}
      Bin edges (\GeVc) & 1.0-2.0 & 2.0-3.0 & 3.0-4.0 & 4.0-5.5 & 5.5-7.5 & 7.5-10.0 & 10.0-13.0 & 13.0-20.0\\
      \hline
      \hline
      1.0-2.0 & 31.572 & -2.731 & -0.520 & 0.007 & -0.008 & -0.004 & -0.002 & -0.002\\
      2.0-3.0 & -2.731 & 100.541 & -9.012 & -1.584 & 0.172 & -0.002 & -0.011 & -0.005\\
      3.0-4.0 & -0.520 & -9.012 & 152.644 & -9.033 & -1.381 & 0.188 & 0.002 & -0.007\\
      4.0-5.5 & 0.007 & -1.584 & -9.033 & 105.664 & -5.738 & -0.907 & 0.062 & -0.003\\
      5.5-7.5 & -0.008 & 0.172 & -1.381 & -5.738 & 43.363 & -1.174 & -0.374 & 0.007\\
      7.5-10.0 & -0.004 & -0.002 & 0.188 & -0.907 & -1.174 & 8.158 & -0.012 & -0.064\\
      10.0-13.0 & -0.002 & -0.011 & 0.002 & 0.062 & -0.374 & -0.012 & 1.704 & -0.036\\
      13.0-20.0 & -0.002 & -0.005 & -0.007 & -0.003 & 0.007 & -0.064 & -0.036 & 0.303
    \end{tabular}
  \end{adjustbox}
  \label{tbl:statcov_pmu_scintillator}
\end{table}

\begin{table}
  \centering
  \renewcommand{\arraystretch}{1.15}
  \caption{Systematic covariance matrix of measured cross section as a function of $p_{\mu}$, in units of $10^{-84}$ $(\cmsq/\GeVc/\mathrm{nucleon})^2$}
  \begin{adjustbox}{max width=\textwidth}
    \begin{tabular}{c||ccccccccc}
      Bin edges (\GeVc) & 1.0-2.0 & 2.0-3.0 & 3.0-4.0 & 4.0-5.5 & 5.5-7.5 & 7.5-10.0 & 10.0-13.0 & 13.0-20.0\\
      \hline
      \hline
      1.0-2.0 & 24.229 & 98.485 & 149.948 & 139.797 & 54.665 & 5.110 & 1.414 & 1.455\\
      2.0-3.0 &  98.485 & 555.404 & 895.910 & 880.907 & 449.443 & 75.602 & 16.967 & 7.919\\
      3.0-4.0 &  149.948 & 895.910 & 1503.741 & 1508.954 & 782.012 & 134.859 & 30.272 & 14.165\\
      4.0-5.5 &  139.797 & 880.907 & 1508.954 & 1712.809 & 994.651 & 199.094 & 43.026 & 19.432\\
      5.5-7.5 &  54.665 & 449.443 & 782.012 & 994.651 & 698.918 & 168.593 & 35.425 & 11.899\\
      7.5-10.0 &  5.110 & 75.602 & 134.859 & 199.094 & 168.593 & 49.562 & 10.143 & 2.618\\
      10.0-13.0 &  1.414 & 16.967 & 30.272 & 43.026 & 35.425 & 10.143 & 2.321 & 0.605\\
      13.0-20.0 &  1.455 & 7.919 & 14.165 & 19.432 & 11.899 & 2.618 & 0.605 & 0.382
    \end{tabular}
  \end{adjustbox}
  \label{tbl:syscov_pmu_scintillator}
\end{table}

\pagebreak
\begin{landscape}
\begin{table}
  \centering
  \renewcommand{\arraystretch}{1.15}
  \caption{Statistical covariance matrix of measured cross section as a function of $p_{\mu,T}$, in units of $10^{-84}$ $(\cmsq/\GeVc/\mathrm{nucleon})^2$}
  \begin{adjustbox}{max width=1.3\textheight}
    \begin{tabular}{c||cccccccccc}
      Bin edges (\GeVc) & 0.00-0.10 & 0.10-0.20 & 0.20-0.30 & 0.30-0.40 & 0.40-0.50 & 0.50-0.60 & 0.60-0.80 & 0.80-1.00 & 1.00-1.25 & 1.25-1.50\\
      \hline
      \hline
      0.00-0.10 & 619.145 & -89.777 & -34.709 & 19.023 & 3.529 & -1.016 & -0.539 & -0.039 & 0.032 & -0.008\\
      0.10-0.20 & -89.777 & 2241.606 & -176.862 & -119.660 & 38.104 & 13.379 & -1.026 & -0.426 & 0.019 & 0.208\\
      0.20-0.30 & -34.709 & -176.862 & 4626.883 & -245.839 & -290.794 & 42.632 & 17.481 & -0.915 & -0.309 & -0.302\\
      0.30-0.40 & 19.023 & -119.660 & -245.839 & 7640.913 & -208.961 & -568.132 & 26.393 & 20.353 & -0.192 & -2.143\\
      0.40-0.50 & 3.529 & 38.104 & -290.794 & -208.961 & 10561.433 & 58.167 & -470.764 & 30.791 & 20.573 & 2.938\\
      0.50-0.60 & -1.016 & 13.379 & 42.632 & -568.132 & 58.167 & 13681.774 & -224.546 & -374.624 & 11.563 & 6.389\\
      0.60-0.80 & -0.539 & -1.026 & 17.481 & 26.393 & -470.764 & -224.546 & 6679.478 & -229.139 & -248.609 & -36.635\\
      0.80-1.00 & -0.039 & -0.426 & -0.915 & 20.353 & 30.791 & -374.624 & -229.139 & 6465.738 & -19.950 & -278.470\\
      1.00-1.25 & 0.032 & 0.019 & -0.309 & -0.192 & 20.573 & 11.563 & -248.609 & -19.950 & 3348.356 & 118.470\\
      1.25-1.50 & -0.008 & 0.208 & -0.302 & -2.143 & 2.938 & 6.389 & -36.635 & -278.470 & 118.470 & 3174.331
    \end{tabular}
  \end{adjustbox}
  \label{tbl:statcov_ptmu_scintillator}
\end{table}
\end{landscape}
\clearpage

\pagebreak
\begin{landscape}
\begin{table}
  \centering
  \renewcommand{\arraystretch}{1.15}
  \caption{Systematic covariance matrix of measured cross section as a function of $p_{\mu,T}$, in units of $10^{-84}$ $(\cmsq/\GeVc/\mathrm{nucleon})^2$}
  \begin{adjustbox}{max width=1.3\textheight}
    \begin{tabular}{c||cccccccccc}
      Bin edges (\GeVc) & 0.00-0.10 & 0.10-0.20 & 0.20-0.30 & 0.30-0.40 & 0.40-0.50 & 0.50-0.60 & 0.60-0.80 & 0.80-1.00 & 1.00-1.25 & 1.25-1.50\\
      \hline
      \hline
      0.00-0.10 & 696.853 & 2001.876 & 2747.522 & 3545.997 & 3407.108 & 3338.105 & 2920.001 & 1737.591 & 980.902 & 625.401\\
      0.10-0.20 & 2001.876 & 6112.805 & 8815.083 & 11833.942 & 12153.920 & 12237.098 & 10720.256 & 6277.406 & 2772.537 & 1624.353\\
      0.20-0.30 & 2747.522 & 8815.083 & 14570.982 & 20276.742 & 22969.359 & 23781.402 & 20758.378 & 12783.896 & 5341.020 & 2443.439\\
      0.30-0.40 & 3545.997 & 11833.942 & 20276.742 & 30440.336 & 35438.503 & 37883.971 & 34188.100 & 21861.375 & 9294.893 & 4766.429\\
      0.40-0.50 & 3407.108 & 12153.920 & 22969.359 & 35438.503 & 47187.230 & 48969.493 & 46266.258 & 31437.600 & 12411.100 & 6938.221\\
      0.50-0.60 & 3338.105 & 12237.098 & 23781.402 & 37883.971 & 48969.493 & 55038.110 & 52273.870 & 36730.434 & 16108.935 & 8909.019\\
      0.60-0.80 & 2920.001 & 10720.256 & 20758.378 & 34188.100 & 46266.258 & 52273.870 & 54946.537 & 42295.359 & 21141.665 & 13756.200\\
      0.80-1.00 & 1737.591 & 6277.406 & 12783.896 & 21861.375 & 31437.600 & 36730.434 & 42295.359 & 38922.008 & 24035.335 & 15968.648\\
      1.00-1.25 & 980.902 & 2772.537 & 5341.020 & 9294.893 & 12411.100 & 16108.935 & 21141.665 & 24035.335 & 19832.334 & 13108.243\\
      1.25-1.50 & 625.401 & 1624.353 & 2443.439 & 4766.429 & 6938.221 & 8909.019 & 13756.200 & 15968.648 & 13108.243 & 10246.530
    \end{tabular}
  \end{adjustbox}
  \label{tbl:syscov_ptmu_scintillator}
\end{table}
\end{landscape}
\clearpage

\begin{table}
  \centering
  \renewcommand{\arraystretch}{1.15}
  \caption{Statistical covariance matrix of measured cross section as a function of $p_{\mu,||}$, in units of $10^{-84}$ $(\cmsq/\GeVc/\mathrm{nucleon})^2$}
  \begin{adjustbox}{max width=\textwidth}
    \begin{tabular}{c||cccccccccc}
      Bin edges (\GeVc) & 1-2 & 2-3 & 3-4 & 4-5 & 5-6 & 6-8 & 8-10 & 10-15 & 15-20\\
      \hline
      \hline
      0-1 & 0.000 & 0.000 & 0.000 & 0.000 & 0.000 & 0.000 & 0.000 & 0.000 & 0.000 & 0.000\\
      1-2 & 0.000 & 38.696 & -3.116 & -0.603 & 0.045 & -0.018 & -0.021 & -0.005 & -0.002 & -0.002\\
      2-3 & 0.000 & -3.116 & 107.204 & -9.339 & -2.530 & 0.270 & 0.006 & -0.025 & -0.009 & -0.004\\
      3-4 & 0.000 & -0.603 & -9.339 & 155.252 & -10.858 & -5.469 & 0.340 & 0.039 & -0.013 & -0.005\\
      4-5 & 0.000 & 0.045 & -2.530 & -10.858 & 168.229 & -6.122 & -4.065 & 0.199 & 0.015 & -0.010\\
      5-6 & 0.000 & -0.018 & 0.270 & -5.469 & -6.122 & 129.927 & -3.285 & -1.757 & 0.031 & 0.007\\
      6-8 & 0.000 & -0.021 & 0.006 & 0.340 & -4.065 & -3.285 & 30.614 & -0.418 & -0.253 & 0.012\\
      8-10 & 0.000 & -0.005 & -0.025 & 0.039 & 0.199 & -1.757 & -0.418 & 7.549 & -0.028 & -0.068\\
      10-15 & 0.000 & -0.002 & -0.009 & -0.013 & 0.015 & 0.031 & -0.253 & -0.028 & 0.811 & -0.037\\
      15-20 & 0.000 & -0.002 & -0.004 & -0.005 & -0.010 & 0.007 & 0.012 & -0.068 & -0.037 & 0.433
    \end{tabular}
  \end{adjustbox}
  \label{tbl:statcov_pzmu_scintillator}
\end{table}

\begin{table}
  \centering
  \renewcommand{\arraystretch}{1.15}
  \caption{Systematic covariance matrix of measured cross section as a function of $p_{\mu,||}$, in units of $10^{-84}$ $(\cmsq/\GeVc/\mathrm{nucleon})^2$}
  \begin{adjustbox}{max width=\textwidth}
    \begin{tabular}{c||cccccccccc}
      Bin edges (\GeVc) & 0-1 & 1-2 & 2-3 & 3-4 & 4-5 & 5-6 & 6-8 & 8-10 & 10-15 & 15-20\\
      \hline
      \hline
      0-1 & 0.000 & 0.000 & 0.000 & 0.000 & 0.000 & 0.000 & 0.000 & 0.000 & 0.000 & 0.000\\
      1-2 & 0.000 & 37.753 & 125.944 & 177.970 & 177.261 & 122.463 & 35.682 & 6.024 & 1.690 & 1.818\\
      2-3 & 0.000 & 125.944 & 639.794 & 983.062 & 965.408 & 741.762 & 333.378 & 60.996 & 14.475 & 8.102\\
      3-4 & 0.000 & 177.970 & 983.062 & 1576.449 & 1574.526 & 1250.471 & 591.070 & 109.569 & 26.231 & 14.337\\
      4-5 & 0.000 & 177.261 & 965.408 & 1574.526 & 1709.123 & 1431.908 & 701.918 & 135.031 & 32.611 & 18.983\\
      5-6 & 0.000 & 122.463 & 741.762 & 1250.471 & 1431.908 & 1291.766 & 695.474 & 141.044 & 32.914 & 17.480\\
      6-8 & 0.000 & 35.682 & 333.378 & 591.070 & 701.918 & 695.474 & 450.583 & 98.792 & 21.269 & 8.802\\
      8-10 & 0.000 & 6.024 & 60.996 & 109.569 & 135.031 & 141.044 & 98.792 & 23.817 & 4.929 & 1.823\\
      10-15 & 0.000 & 1.690 & 14.475 & 26.231 & 32.611 & 32.914 & 21.269 & 4.929 & 1.196 & 0.489\\
      15-20 & 0.000 & 1.818 & 8.102 & 14.337 & 18.983 & 17.480 & 8.802 & 1.823 & 0.489 & 0.424
    \end{tabular}
  \end{adjustbox}
  \label{tbl:syscov_pzmu_scintillator}
\end{table}

\pagebreak
\begin{landscape}
\begin{table}
  \centering
  \renewcommand{\arraystretch}{1.15}
  \caption{Statistical covariance matrix of measured cross section as a function of $Q^2$, in units of $10^{-84}$ $(\cmsq/\GeVsqcsq/\mathrm{nucleon})^2$}
  \begin{adjustbox}{max width=1.3\textheight}
    \begin{tabular}{c||cccccccccccc}
      Bin edges (\GeVsqcsq) & 0.000-0.025 & 0.025-0.050 & 0.050-0.100 & 0.100-0.200 & 0.200-0.300 & 0.300-0.400 & 0.400-0.500 & 0.500-0.700 & 0.700-1.000 & 1.000-1.300 & 1.300-2.000 & 2.000-3.000\\
      \hline
      \hline
      0.000-0.025 & 20299.500 & -1638.420 & -762.736 & 139.176 & 32.707 & -5.631 & -4.562 & -0.565 & -0.035 & 0.104 & -0.029 & -0.003\\
      0.025-0.050 & -1638.420 & 37611.060 & -422.985 & -681.221 & 118.006 & 59.282 & 5.550 & -2.878 & -0.884 & -0.260 & 0.284 & -0.036\\
      0.050-0.100 & -762.736 & -422.985 & 18064.868 & -559.215 & -459.842 & 71.507 & 55.678 & 7.711 & -1.075 & -0.557 & -0.084 & 0.352\\
      0.100-0.200 & 139.176 & -681.221 & -559.215 & 9441.833 & -294.226 & -564.187 & -61.741 & 37.035 & 6.760 & 0.080 & -0.381 & -1.016\\
      0.200-0.300 & 32.707 & 118.006 & -459.842 & -294.226 & 11163.797 & 327.543 & -654.435 & -168.251 & 19.621 & 6.778 & 1.134 & 1.180\\
      0.300-0.400 & -5.631 & 59.282 & 71.507 & -564.187 & 327.543 & 10064.136 & 814.258 & -434.477 & -102.997 & 11.796 & 2.370 & -1.567\\
      0.400-0.500 & -4.562 & 5.550 & 55.678 & -61.741 & -654.435 & 814.258 & 8398.012 & 349.019 & -266.213 & -49.034 & 1.440 & 0.137\\
      0.500-0.700 & -0.565 & -2.878 & 7.711 & 37.035 & -168.251 & -434.477 & 349.019 & 4565.684 & 25.471 & -159.017 & -26.040 & -3.186\\
      0.700-1.000 & -0.035 & -0.884 & -1.075 & 6.760 & 19.621 & -102.997 & -266.213 & 25.471 & 2389.509 & 92.030 & -69.423 & -20.023\\
      1.000-1.300 & 0.104 & -0.260 & -0.557 & 0.080 & 6.778 & 11.796 & -49.034 & -159.017 & 92.030 & 1561.242 & 45.567 & -35.555\\
      1.300-2.000 & -0.029 & 0.284 & -0.084 & -0.381 & 1.134 & 2.370 & 1.440 & -26.040 & -69.423 & 45.567 & 445.043 & 8.608\\
      2.000-3.000 & -0.003 & -0.036 & 0.352 & -1.016 & 1.180 & -1.567 & 0.137 & -3.186 & -20.023 & -35.555 & 8.608 & 273.137
    \end{tabular}
  \end{adjustbox}
  \label{tbl:statcov_q2_scintillator}
\end{table}
\end{landscape}
\clearpage

\pagebreak
\begin{landscape}
\begin{table}
  \centering
  \renewcommand{\arraystretch}{1.15}
  \caption{Systematic covariance matrix of measured cross section as a function of $Q^2$, in units of $10^{-84}$ $(\cmsq/\GeVsqcsq/\mathrm{nucleon})^2$}
  \begin{adjustbox}{max width=1.3\textheight}
    \begin{tabular}{c||cccccccccccc}
      Bin edges (\GeVsqcsq) & 0.000-0.025 & 0.025-0.050 & 0.050-0.100 & 0.100-0.200 & 0.200-0.300 & 0.300-0.400 & 0.400-0.500 & 0.500-0.700 & 0.700-1.000 & 1.000-1.300 & 1.300-2.000 & 2.000-3.000\\
      \hline
      \hline
      0.000-0.025 & 54579.490 & 51824.702 & 44495.975 & 39165.930 & 29233.638 & 24650.321 & 17356.139 & 15343.562 & 8657.449 & 5749.085 & 2263.147 & 952.220\\
      0.025-0.050 & 51824.702 & 55427.561 & 47653.409 & 46311.401 & 36674.476 & 31910.329 & 24117.018 & 20408.037 & 12292.897 & 6817.861 & 2831.073 & 1192.574\\
      0.050-0.100 & 44495.975 & 47653.409 & 49587.025 & 44921.119 & 37551.407 & 34006.807 & 24077.594 & 20540.552 & 12042.028 & 6580.033 & 2374.924 & 669.456\\
      0.100-0.200 & 39165.930 & 46311.401 & 44921.119 & 47192.402 & 41403.993 & 37033.782 & 29025.137 & 24827.192 & 16051.786 & 8202.527 & 3761.731 & 1551.915\\
      0.200-0.300 & 29233.638 & 36674.476 & 37551.407 & 41403.993 & 42086.849 & 35820.630 & 30170.754 & 25166.509 & 17242.479 & 8323.571 & 3830.093 & 1707.597\\
      0.300-0.400 & 24650.321 & 31910.329 & 34006.807 & 37033.782 & 35820.630 & 34416.030 & 26053.298 & 23852.695 & 16087.347 & 8574.149 & 4103.922 & 1690.413\\
      0.400-0.500 & 17356.139 & 24117.018 & 24077.594 & 29025.137 & 30170.754 & 26053.298 & 23423.629 & 19320.257 & 13940.740 & 6821.947 & 3595.563 & 1603.606\\
      0.500-0.700 & 15343.562 & 20408.037 & 20540.552 & 24827.192 & 25166.509 & 23852.695 & 19320.257 & 19288.435 & 14094.099 & 8023.453 & 4578.884 & 2104.599\\
      0.700-1.000 & 8657.449 & 12292.897 & 12042.028 & 16051.786 & 17242.479 & 16087.347 & 13940.740 & 14094.099 & 11220.843 & 6490.984 & 4104.328 & 1945.594\\
      1.000-1.300 & 5749.085 & 6817.861 & 6580.033 & 8202.527 & 8323.571 & 8574.149 & 6821.947 & 8023.453 & 6490.984 & 4845.674 & 3040.244 & 1491.591\\
      1.300-2.000 & 2263.147 & 2831.073 & 2374.924 & 3761.731 & 3830.093 & 4103.922 & 3595.563 & 4578.884 & 4104.328 & 3040.244 & 2221.855 & 1094.444\\
      2.000-3.000 & 952.220 & 1192.574 & 669.456 & 1551.915 & 1707.597 & 1690.413 & 1603.606 & 2104.599 & 1945.594 & 1491.591 & 1094.444 & 613.832
    \end{tabular}
  \end{adjustbox}
  \label{tbl:syscov_q2_scintillator}
\end{table}
\end{landscape}
\clearpage

\begin{table}
  \centering
  \renewcommand{\arraystretch}{1.15}
  \caption{Statistical covariance matrix of measured cross section as a function of $\theta_{\mu}$, in units of $10^{-84}$ $(\cmsq/degree/\mathrm{nucleon})^2$}
  \begin{adjustbox}{max width=\textwidth}
    \begin{tabular}{c||ccccccccccccccc}
      Bin edges (degree) & 0-1 & 1-2 & 2-3 & 3-4 & 4-5 & 5-6 & 6-7 & 7-8 & 8-9 & 9-10 & 10-11 & 11-12 & 12-14 & 14-16 & 16-20\\
      \hline
      \hline
      0-1 & 5.419 & -0.919 & -0.113 & 0.097 & -0.008 & -0.003 & 0.000 & 0.000 & 0.000 & 0.000 & 0.000 & 0.000 & 0.000 & 0.000 & 0.000\\
      1-2 & -0.919 & 18.342 & -1.851 & -0.224 & 0.175 & -0.014 & -0.005 & 0.001 & 0.000 & 0.000 & 0.000 & 0.000 & 0.000 & 0.000 & 0.000\\
      2-3 & -0.113 & -1.851 & 31.636 & -2.729 & -0.350 & 0.256 & -0.018 & -0.010 & 0.001 & 0.000 & 0.000 & 0.000 & 0.000 & 0.000 & -0.001\\
      3-4 & 0.097 & -0.224 & -2.729 & 42.097 & -3.571 & -0.520 & 0.314 & -0.020 & -0.012 & 0.001 & 0.000 & 0.000 & 0.000 & 0.000 & 0.000\\
      4-5 & -0.008 & 0.175 & -0.350 & -3.571 & 53.498 & -4.264 & -0.656 & 0.399 & -0.014 & -0.015 & 0.000 & 0.000 & 0.000 & -0.002 & 0.000\\
      5-6 & -0.003 & -0.014 & 0.256 & -0.520 & -4.264 & 60.081 & -4.386 & -0.862 & 0.409 & -0.002 & -0.018 & 0.000 & -0.001 & 0.000 & 0.000\\
      6-7 & 0.000 & -0.005 & -0.018 & 0.314 & -0.656 & -4.386 & 61.006 & -4.864 & -0.890 & 0.428 & 0.002 & -0.019 & -0.003 & 0.001 & 0.000\\
      7-8 & 0.000 & 0.001 & -0.010 & -0.020 & 0.399 & -0.862 & -4.864 & 66.797 & -4.664 & -1.093 & 0.505 & 0.018 & -0.014 & 0.000 & 0.000\\
      8-9 & 0.000 & 0.000 & 0.001 & -0.012 & -0.014 & 0.409 & -0.890 & -4.664 & 62.892 & -4.402 & -1.390 & 0.514 & 0.007 & -0.008 & 0.000\\
      9-10 & 0.000 & 0.000 & 0.000 & 0.001 & -0.015 & -0.002 & 0.428 & -1.093 & -4.402 & 64.722 & -4.527 & -1.673 & 0.325 & -0.005 & -0.004\\
      10-11 & 0.000 & 0.000 & 0.000 & 0.000 & 0.000 & -0.018 & 0.002 & 0.505 & -1.390 & -4.527 & 62.300 & -4.030 & -0.658 & 0.196 & -0.009\\
      11-12 & 0.000 & 0.000 & 0.000 & 0.000 & 0.000 & 0.000 & -0.019 & 0.018 & 0.514 & -1.673 & -4.030 & 62.001 & -3.073 & -0.135 & 0.072\\
      12-14 & 0.000 & 0.000 & 0.000 & 0.000 & 0.000 & -0.001 & -0.003 & -0.014 & 0.007 & 0.325 & -0.658 & -3.073 & 27.615 & -1.861 & 0.043\\
      14-16 & 0.000 & 0.000 & 0.000 & 0.000 & -0.002 & 0.000 & 0.001 & 0.000 & -0.008 & -0.005 & 0.196 & -0.135 & -1.861 & 30.059 & -1.496\\
      16-20 & 0.000 & 0.000 & -0.001 & 0.000 & 0.000 & 0.000 & 0.000 & 0.000 & 0.000 & -0.004 & -0.009 & 0.072 & 0.043 & -1.496 & 16.002
    \end{tabular}
  \end{adjustbox}
  \label{tbl:statcov_thetamu_deg_scintillator}
\end{table}

\begin{table}
  \centering
  \renewcommand{\arraystretch}{1.15}
  \caption{Systematic covariance matrix of measured cross section as a function of $\theta_{\mu}$, in units of $10^{-84}$ $(\cmsq/degree/\mathrm{nucleon})^2$}
  \begin{adjustbox}{max width=\textwidth}
    \begin{tabular}{c||ccccccccccccccc}
      Bin edges (degree) & 0-1 & 1-2 & 2-3 & 3-4 & 4-5 & 5-6 & 6-7 & 7-8 & 8-9 & 9-10 & 10-11 & 11-12 & 12-14 & 14-16 & 16-20\\
      \hline
      \hline
      0-1 & 8.254 & 19.794 & 28.715 & 30.238 & 31.369 & 28.338 & 25.117 & 25.939 & 19.941 & 17.580 & 15.856 & 12.465 & 13.396 & 11.435 & 7.590\\
      1-2 & 19.794 & 57.898 & 82.888 & 97.521 & 99.589 & 100.838 & 92.214 & 86.960 & 76.284 & 66.123 & 58.684 & 50.616 & 44.467 & 33.909 & 27.073\\
      2-3 & 28.715 & 82.888 & 128.568 & 151.948 & 164.792 & 165.935 & 151.200 & 144.875 & 126.304 & 112.346 & 98.127 & 81.245 & 74.270 & 56.663 & 39.564\\
      3-4 & 30.238 & 97.521 & 151.948 & 199.158 & 217.048 & 230.289 & 210.985 & 202.842 & 183.248 & 164.952 & 143.281 & 118.760 & 101.311 & 75.215 & 54.277\\
      4-5 & 31.369 & 99.589 & 164.792 & 217.048 & 255.361 & 267.786 & 246.888 & 239.434 & 215.350 & 199.952 & 169.593 & 133.233 & 118.808 & 90.542 & 60.155\\
      5-6 & 28.338 & 100.838 & 165.935 & 230.289 & 267.786 & 296.698 & 272.710 & 260.050 & 243.745 & 226.339 & 189.592 & 153.554 & 130.240 & 97.035 & 70.814\\
      6-7 & 25.117 & 92.214 & 151.200 & 210.985 & 246.888 & 272.710 & 272.328 & 246.434 & 241.702 & 225.125 & 193.539 & 157.076 & 134.876 & 101.724 & 78.209\\
      7-8 & 25.939 & 86.960 & 144.875 & 202.842 & 239.434 & 260.050 & 246.434 & 250.930 & 229.667 & 215.860 & 187.274 & 149.822 & 133.096 & 103.248 & 72.891\\
      8-9 & 19.941 & 76.284 & 126.304 & 183.248 & 215.350 & 243.745 & 241.702 & 229.667 & 227.731 & 212.034 & 184.319 & 151.740 & 129.502 & 97.594 & 75.432\\
      9-10 & 17.580 & 66.123 & 112.346 & 164.952 & 199.952 & 226.339 & 225.125 & 215.860 & 212.034 & 205.163 & 174.519 & 142.998 & 124.395 & 96.578 & 73.343\\
      10-11 & 15.856 & 58.684 & 98.127 & 143.281 & 169.593 & 189.592 & 193.539 & 187.274 & 184.319 & 174.519 & 157.915 & 127.750 & 111.485 & 87.554 & 64.111\\
      11-12 & 12.465 & 50.616 & 81.245 & 118.760 & 133.233 & 153.554 & 157.076 & 149.822 & 151.740 & 142.998 & 127.750 & 115.074 & 95.208 & 72.036 & 58.750\\
      12-14 & 13.396 & 44.467 & 74.270 & 101.311 & 118.808 & 130.240 & 134.876 & 133.096 & 129.502 & 124.395 & 111.485 & 95.208 & 86.696 & 68.253 & 52.839\\
      14-16 & 11.435 & 33.909 & 56.663 & 75.215 & 90.542 & 97.035 & 101.724 & 103.248 & 97.594 & 96.578 & 87.554 & 72.036 & 68.253 & 60.592 & 43.967\\
      16-20 & 7.590 & 27.073 & 39.564 & 54.277 & 60.155 & 70.814 & 78.209 & 72.891 & 75.432 & 73.343 & 64.111 & 58.750 & 52.839 & 43.967 & 42.846
    \end{tabular}
  \end{adjustbox}
  \label{tbl:syscov_thetamu_deg_scintillator}
\end{table}